\documentclass[a4paper,11pt]{article}
\pdfoutput=1 % if your are submitting a pdflatex (i.e. if you have
\usepackage{jcappub} % for details on the use of the package, please
\usepackage[T1]{fontenc} % if needed
\usepackage{multirow}
\usepackage{makecell}

\title{\boldmath Optimizing spectral-matching component separation for measuring CMB B-mode polarization with a satellite mission}

\author[a,1]{Hoang Viet Tran,\note{Corresponding author.}}
\author[b,c,d]{Guillaume Patanchon,}
\author[a,e]{Michele Citran,}
\author[a]{and Benjamin Beringue.}

\affiliation[a]{Université Paris Cité, CNRS, Astroparticule et Cosmologie, F-75013 Paris, France}
\affiliation[b]{ILANCE, CNRS – University of Tokyo International Research Laboratory, Kashiwa, Chiba 277-
8582, Japan}
\affiliation[c]{Université Paris Cité, F-75006 Paris, France}
\affiliation[d]{Kavli Institute for the Physics and Mathematics of the Universe (Kavli IPMU, WPI), UTIAS, The University of Tokyo, Kashiwa, Chiba 277-8583, Japan}
\affiliation[e]{Université Paris-Saclay, CNRS/IN2P3, IJCLab, 91405 Orsay, France}
\emailAdd{tran@apc.in2p3.fr}
\abstract{
The measurement of primordial CMB B-mode polarization is fundamentally limited by our ability to separate this faint signal from bright and complex Galactic foregrounds. Spectral Matching Independent Component Analysis (SMICA) offers a flexible, semi-blind approach to this foreground-cleaning problem. However, its standard implementation relies on a physically unrealistic assumption of non-spatially varying foreground emission, reducing the efficiency of the method. In this work, we develop two complementary strategies to address this limitation at the component-separation level. First, we augment SMICA with effective foreground components beyond the physical dust and synchrotron degrees of freedom. Second, we extend SMICA to operate independently on subsets of the sky, better capturing local variations in the foreground emission. Beyond component separation, we further explore a likelihood-level marginalization procedure using internally reconstructed foreground templates. We demonstrate the performance of the pipeline on simulated LiteBIRD-like observations, using foreground models of increasing complexity based on \texttt{d1s1} and \texttt{d10s5}. Combining our component-separation improvements with template marginalization brings the recovered tensor-to-scalar ratio $r$ to effectively unbiased levels, with statistical uncertainties on the order of $10^{-3}$ or below.
}

\begin{document}
\maketitle
\flushbottom

\section{Introduction}
\label{sec:intro}

The cosmic microwave background (CMB) provides a unique snapshot of the Universe at the time of recombination and constitutes one of the most powerful probes of fundamental cosmology. In addition to its temperature anisotropies, the CMB is linearly polarized, and this polarization field can be decomposed into parity-even $E$-modes and parity-odd $B$-modes. The detection of primordial $B$-mode polarization in the cosmic microwave background (CMB) remains one of the frontiers of modern cosmology. Such a detection would provide compelling evidence for a cosmological inflationary epoch, while a measurement of the tensor-to-scalar ratio $r$, defined as the ratio of the primordial tensor and scalar perturbation amplitudes, would offer a unique observational window into the physics of the primordial Universe. The main challenge lies in the extremely small amplitude of the primordial B-mode signal, which is expected to be of the order of nanokelvin or below. The current strongest observational constraint is $r\le0.032$ at the 95\% confidence level with the reanalysis of BICEP and Planck data~\cite{Tristram_2022}. Motivated by the prospect of probing new physics, a broad experimental effort has been established to look for this B-mode polarization. This includes ground-based observatories such as BICEP/Keck~\cite{BICEPKeck_2021}, the Simons Observatory\cite{SO_2019}, SPT-3G~\cite{SPT3G_2014} and ACT~\cite{ACT_2020}, balloon-borne experiments such as SPIDER~\cite{Fraisse_2013} and LSPE-SWIPE~\cite{deBernardis_2012}, and satellite missions including the forthcoming LiteBIRD~\cite{LiteBIRD_2022}.\\
\indent The extremely low amplitude of the primordial B-mode signal imposes stringent requirements on both instrumental sensitivity and the control of systematic effects. Achieving such precision requires not only detectors with unprecedented sensitivity but also accurate mitigation of instrumental and astrophysical contaminants. Among these, polarized Galactic foreground emission constitutes one of the principal obstacles to a robust measurement of primordial B-modes. Thermal dust and synchrotron emission dominate the microwave sky over a broad range of frequencies and angular scales, exceeding this faint cosmological signal by several orders of magnitude. If not properly accounted for, residual foreground contamination can lead to significant biases in the inferred tensor-to-scalar ratio, especially if we're targeting the large-scale polarization.\\
\indent To address the challenge of foreground contamination, a broad class of techniques collectively known as component separation has been developed. Multi-frequency observations by the Planck satellite revealed the complex and spatially varying nature of foreground emissions, which, together with the increasing sensitivity of CMB polarization experiments, places increasingly stringent requirements on their separation from the CMB. These challenges make component separation an active area of research. Existing approaches can be broadly categorized into three families: blind, parametric, and template-based methods. Blind methods, such as NILC\cite{Delabrouille_2008, Carones_2023}, SMICA\cite{Delabrouille_2003, Cardoso_2008} or Minimally-Informed\cite{Leloup_2023, Morshed_2024}, exploit the statistical properties of multifrequency observations while making minimal assumptions about the foreground emission. Parametric methods, including Commander\cite{Eriksen_2004, Eriksen_2008}, FGBuster\cite{Stompor_2009, Rizzieri_2026}, and Moment expansion\cite{Chluba_2017, Vacher_2023}, explicitly model the frequency dependence of the foregrounds through a set of physical parameters that are fitted to the data. Finally, template-based approaches, such as SEVEM\cite{MartinezGonzalez_2003} or Delta-map\cite{Minami_2023}, construct internal foreground templates from multifrequency observations and subtract them from the data. Each of these approaches offers distinct advantages and limitations, reflecting different compromises between model dependence, flexibility, and robustness against foreground complexity.\\
\indent SMICA, one of the successful reference methods of component separation during the Planck experiment\cite{Planck2018_IV}, is peculiar for its flexibility. Despite generally being classified as blind, the method, in fact, does model the foreground explicitly. This allows it to "customize" its degree of blindness, from fully relying on statistical properties to relying on a parametric model or somewhere in between. This work will focus on adapting SMICA to these new contexts of polarization measurement. However, SMICA in its simplest form assumes that the emission of each component is constant across the sky, which is inadequate for realistic foreground models exhibiting small but significant spatial variability in their frequency dependence. The resulting mismatch from this assumption leads to residuals that cannot be removed simply by increasing observational sensitivity. Therefore, we investigate a strategy to negate this main limitation. We increase the effective dimensionality of the foreground covariance and incorporate additional effective components beyond the underlying physical emissions. Also, we extend SMICA to operate independently on subsets of the sky, enabling the method to adapt to spatially varying emission while preserving the statistical framework of covariance matching. The resulting local component estimates by SMICA are subsequently combined into full-sky maps with linear combinations.\\
\indent We assess the performance of this framework using simulated observations from a LiteBIRD-like~\cite{LiteBIRD_2022} mission containing CMB, instrumental noise, and multiple PySM~\cite{PySM4_2026} polarized Galactic foreground models of increasing complexity. The impact of the proposed method is quantified both in terms of the residuals and their propagation into the cosmological likelihood for the tensor-to-scalar ratio. This includes methodological biases as well as those induced by foreground complexity.\\
\indent The paper is organized as follows. Section \ref{sec:fgs_model} introduces the astrophysical foregrounds that contaminate our observation of the CMB. Section \ref{sec:comp_sep} presents the proposed component separation pipeline in detail, from the core formalism of SMICA, the possible augmentations to the method, and the steps to measuring the tensor-to-scalar ratio. Section \ref{sec:results} shows the stage-by-stage improvement in performance of the method and highlights its main results on simulated LiteBIRD-like~\cite{LiteBIRD_2022} observations. Additional points are discussed in Section \ref{sec:discussions}, including the maximum-likelihood approach in relation to the proposed pipeline and a possible physical interpretation of extra components in SMICA. Finally, Section \ref{sec:conclusions} summarizes the main findings and concludes the paper.

\section{The polarized astrophysical foregrounds}\label{sec:fgs_model}
% \begin{table}
% \centering
% \begin{tabular}{| p{0.1\textwidth} | p{0.35\textwidth} | p{0.35\textwidth}|}
% \hline
% Model & Frequency Scaling & Emission Template \\
% \hline
% d0 &a
% Fixed spectral index $\beta_d=1.54$ and dust temperature $T_d=20\,\mathrm{K}$ across the sky.
% &
% Planck-based templates: intensity at 545\,GHz and polarization at 353\,GHz. Polarization maps include synthetic small-scale fluctuations.
% \\
% \hline
% d1 &
% Spatially varying $\beta_d(\mathbf{\hat{p}})$ and $T_d(\mathbf{\hat{p}})$ from the Planck Commander analysis.
% &
% Similar to d0
% \\
% \hline
% d10 &
% Spatially varying $\beta_d(\mathbf{\hat{p}})$ and $T_d(\mathbf{\hat{p}})$ based on Planck 2018 GNILC products.
% &
% GNILC dust templates with reduced CIB contamination and high-resolution small-scale structure.
% \\
% \hline
% s0 &
% Fixed synchrotron spectral index $\beta_s=-3.0$ over the entire sky.
% &
% Intensity template from the Haslam 408\,MHz map and polarization templates from WMAP 23\,GHz $Q/U$ maps. Includes added small-scale fluctuations.
% \\
% \hline
% s1 &
% Spatially varying spectral index $\beta_s(\mathbf{\hat{p}})$, derived from Haslam 408\,MHz and WMAP 23\,GHz observations.
% &
% Same as s0
% \\
% \hline
% s5 &
% Spatially varying $\beta_s(\mathbf{\hat{p}})$, enhanced using S-PASS information and additional small-scale fluctuations.
% &
% Same large-scale templates as s1 supplemented with high-resolution structure generated using the \texttt{logpoltens} formalism.
% \\
% \hline
% \end{tabular}
% \caption{Summary of PySM thermal dust and synchrotron models.}
% \label{tab:fgs_models}
% \end{table}
\indent Before introducing the component-separation framework, we first describe the polarized astrophysical foregrounds that constitute the main source of contamination in CMB polarization measurements. Thermal dust emission and synchrotron radiation are the two dominant polarized foreground components over the frequency range relevant for CMB experiments. Several other effects also contribute to observations in the microwave range, including free--free emission, anomalous microwave emission (AME), and CO molecular lines. Free--free emission is effectively unpolarized, while AME and CO lines are expected to have only weak polarized contributions~\cite{Planck_2013_XIII, Planck_2015_XXV}.\\
\indent The polarized thermal dust emission is commonly modeled by an emission template
$s_{d,\nu_0}(\hat{\mathbf{p}})$ at a reference frequency $\nu_0$, scaled to different
frequencies by a modified blackbody (MBB) law with spectral index
$\beta_d(\hat{\mathbf{p}})$ and dust temperature $T_d(\hat{\mathbf{p}})$:
\begin{equation}\label{eq:dust_model}
    y_{d,\nu}(\hat{\mathbf{p}},\beta_d,T_d)
    =
    s_{d,\nu_0}(\hat{\mathbf{p}})
    \left(\frac{\nu}{\nu_0}\right)^{\beta_d(\hat{\mathbf{p}})}
    \frac{\mathrm{B}_{\nu}(T_d(\hat{\mathbf{p}}))}
    {\mathrm{B}_{\nu_0}(T_d(\hat{\mathbf{p}}))}.
\end{equation}
Here, $\mathrm{B}_\nu(T)$ is the blackbody spectrum at frequency $\nu$ and temperature
$T$. Similarly, the synchrotron component is characterized by a power law of a single parameter $\beta_s (\mathbf{\hat{p}})$ for frequency scaling of an emission template $ s_{s, \nu_0} (\mathbf{\hat{p}})$ at reference frequency $\nu_0$:
\begin{equation}\label{eq:sync_model}
    y_{s, \nu}(\mathbf{\hat{p}}, \beta_s) = s_{s, \nu_0}(\mathbf{\hat{p}})  \bigg( \dfrac{\nu}{\nu_0}\bigg)^{\beta_s(\mathbf{\hat{p}})}.
\end{equation}
For both components, $\hat{\mathbf{p}}$ denotes the sky direction, emphasizing that the spectral parameters can vary across the sky. This is particularly important for component separation: as these parameters vary, the same physical foreground component can exhibit different spectral energy distributions (SEDs) in different Galactic regions. Moreover, these laws should be regarded as approximate descriptions of the physical emissions. Synchrotron spectra may exhibit spectral curvature, while the dust SED can depart from a single modified blackbody; these are owing to additional physical effects or variations along the line of sight~\cite{Pelgrims_2021, Kogut_2012}. Such spatial and spectral complexity introduces additional complications to the separation of foregrounds from the CMB.\\
\indent Given their dominant contribution to polarized emission, only thermal dust and synchrotron radiation are considered throughout this work. The simulated foregrounds are generated with the \texttt{PySM} package~\cite{PySM4_2026}, which provides a range of foreground models constructed from observationally motivated templates and spectral properties. We consider three pairs of dust and synchrotron models, denoted by \texttt{dx} and \texttt{sx}, respectively,\footnote{The specific properties of each \texttt{PySM} dust and synchrotron model are described at \url{https://pysm3.readthedocs.io/en/latest/models.html}.} spanning increasing levels of foreground complexity:
\begin{itemize}
\item \texttt{d0s0}: A simplified foreground model with fixed spectral parameters across the sky. This case is primarily used to verify and validate the component-separation pipeline under idealized conditions. This model is already disfavoured by polarization observations.
\item \texttt{d1s1}: A more realistic foreground model incorporating spatially varying spectral indices and emission templates derived from observational data. This case provides a representative test of the pipeline under moderate foreground complexity.
\item \texttt{d10s5}: A more complex foreground model featuring advanced dust and synchrotron templates, spatially varying spectral parameters, and additional small-scale fluctuations. This case is designed to assess the robustness of the pipeline under more challenging foreground conditions.
\end{itemize}
\section{The framework for component separation}\label{sec:comp_sep}
\subsection{Spectral Matching Independent Component Analysis}\label{subsec:smica}
\textbf{S}pectral \textbf{M}atching \textbf{I}ndependent \textbf{C}omponent \textbf{A}nalysis (\textbf{SMICA})~\cite{Delabrouille_2003, Cardoso_2008} was one of the reference methods of component separation for the Planck~\cite{Planck2018_IV} mission. SMICA jointly estimates the frequency emission laws and the angular power spectra of statistically independent sky components by exploiting their spectral diversity (non-proportionality of power spectra) across the available frequency channels. In this framework, the observation of the sky at a certain frequency $\nu$ can be expressed as a noisy linear combination of $N_c$ components:
\begin{equation}
    y_\nu(\mathbf{\hat{p}}) = \sum_{j=1}^{N_c}A_{\nu j}s_j(\mathbf{\hat{p}}) + n_{\nu}(\mathbf{\hat{p}}),
\end{equation}
with $s_j$ being the emission template of the $j^{th}$ component. The coefficient $A_{\nu j}$ encodes the emission laws, while $n_\nu$ represents the noise. Suppose we observe the polarized sky at $N_f$ frequencies; the model can be written in a condensed matrix formalism in real space as:
\begin{equation}
\label{eq:linmod_real}
    \mathbf{y}(\mathbf{\hat{p}}) = \mathbf{A}\mathbf{s}(\mathbf{\hat{p}}) + \mathbf{n}(\mathbf{\hat{p}}),
\end{equation}
where $\mathbf{y}$ represents the observed $Q/U$ polarization maps (or scalar $E/B$ maps), $\mathbf{s}$ the underlying polarized components, and $\mathbf{n}$ the instrumental noise maps. The matrix $\mathbf{A}$ has dimensions $N_f \times N_c$ (number of frequencies times number of components), and describes the frequency response of each component, commonly referred to as the mixing matrix. In harmonic space, the same model applies to the polarization modes $X\in\{E,B\}$:
\begin{equation}
    \mathbf{y}^{X}_{\ell m} = \mathbf{A}\mathbf{s}^{X}_{\ell m} + \mathbf{n}^{X}_{\ell m},
    \label{eq:linmod}
\end{equation}
The same mixing matrix applies to both $E$- and $B$-modes since it describes only the frequency dependence of the emission, which is independent of the harmonic mode. We further assume that the instrumental noise properties are identical for the different polarization modes, allowing the same noise to be used for both $E$- and $B$-mode analyses. This linear model is shared by most component separation methods in the literature.\\
\indent From this point onward, we consider a single polarization mode $X$ independently and work with its corresponding auto-spectral density matrix $XX$, rather than the full polarization covariance involving $EE$, $EB$, and $BB$. The motivation for restricting the analysis to a single mode will be discussed in a later section. Under the first assumption of SMICA, namely the Gaussianity of the process $\mathbf{y}$, all the statistical information is contained in the $N_f \times N_f$ auto-spectral density matrix of the selected mode $X \in \{E,B\}$:
\begin{equation}
    \mathbf{R}^{XX}_{\mathbf{Y}}(q) = \dfrac{1}{n_q}\sum_{\ell \in D_q}\sum_m
    \left\langle \mathbf{y}_{\ell m}^{X}\mathbf{y}_{\ell m}^{X,\dagger}\right\rangle .
\end{equation}
This $\mathbf{R}^{XX}_{\mathbf{Y}}(q)$ matrix is also often referred to as the model covariance matrix. Here, we have binned this covariance into $D_q$ bins, each containing $n_q$ modes. In this paper, this binning scheme is fixed to multipole intervals of width $\Delta\ell=5$, except for the first bin, which combines multipoles from $\ell=2$ to $\ell=32$\footnote{The broader first bin groups the poorly sampled largest scales, which are not accessible in the subsequent local analysis on small patches; we have verified that the estimation of $\mathbf{A}$, the main parameter used in the subsequent analysis, is relatively insensitive to this binning choice.}.\\
\indent To further define the model, another set of assumptions is implemented: the mixing matrix $\mathbf{A}$ is constant across the sky. Note that we do not impose a physical model on the mixing matrix, making SMICA relatively robust to deviations from the typically assumed foreground emission laws. Under these assumptions, using equation~\eqref{eq:linmod}, we can express the covariance matrix $\mathbf{R}^{XX}_{\mathbf{Y}}$ as follows:
\begin{equation}\label{eq:smica1}
    \mathbf{R}^{XX}_{\mathbf{Y}}(q) = \mathbf{A}\mathbf{R}^{XX}_{\mathbf{S}}(q)\mathbf{A}^T + \mathbf{R}_{\mathbf{N}}(q).
\end{equation}
Here, we have defined the source covariance $\mathbf{R}^{XX}_{\mathbf{S}}$ (dimension $N_c \times N_c$), where all components are uncorrelated, and the noise covariance $\mathbf{R}_{\mathbf{N}}$ (dimension $N_f \times N_f$) with no inter-channel correlation:
\begin{align}
    \mathbf{R}^{XX}_{\mathbf{S}}(q) = \dfrac{1}{n_q}\sum_{\vec{\ell}\in D_q} 
    \langle \mathbf{s}^{X}_{\ell m} \mathbf{s}^{X,\dagger}_{\ell m} \rangle 
    = \mathrm{diag}(C^{XX}_1(q),...,C^{XX}_{N_c}(q)),\label{eq:source_cov}\\
    \mathbf{R}_{\mathbf{N}}(q) = \dfrac{1}{n_q}\sum_{\vec{\ell}\in D_q} 
    \langle \mathbf{n}_{\ell m} \mathbf{n}^{\dagger}_{\ell m} \rangle 
    = \mathrm{diag}(\sigma_1(q)^2,...,\sigma_{N_f}(q)^2).
\end{align}
In principle, the noise model introduced above allows for one free parameter per channel and multipole bin $q$. However, since we only consider beam-deconvolved isotropic white noise (addressed later on), we assume a fixed shape of $1/b_\ell^2$ with $b_\ell$ being the respective channel beam response and fit only a single amplitude parameter for each channel. From the data, we can compute the corresponding observed set of covariance $\mathbf{\widehat{R}}^{XX}_{\mathbf{Y}}(q)$ - often denoted "empirical covariance matrices":
\begin{equation}
    \mathbf{\widehat{R}}^{XX}_{\mathbf{Y}}(q) = \dfrac{1}{n_q}\sum_{\vec{\ell}\in D_q}
    \mathbf{y}^{X}_{\ell m}\mathbf{y}^{X,\dagger}_{\ell m}.
\end{equation}
It can be proven that, under the maximum likelihood principle~\cite{Delabrouille_2003}, the best set of parameters $\theta = (\mathbf{A}, \mathbf{R}^{XX}_{\mathbf{S}}, \mathbf{R}_{\mathbf{N}})$ can be obtained by matching $\mathbf{R}^{XX}_{\mathbf{Y}}$ to $\mathbf{\widehat{R}}^{XX}_{\mathbf{Y}}(q)$ through minimizing:
\begin{equation}
    \phi(\theta) = \sum_{q=1}^Q \phi_q(\theta) = \sum_{q=1}^Q n_q D(\mathbf{\widehat{R}}^{XX}_{\mathbf{Y}},\mathbf{R}^{XX}_{\mathbf{Y}}(q,\theta)),
    \label{eq:smica_lklh}
\end{equation}
with the $D$ operator being the Kullback-Leibler divergence between two positive $n \times n$ matrices:
\begin{equation}
    D(\mathbf{R_1},\mathbf{R_2}) = \mathrm{trace}(\mathbf{R_1} \mathbf{R_2}^{-1}) - \log{\det(\mathbf{R_1} \mathbf{R_2}^{-1})} - n.
\end{equation}
This divergence is always greater than or equal to zero, with equality holding when $\mathbf{R_1} = \mathbf{R_2}$. Therefore, upon minimizing this likelihood, the model parameters are adjusted to reproduce the empirical covariance matrices across all multipole bins. \\ 
\indent Under the assumption of independent sources Eq.~\eqref{eq:source_cov}, the separation of the components in SMICA relies on their spectral diversity. If the power spectra of two or more sources are proportional across all multipole bins, the factorization $\mathbf{A}\mathbf{R_S}(q)\mathbf{A}^{T}$ is not unique: $\mathbf{A}$ and $\mathbf{R}_s(q)$ can be transformed by a $q$-independent scaling and rotation matrix $\mathbf{T}$,
\begin{equation}
\mathbf{A}\rightarrow\mathbf{A}\mathbf{T},\qquad
\mathbf{R_S}(q)\rightarrow\mathbf{T}^{-1}\mathbf{R_S}(q)\mathbf{T}^{-T},
\end{equation}
without breaking the assumed diagonality of $
\mathbf{R_S}$ or changing the model covariance $\mathbf{R_Y}$. In this case, the individual components can only be separated up to this $\mathbf{T}$ matrix. Conversely, when the source power spectra are non-proportional, as is the case between the CMB and foregrounds, such a transformation would require a rotation matrix $\mathbf{T}(q)$ that acts differently for each $q$ bin. Since SMICA assumes a $q$-independent mixing matrix $\mathbf{A}$, this transformation is not allowed, and components can therefore be uniquely identified, up to normalization and permutation ambiguities. This is the core working principle of the class of ICA methods.\\
\indent The optimized model parameters can then either be interpreted directly as the best-fit description of the sky components, which is optimal in the case of full-sky coverage, or used as intermediate quantities to construct a set of component-separation weights. In this work, as we will divide the sky into small regions, we adopt the latter approach and use the estimated mixing matrix together with the noise covariance to construct the Generalized Least-Squares (GLS) estimator~\cite{Delabrouille_2003, Tegmark_1997}, which provides an estimate of the component maps:
\begin{equation}
    \hat{\mathbf{s}}
    =
    (\mathbf{A}^{T}\mathbf{R_N}^{-1}\mathbf{A})^{-1}
    \mathbf{A}^{T}\mathbf{R_N}^{-1}\mathbf{y}
    =
    \mathbf{W}\mathbf{y}.
    \label{eq:gls}
\end{equation}
The motivation for this choice, as well as the rationale behind adopting this specific weighting scheme, will be further discussed in Section~\ref{subsec:pipeline}. Since we work with a single polarization mode $X\in\{E,B\}$, the input data $\mathbf{y}$ and reconstructed components $\hat{\mathbf{s}}$ are either expressed in harmonic space or represented as the corresponding scalar $E$- or $B$-mode maps.\\
\indent Going forward with this weighting-based formalism, for each realization $i$ of our simulations, the residual after component separation is defined as the difference between the reconstructed and true component maps,
\begin{align}
    \mathbf{m}_{\rm res}^{(i)}
    &=
    \hat{\mathbf{s}}^{(i)}
    -
    \mathbf{s}^{(i)}
    =
    \left(
        \hat{\mathbf{W}}^{(i)}\mathbf{A}
        -
        \mathbf{I}
    \right)\mathbf{s}^{(i)}
    +
    \hat{\mathbf{W}}^{(i)}\mathbf{n}^{(i)}\\
    &=
    \left(
        \overline{\mathbf{W}}\mathbf{A}
        -
        \mathbf{I}
    \right)
    \mathbf{s}^{(i)}
    +
    \overline{\mathbf{W}}\mathbf{n}^{(i)}
    +
    \delta\mathbf{W}^{(i)}\mathbf{y}^{(i)},
    \label{eq:res_map}
\end{align}
where, in the last equality, we have decomposed the estimated component-separation weights into their mean and statistical fluctuation,
\begin{equation}
    \hat{\mathbf{W}}^{(i)}
    =
    \overline{\mathbf{W}}
    +
    \delta\mathbf{W}^{(i)},
    \qquad
    \overline{\mathbf{W}}
    \equiv
    \left\langle
        \hat{\mathbf{W}}^{(i)}
    \right\rangle,
    \label{eq:W_decomposition}
\end{equation}
such that $\langle\delta\mathbf{W}^{(i)}\rangle=0$, with
$\langle\cdot\rangle$ denoting an average over realizations. The three terms
in the final line of Eq.~\eqref{eq:res_map} have distinct contributions to the
residual. The first, $(\overline{\mathbf{W}}\mathbf{A}-\mathbf{I})\mathbf{s}^{(i)}$, describes the residual that is associated with a mismatch between the mean reconstruction operator and the true solution, for which $\mathbf{W}_{\rm true}\mathbf{A}=\mathbf{I}$. It therefore persists on average whenever $\overline{\mathbf{W}}\mathbf{A}\neq\mathbf{I}$ and vanishes for an exact model. The second, $\overline{\mathbf{W}}\mathbf{n}^{(i)}$, is the instrumental noise propagated through the component-separation weights and contributes to the statistical uncertainty, with zero mean as $\langle\mathbf{n}^{(i)}\rangle=0$. Finally, $\delta\mathbf{W}^{(i)}\mathbf{y}^{(i)}$ describes the additional statistical fluctuations in the reconstructed components induced by the propagation of the uncertainties of the fitted parameters. We can therefore identify the systematic residual with the part that persists after averaging over realizations,
\begin{align}
    \mathbf{m}_{\rm sys}
    =
    \left\langle
    \left(
        \overline{\mathbf{W}}\mathbf{A}
        -
        \mathbf{I}
    \right)
    \mathbf{s}^{(i)}
    \right\rangle
    =
    \left\langle
        \mathbf{m}_{\rm res}^{(i)}
    \right\rangle,
    \label{eq:sys_map}
\end{align}
and define the statistical residual for each realization as the remaining terms,
\begin{align}
    \mathbf{m}_{\rm stat}^{(i)}
    =
    \overline{\mathbf{W}}\mathbf{n}^{(i)}
    +
    \delta\mathbf{W}^{(i)}\mathbf{y}^{(i)}
    =
    \mathbf{m}_{\rm res}^{(i)}
    -
    \left\langle
        \mathbf{m}_{\rm res}^{(i)}
    \right\rangle.
    \label{eq:stat_map}
\end{align}
From these definitions, we can compute the corresponding total, systematic, and statistical residual power spectra:
\begin{align}
    C_{\ell}^{\rm res,(i)}
    &=
    \mathcal{C}_{\ell}
    \left[
        \mathbf{m}_{\rm res}^{(i)}
        \mathbf{m}_{\rm res}^{(i),\dagger}
    \right],
    \label{eq:res_def}\\
    C_{\ell}^{\rm sys}
    &=
    \mathcal{C}_{\ell}
    \left[
        \mathbf{m}_{\rm sys}
        \mathbf{m}_{\rm sys}^{\dagger}
    \right],
    \label{eq:sys_def}\\
    C_{\ell}^{\rm stat,(i)}
    &=
    \mathcal{C}_{\ell}
    \left[
        \mathbf{m}_{\rm stat}^{(i)}
        \mathbf{m}_{\rm stat}^{(i),\dagger}
    \right].
    \label{eq:stat_def}
\end{align}

Writing $\hat{\mathbf{s}}^{(i)}=\mathbf{s}^{(i)}+\mathbf{m}^{(i)}_{\rm res}$, the recovered
angular power spectrum becomes
\begin{align}
    \tilde{C}_\ell^{(i),\,\rm GLS}
    &=
    C_\ell^{(i)}
    +
    2\mathcal{C}_\ell
    \left[
        \mathbf{s}^{(i)}\mathbf{m}_{\rm res}^{(i),\dagger}
    \right]
    +
    C_{\ell}^{\rm res,(i)}
    \label{eq:cl_gls}
\end{align}
The final term, the auto-spectrum of the residual, contains both systematic and statistical contributions and is strictly positive, thereby adding directly to the true $C_\ell^{(i)}$. Even when the model is exact, such that $C_\ell^{\rm sys}=0$, the statistical contribution $C_\ell^{\rm stat}$ remains and must be taken into account. The cross-spectrum,
$\mathcal{C}_\ell[\mathbf{s}^{(i)}\mathbf{m}_{\rm res}^{(i),\dagger}]$, is more subtle. Unlike the residual auto-spectrum, its contribution is not positive definite and can either increase or decrease the recovered power. Predicting this term is also more difficult, as it depends on the particular underlying sky realization $\mathbf{s}^{(i)}$. In principle, since the underlying components (CMB, foregrounds and noise) are assumed to be mutually uncorrelated on average, this contribution is expected to be small and is not usually modeled explicitly. However, as the reconstruction operator $\hat{\mathbf{W}}$ is itself estimated from the same data realization, it can therefore inherit accidental correlations present in the empirical covariance. These are subsequently propagated into the reconstruction and can induce a non-negligible cross-spectrum. In fact, when this occurs, part of the particular signal realization aligned with these accidental correlations can be attenuated, inducing a negative correlation between the residual and the true signal and appearing, on average, as a suppression at the power-spectrum level. This effect is typically known as the ILC bias~\cite{Delabrouille_2008, Coulton_2024} and we'll revisit this point several times within this paper.
\indent To infer cosmological parameters, the recovered angular power spectrum $\tilde{C}_\ell^{\rm GLS}$, together with its statistical contributions, must be propagated into a likelihood model. The treatment adopted in this work is presented in Section~\ref{subsec:likelihood_r}.
\subsection{Effective degree of freedom of the foregrounds}\label{subsec:svd}
\begin{figure}
\centering 
\includegraphics[width=0.85\linewidth]{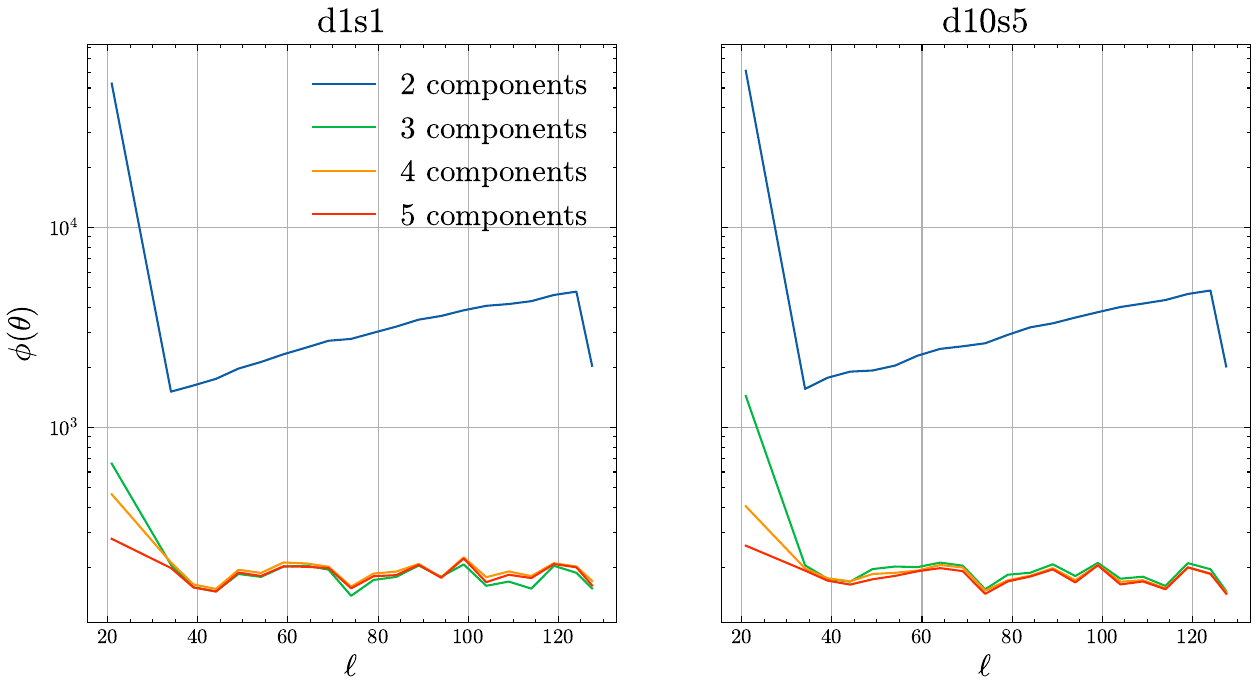}
\caption{The SMICA $\phi_q(\theta)$ goodness-of-fit applied on LiteBIRD-like data with a single realization of CMB, noise and foregrounds $\texttt{d1s1}$ (left) and $\texttt{d10s5}$ (right). The values are shown for each multipole bins and for models with an increasing number of foreground components. In both cases, increasing the model from two to three components results in a substantial improvement in the goodness of fit, while adding a fourth or fifth component provides small improvements, primarily in the first bin.}\label{fig:phi_theta}
\end{figure}

For convenience, as we have assumed decorrelation between the components, the $\mathbf{A}\mathbf{R_S}\mathbf{A}^T$ term can be split into CMB and foreground terms, allowing the modeled covariance to be rewritten as follows:
\begin{align}
    \mathbf{A} &= \begin{pmatrix}
        \mathbf{a_c} & \mathbf{A_f}
    \end{pmatrix},\\
    \mathbf{R_S} &= \begin{pmatrix}
        \mathbf{C} & 0\\
        0 & \mathbf{P_f}
    \end{pmatrix},\\
    \mathbf{R_Y} &= \mathbf{a_c} \mathbf{a_c}^T \mathbf{C}  +\mathbf{A_f P_f} \mathbf{A_f}^T + \mathbf{R_N}.\label{eq:model_cov_split}
\end{align}
Here, we have dropped the $XX$ polarization mode notation for simplicity's sake. The CMB is modeled as a single component with angular power spectrum $\mathbf{C}$ and its mixing vector $\mathbf{a}_{\rm c}$ fixed to a unit vector. This relies on the fact that the CMB is a perfect blackbody: when the observations are converted into CMB temperature units, it has the same amplitude in all frequency channels, allowing us to fix its mixing matrix a priori. The foreground term is modeled with harmonic covariance $\mathbf{P_f}$ of a set dimension $N_d \times N_d$ (number of foreground components) and with mixing matrix $\mathbf{A_f}$ of dimension $N_f \times N_d$. \\
\indent In SMICA, the number of foreground components $N_d$ is not a fixed parameter. In fact, the structure of the modeled covariance allows for as total many components as there are observing frequencies. In this case, we would be effectively fitting for the CMB plus foreground of dimension $N_d = N_f - 1$. Running SMICA in this regime would be close to performing an ILC component separation where the foreground properties are not modeled explicitly, and the algorithm simply eliminates any residual frequency-dependent structure. This mode of running SMICA will not be the main focus of the paper; instead, we take the direction in which an optimal number of components is chosen for each run.\\
\indent A way to determine the number of effective foreground components required by the model is to use the value of the $\phi_q(\theta)$ function in Eq.~\eqref{eq:smica_lklh} as a measure of goodness of fit for each multipole bin~\cite{Patanchon_2004}. Starting from a minimal number of components, we progressively increase the number of foreground components and monitor this value in each bin. The result of this procedure is shown in Figure~\ref{fig:phi_theta}. We observe a consistent behaviour for both $\texttt{d1s1}$ and $\texttt{d10s5}$. A model with only two foreground components provides a poor description of the data, as indicated by the relatively large values of $\phi_q(\theta)$. Increasing the model to three components leads to a substantial improvement, with a pronounced reduction in $\phi_q(\theta)$ across all bins. In contrast, adding a fourth or fifth component produces smaller improvements, which are confined to the first bin. This suggests that three foreground components are sufficient to capture the dominant structure of the foreground emission in both cases. This conclusion can also be supported by testing  range of configurations and examining the contribution of each additional component.\\
\begin{figure}
\centering 
\includegraphics[width=0.85\linewidth]{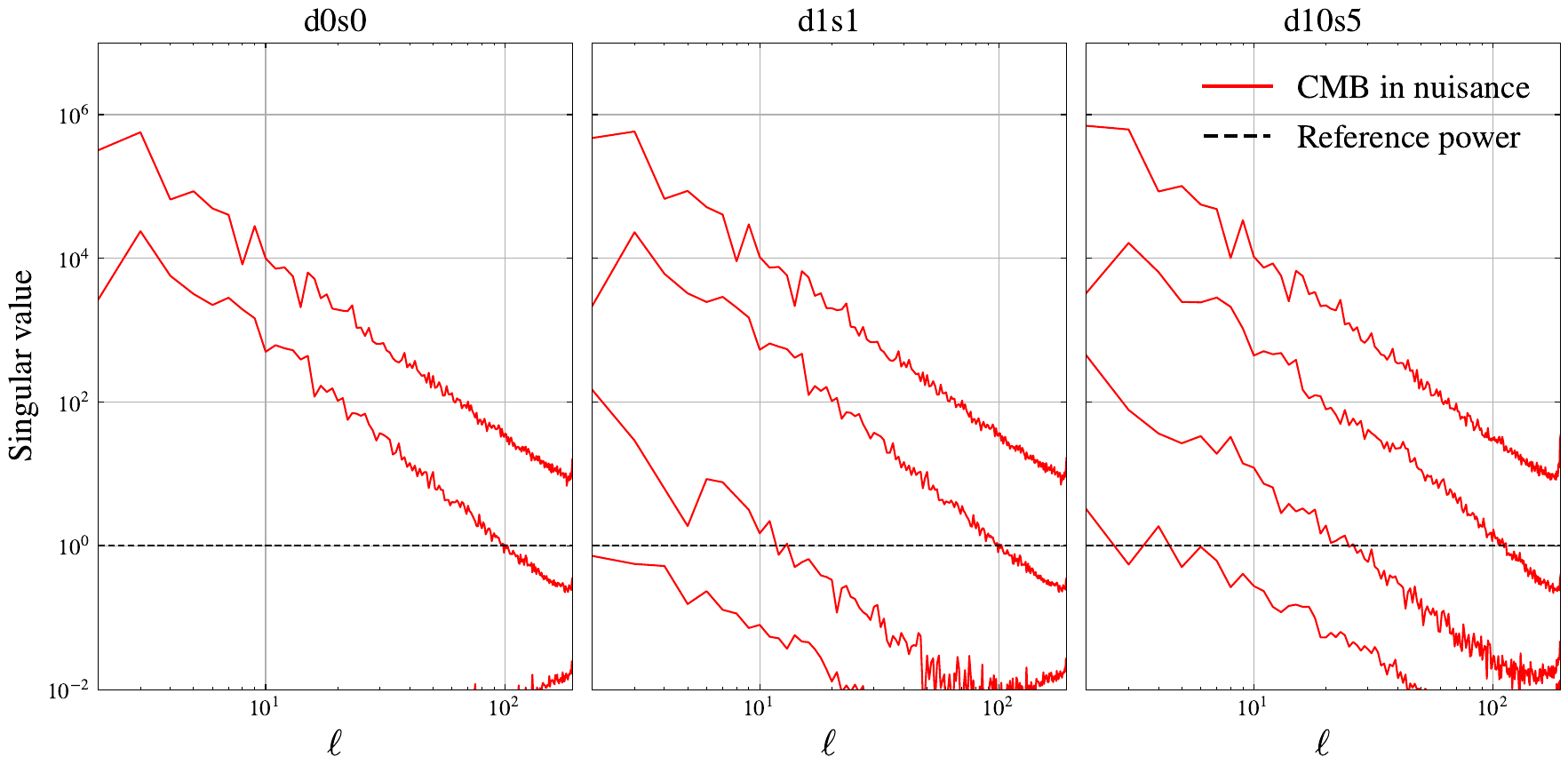}
\caption{Results of performing a singular value decomposition on the whitened covariance $ \mathbf{\tilde{R}_Y}$. The curves have an offset of 1, and the plot is on a log scale to highlight the values that are much more significant than one. The dashed black line highlights the magnitude-1 singular values; modes lying close to this line have power equivalent to the CMB + noise. The computed singular values form exactly 2 components for \texttt{d0s0} and 3 strong plus 1 weak component for the two more complex models \texttt{d1s1} and \texttt{d10s5}.}\label{fig:svd}
\end{figure}
\indent An alternative way to determine the effective dimensionality of the foreground covariance is to follow the blind, data-dependent procedure of Refs.~\cite{Remazeilles_2011,Carones_2024}. We first construct the noise-whitened covariance
\begin{align}
    \mathbf{\widetilde{R}_Y}
    &=
    (\mathbf{R_C+R_N})^{-1/2}\mathbf{\widehat{R}_Y}
    (\mathbf{R_C+R_N})^{-1/2} \nonumber\\
    &\simeq
    (\mathbf{R_C+R_N})^{-1/2}\mathbf{R_f}
    (\mathbf{R_C+R_N})^{-1/2}+\mathbb{I},
    \label{eq:whit_cov}
\end{align}
with $\mathbf{R_C}=\mathbf{a_c}\mathbf{a_c}^T\mathbf{C}$ and
$\mathbf{R_f}=\mathbf{A_fP_fA_f}^T$. We then perform a singular-value decomposition of $\mathbf{\widetilde{R}_Y}$ in each multipole bin, with singular values above unity identifying foreground modes that are significant with respect to the CMB and noise contribution, highlighting the effective dimensionality of the foreground covariance. \\
\indent The results for the three foreground models in simulated LiteBIRD-like observations are shown in Figure~\ref{fig:svd}. For the simplified \texttt{d0s0} model, the assumption of a spatially constant mixing matrix is exact, and the SVD correspondingly shows two dominant foreground modes, as expected. In contrast, both \texttt{d1s1} and \texttt{d10s5} exhibit a clear third mode over most multipoles, despite still containing only two physical components. This reflects the increased effective dimensionality induced by spatially varying spectral parameters. We do observe an additional structure that has relevant power at the largest scales. However, given its low amplitude, interpreting it as an additional component is not so robust. Note that this procedure requires knowledge of the CMB covariance $\mathbf{R_C}$ and therefore cannot be directly applied to real data, where only $\mathbf{R_N}$ is known; this will be discussed further in Appendix~\ref{appendix:a}.\\
\indent The potential of running SMICA with an increased number of components has been demonstrated in~\cite{Steier_2026}. In the context of a ground-based experiment targeting small angular scales over a limited sky fraction, this increased model flexibility enabled the recovery of an unbiased tensor-to-scalar ratio on simulated observations.
\subsection{SMICA on subsets of pixel}\label{subsec:patches_formalism}
\begin{figure}
    \centering
    \includegraphics[width=1.0\linewidth]{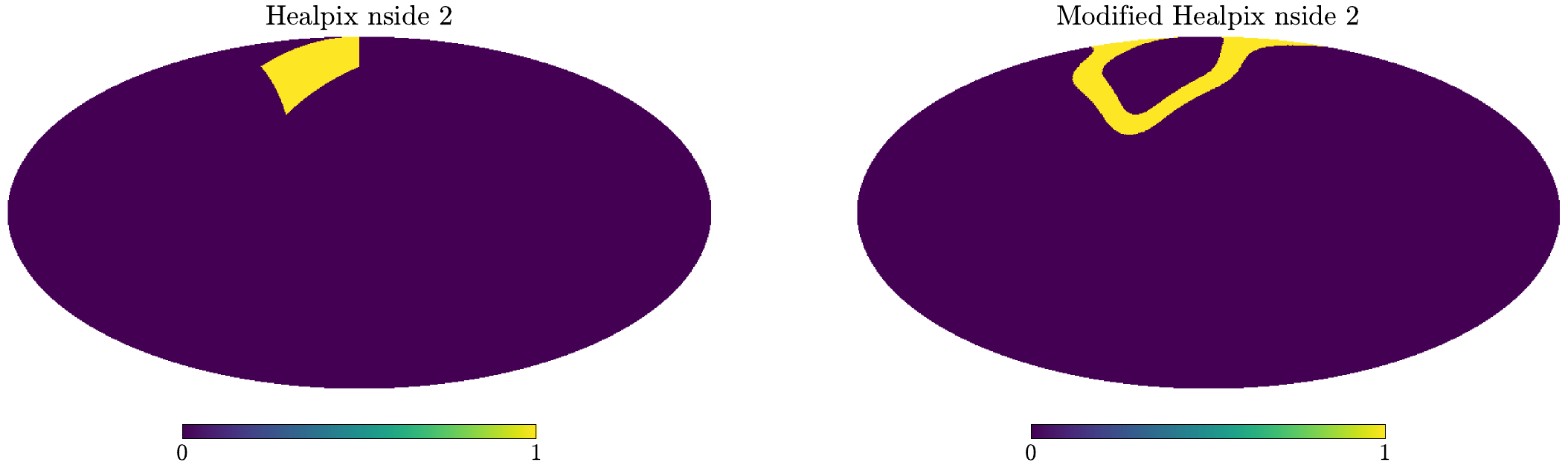}
    \caption{An example patch in use for covariance estimation of the HEALPix (left) and modified-HEALPix (right) of $N_{\rm side} = 2$ pipeline. The modified-HEALPix patch is simply the original super-pixel extended uniformly to cover an area of $2\%$ of the sky after subtracting itself.}
    \label{fig:patches}
\end{figure}

As most of our biases will be coming from the unmodeled spatial variability of the mixing matrix $\mathbf{A}$, we open up the model to include this variation. In principle, the mixing matrix varies across the sky, and in the most general case, can take on as many distinct values as there are observed sky positions. However, the foregrounds exhibit coherent physical structures morphologically~\cite{Puglisi_2022}, which means that the assumption of a constant mixing matrix can still hold locally within certain regions of the sky. This insight motivates the idea of performing component separation on subsets of pixels, or patches that subdivide the sky into regions within which the assumption of a uniform mixing matrix is roughly correct. This technique has been applied and proven to be effective in counteracting the effect of complex foregrounds for both the parametric method of FGBuster~\cite{Rizzieri_2026}\cite{Kabalan_2026} and the blind method of NILC~\cite{Carones_2023}. \\
\indent In the context of SMICA, since the assumption of a coherent $\mathbf{A}$ is key to representing the covariance through Eq.~\eqref{eq:model_cov_split}, the component separation should be effective locally wherever this holds approximately. We can therefore imagine partitioning the sky into $N_k$ pixel subsets and run SMICA independently on each one to retrieve a locally fitted mixing matrix and spectra. By restricting the domain in this way, the model can effectively capture the variability that would otherwise be unrepresented in a global analysis.\\
\indent However, because SMICA is performed independently in each subset, we lose the ability to directly infer a single ``best-fit’’ CMB power spectrum. Unlike the global analysis, where a uniform CMB spectrum is obtained by optimizing the likelihood in Eq.~\eqref{eq:smica_lklh}, each local SMICA run yields its own estimate of the CMB pseudo-spectrum, with varying levels of contamination depending on how well the assumption of constant $\mathbf{A}$ holds within each region. This prevents us from simply aggregating the local spectra, which are biased estimates given the partial sky coverage, particularly for the scales larger than the size of the patch, into a consistent full-sky estimate. \\
\indent Consequently, we adopt the second direction described by Eq.~\eqref{eq:gls}, using the locally fitted parameters to construct a set of component-separation weights for each pixel subset and reconstruct the CMB map locally. These locally reconstructed maps $\hat{\mathbf{W}}^{k}\mathbf{y}^{k}$ are combined to obtain the final estimated component maps (which are later used for power spectrum estimation):
\begin{align}
    \mathbf{\hat{s}}
    &=
    \sum_{k=1}^{N_k}
    \hat{\mathbf{W}}^{k}
    \mathbf{y}^{k},
\end{align}
This multi-patch reconstruction allows for the local adaptability of the subsets while enabling a coherent global CMB map to be formed, even in the presence of significant variability in the mixing matrix. In this approach, we will infer cosmological parameters from this reconstructed CMB map, using an external likelihood demonstrated Section \ref{subsec:likelihood_r}.\\
\indent In this newly defined implementation of SMICA, we construct our subsets based on HEALPix\footnote{\url{http://healpix.sourceforge.net}}\cite{Gorski_2005} pixels at low resolution, specifically $N_{\mathrm{side}} = 2$ and $N_{\mathrm{side}} = 4$, corresponding to 48 and 192 patches across the sky. This choice of pixel subsets is purely arbitrary and is simplistic enough for the purpose of demonstrating the feasibility of the method, which can be further refined by imposing a physically motivated division of the sky~\cite{Carones_2023, Puglisi_2022}. \\
\indent However, one notable feature of this implementation is the emergence of the ILC bias~\cite{Delabrouille_2008, Coulton_2024}. As discussed in Section~\ref{subsec:smica}, this arises primarily because the covariance is estimated over increasingly smaller sky regions, reducing the number of available modes and making the covariance estimate more susceptible to statistical fluctuations. These fluctuations can induce spurious correlations between the components, which are then inherited by the reconstruction operator $\hat{\mathbf{W}}$. Specifically, in Eq.~\eqref{eq:cl_gls}, such correlations can make the cross-spectrum term non-negligible, resulting in a biased suppression of the recovered signal. We provide a more detailed discussion of this effect in Section~\ref{subsec:ilcbias}. In principle, the ILC bias could be corrected analytically or by using Monte Carlo simulations, described in~\cite{Delabrouille_2008}. In practice, however, implementing and validating these corrections is non-trivial, and assuming that they can be modeled accurately may be overly optimistic in the context of real data. \\
\indent Therefore, we consider a strategy that instead mitigates the effect directly at the component separation stage. To achieve this goal, we have adopted a procedure heavily inspired by~\cite{Coulton_2024}, where the harmonic covariance, and subsequently the parameters, are estimated on an extended region that excludes the subset on which the component separation weights are applied. Doing this effectively "de-correlates" these weights $\mathbf{\hat{W}}$ from the data it is applied to, thereby eliminating the cross-term that causes the ILC bias. This approach, shown in Figure  \ref{fig:patches}, will be denoted as "modified-HEALPix" from this point onward. The size of the inner region to be cut and the area covered are free parameters that can be adjusted to control the level of ILC bias in the method. For this specific implementation, in both cases we cut out the center region equal to the original patch, while the area of the ring is extended to cover roughly $2\%$ and $1\%$ of the sky in the $N_{\rm side} = 2$ and $N_{\rm side} = 4$ cases, respectively. Note that, although the covariance is computed on the ring-like regions, the resulting weights are still applied to the standard HEALPix patches to reconstruct the full-sky maps. This mismatch is expected to degrade the performance of the method.
\subsection{Masking strategy}\label{subsec:masking}
Similar to many other pipelines, we apply a mask to the Galactic plane, where the emission is particularly bright and complex, to reduce foreground contamination. In this paper, we implement two masks, one before the computation of the empirical covariance (pre-compsep) in selected cases, and one after component separation (post-compsep) to compute the final CMB spectra. The pre-compsep mask is simply the Planck GAL60 mask\footnote{Planck Legacy Archive, \url{http://pla.esac.esa.int/pla/aio/product-action?MAP.MAP_ID=HFI_Mask_GalPlane-apo0_2048_R2.00.fits}.}~\cite{Planck2018_XII}, retaining $60\%$ of the sky. The choice is purely arbitrary, and optimization of this mask is not done for this work, serving the simple purpose of ensuring that the bright pixels in the galactic center do not leak into the computation of our harmonic covariance. \\
\indent The post-component-separation mask can, in principle, be optimized to exclude regions in which the component separation performs poorly. In this work, we employ the Planck GAL40 mask for most of the results presented in this paper. This choice is relatively pessimistic, as the mask was constructed from intensity data and is not specifically optimized for the residuals of our polarization component-separation method. It nevertheless represents a realistic situation in which an optimal post-component-separation mask cannot be directly constructed and an external mask must be adopted, at the cost of reducing the available sky fraction. In Section \ref{subsec:ideal_mask}, for a few selected cases, we also construct an idealized mask by thresholding the foreground residuals after component separation and retaining $50\%$ of the sky. This provides an estimate of the maximum performance achievable by optimally removing the regions with the largest residuals, but cannot be implemented in practice, as the residual level is not directly known from the data.
\subsection{Cosmological likelihood for the tensor-to-scalar ratio}
\label{subsec:likelihood_r}
\indent To assess the effectiveness of our component separation method on the measurement of the tensor-to-scalar ratio $r$, we propagate the residuals after component separation through the use of an Inverted Wishart likelihood corrected by the sky fraction $f_{\rm sky}$\cite{Hamimeche_2008}:
\begin{equation}
-2\ln \mathcal{L}^{(i)}_{\bf IW}(r)
= f_{\rm sky} \sum_{\ell} (2\ell+1)
\left[\frac{C^{\rm total, (i)}_{\ell}}{C^{\rm model, (i)}_{\ell}(r) } + \ln C^{\rm model, (i)}_{\ell}(r) -\frac{2\ell-1}{2\ell+1}\ln C^{\rm total, (i)}_{\ell} 
\right],
\label{eq:cosmo_like}
\end{equation}
The quantity $C^{\rm total, (i)}_{\ell}$ denotes the observed angular power spectrum of the CMB map after component separation and is therefore analogous to the recovered spectrum $\tilde{C}_\ell^{\rm GLS, (i)}$ in Eq.~\eqref{eq:cl_gls}. For each realization, and assuming $r=0$, we write it as
\begin{equation}
C^{\rm total,(i)}_{\ell}
= C^{\rm lens}_{\ell}
+ C^{\rm res,(i)}_{\ell}
= C^{\rm lens}_{\ell}
+ C^{\rm stat,(i)}_{\ell}
+ C^{\rm sys}_{\ell},
\label{eq:cl_total}
\end{equation}
where $C^{\rm lens}_{\ell}$ is the lensing B-mode spectrum, while $C^{\rm res,(i)}_{\ell}$, $C^{\rm sys}_{\ell}$, and $C^{\rm stat,(i)}_{\ell}$ denote the total, systematic, and statistical
residual power spectra defined in Section $\ref{subsec:smica}$. Using these definitions, the individual residual contributions can be quantified exactly in our simulations, where the input sky components are known. In the case of real data, however, they must instead be estimated indirectly. The statistical residual can be accurately predicted using Monte Carlo simulations and Fisher forecasting~\cite{Rizzieri_2026}, whereas estimating the systematic residual requires accurate knowledge of the true foreground emission and is therefore considerably more challenging. We consequently include the statistical residual in the model angular power spectrum $C^{\rm model, (i)}_{\ell}(r)$, while leaving any systematic contribution unaccounted for:
\begin{equation}
C^{\rm model, (i)}_{\ell}(r)
= rC^{\rm prim}_{\ell}
+ C^{\rm lens}_{\ell}
+ C^{\rm stat,(i)}_{\ell}.
\label{eq:cl_model}
\end{equation}
$C^{\rm prim}_{\ell}$ is the template for the target primordial B-mode spectrum, scaled by the tensor-to-scalar ratio $r$. The statistical residual is thus propagated into the uncertainty of the $r$ estimate, whereas an unmodeled $C^{\rm sys}_{\ell}$, if non-negligible, can bias the inferred value of $r$. In writing
Eq.~\eqref{eq:cl_total} and Eq.~\eqref{eq:cl_model}, we neglect the cross-spectrum term discussed previously. This is because if the component-separation procedure performs as intended, its contribution should be small.\\
\indent With these definitions, the likelihood is evaluated over an ensemble of realizations, from which we derive statistical constraints on the resulting average posterior distribution of $r$. We characterize this distribution using the following quantities of interest:
\begin{itemize}
    \item The peak of the average posterior, which we define as the bias on $r$ (as $r_{true} = 0$):
    \begin{equation}
        r_{\rm bias} = \arg\max_{r}\langle \mathcal{L}_{\bf IW}(r) \rangle.
    \end{equation}

    \item The $16\%$ and $84\%$ quantiles, denoted $r_{\rm lower}$ and $r_{\rm upper}$:
    \begin{equation}
        \int_{-\infty}^{r_{\rm lower}} \langle \mathcal{L}_{\bf IW}(r) \rangle\,{\rm d}r = 0.16,
        \qquad
        \int_{-\infty}^{r_{\rm upper}} \langle \mathcal{L}_{\bf IW}(r) \rangle\,{\rm d}r = 0.84,
    \end{equation}
     which define the central $68\%$ interval of width $r_{\rm upper}-r_{\rm lower}$.

    \item The one-sided $68\%$ upper limit measured from $r=0$, denoted $r_{68\,(\rm from\,0)}$:
    \begin{equation}
        \int_{0}^{r_{68\,(\rm from\,0)}} \langle \mathcal{L}_{\bf IW}(r) \rangle\,{\rm d}r = 0.68.
    \end{equation}
\end{itemize}
When a significant bias is present, we report $r_{\rm bias}$, $r_{\rm lower}$ and $r_{\rm upper}$ ($|r_{\rm bias} - r_{\rm lower}|$ and $|r_{\rm bias} - r_{\rm upper}|$ in the plots) to characterize the location and width of the central $68\%$ interval. When the true value, $r_{\rm true}=0$, lies within $1\sigma$ of $r_{\rm bias}$, we also show the one-sided upper limit $r_{68\,(\rm from\,0)}$. Note that this last quantity might become less reliable if a significant area of the posterior lies below zero, which can be induced by, for example, the ILC bias.\\
\indent To get rid of the systematic bias induced by a nonzero $C_\ell^{\rm sys}$, at the likelihood level, one can implement a marginalization procedure~\cite{Carones_2025, Errard_2019} over some template emission $C_\ell^{\mathrm{temp}}$ that can replicate well the shape of the systematic residual. The modeled power spectrum is then parameterized by two parameters, the tensor-to-scalar ratio $r$ and the template amplitude $\alpha_\mathrm{temp}$:
\begin{equation}
C^{\rm model, (i)}_{\ell}(r)
= rC^{\rm prim}_{\ell}
+ C^{\rm lens}_{\ell}
+ C^{\rm stat, (i)}_{\ell}
+ \alpha_{\rm temp}C^{\rm temp, (i)}_{\ell}
\label{eq:cl_model_marg}
\end{equation}
The added term, if sufficiently well constructed with respect to the true systematic residual, should be able to absorb the bias on our estimation. In this work, we demonstrate this procedure for a few selected cases, utilizing the internally reconstructed foreground templates within SMICA. This is based on a loose assumption that foreground residual spectra should still scale with the actual foreground signal spectra.\\
\indent The presented likelihood model is only valid in the case of full-sky signal-dominated power spectra~\cite{Hamimeche_2008}, which is not exactly true in this context. The mask applied post-component-separation will create mode couplings~\cite{Hivon_2002}, and residuals are expected to have significant contributions in the higher complexity foregrounds. Therefore, the numerical values reported might not reflect the true, most accurate performance diagnostic. However, we still choose to adopt this likelihood model as it is first simple to implement, it does not overcomplicate this study into exploring this degree of freedom, and for the goal of showing gains/losses between the different cases, it is fully sufficient. 
\subsection{Pipeline overview}\label{subsec:pipeline}
\begin{figure}
    \centering
    \includegraphics[width=1.0\linewidth]{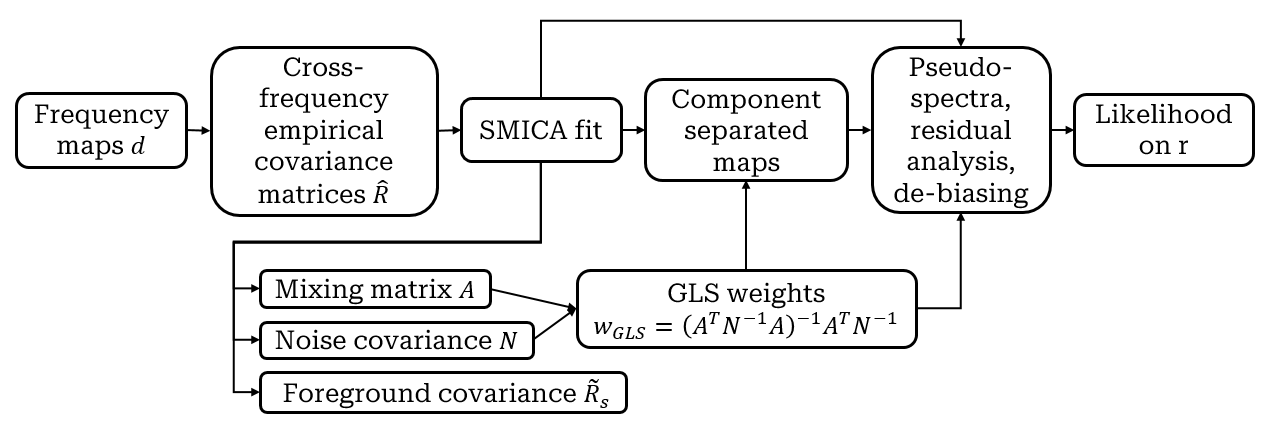}
    \caption{Overall pipeline for a B-mode optimized SMICA-based component separation}
    \label{fig:pipeline}
\end{figure}

Figure \ref{fig:pipeline} illustrates the complete component separation pipeline.\\
\indent The frequency maps are first deconvolved from their respective beams and then converted from $Q/U$ to scalar $E/B$ maps. This conversion is performed through a spin-2 full-sky spherical harmonic transform of the polarization maps, from which the $E$- and $B$-mode harmonic coefficients are extracted. These coefficients are then transformed back into map space, yielding spin-0 $E$- and $B$-mode fields. The pre-component separation mask and the patch mask are subsequently applied to these scalar fields, avoiding the need for an additional $E/B$ purification step. However, strictly speaking, this choice is not physically well-motivated: while Galactic foregrounds are local in the $Q/U$ maps, the scalar $E/B$ fields are non-local on the sphere, so a mask on the $B$ map does not directly correspond to a localized region of foreground emission. This is particularly relevant when considering masks and sky-partitioning based on external Galactic tracers. Nevertheless, we find that the use of masks on scalar $E/B$ fields still performs well in reducing foreground contamination. The harmonic covariance matrices used by SMICA are computed with \texttt{anafast} from the \texttt{healpy}\cite{Zonca_2019} package. \\
\indent This approximation of pseudo-spectra also motivates our choice of the GLS estimator over, for example, the Wiener filter:
\begin{equation}
    \mathbf{\hat{s}} =
    (\mathbf{A}^T \mathbf{R_N}^{-1} \mathbf{A}
    + \mathbf{R_S}^{-1})^{-1}
    \mathbf{A}^T \mathbf{R_N}^{-1}\mathbf{y}.
\end{equation}
In principle, since the signal covariance $\mathbf{R_S}$ is also estimated by SMICA, the Wiener filter could provide a more optimal reconstruction. However, the estimate of $\mathbf{R_S}$ is considerably less reliable under our pseudo-spectrum treatment, particularly at the largest angular scales. This limitation is more serious for the SMICA on subsets pipeline, where the finite patch size prevents accurate reconstruction of the largest modes. For these reasons, we adopt the GLS estimator throughout this work and apply it consistently across all analyses. We have checked that when this method is applied to full-sky LiteBIRD configurations, the results are still close to optimal, as discussion in section \ref{subsec:smica_gls_lklh}.\\
\indent As described earlier, we work with a single polarization mode rather than the full joint $E/B$ covariance. This is motivated mostly by practical considerations. Treating each mode independently allows the SMICA formalism to be applied directly, avoiding the additional complexity of modeling the full $EE$, $EB$, and $BB$ structure. Furthermore, constructing the scalar $E$- and $B$-mode maps before the covariance estimation allows the $E/B$ decomposition and possible leakage mitigation to be handled separately from the component-separation procedure, avoiding the need to introduce an additional purification step within the SMICA framework. This is not to say that modelling the full joint-$E/B$ covariance is without its merit. If such a structure is in place, we may relax the assumption of the CMB column of the mixing matrix, as, unlike the weaker $B$-mode, the strong $E$-mode signal is able to maintain this column within the routine. This may be particularly useful for self-calibrating the CMB channels in the presence of systematic effects. However, this is out of the scope of this paper and is left for future studies.\\
\indent All of the SMICA runs are performed with $N_d = 3$ unless specified otherwise. We impose physically motivated constraints on the foreground mixing matrix to partially lift the normalization and permutation ambiguities, as well as the degeneracy associated with proportional power spectra, as is the case for dust and synchrotron:
\begin{equation}\label{eq:mixmat_constraints}
    \mathbf{A} = \begin{pmatrix}
        1 & 0 & 1 & A_{\nu_1}^3\\
        1 & A_{\nu_2}^1 & A_{\nu_2}^2 & A_{\nu_2}^3\\
        ... & ... & ... & .... \\
        1 & A_{\nu_{n-1}}^1 & A_{\nu_{n-1}}^2 & A_{\nu_{n-1}}^3\\
        1 & 1 & 0 &  A_{\nu_{n}}^3
    \end{pmatrix}.
\end{equation}
The $A_{\nu_x}^y$ indicates free elements to be fitted by SMICA. These constraints only fix the foreground basis and do not affect the CMB component. This structure imposes that the first foreground component will contain dust primarily, the second mostly synchrotron, while the third and beyond are left free to capture remaining contamination. The signal covariance $\mathbf{R_S}$ is assumed to have no cross-terms; that is, all Galactic components are uncorrelated. This is not a good assumption, as we know correlations are there between the foreground components. However, we have examined that the inclusion of this aspect only causes marginal changes to the tensor-to-scalar estimate. Further discussion on the inclusion of foreground cross-spectra and additional columns of $\mathbf{A}$ is presented in \ref{subsec:moment_expansion}. \\
\indent One note on the SMICA model \eqref{eq:smica1}: by confining the description of the covariance to harmonic space, we have assumed that the components follow Gaussian statistics. This is well-justified for the CMB but is strictly false for the foregrounds, as they exhibit strong non-Gaussianity and spatial non-stationarity. This means that information on higher-order statistics of the foregrounds are not captured. However, a previous work \cite{Citran_2026} has demonstrated the feasibility of exploiting this information and provided insights into its implications. We therefore leave this aspect beyond the scope of the present study.\\
\indent The main routine used to determine the best-fit parameters within SMICA employs a preconditioned conjugate gradient algorithm. The corresponding core implementation is publicly available in the SMICA GitLab repository\footnote{\url{https://gitlab.in2p3.fr/smica/smica}}.
\section{Results}\label{sec:results}
To gauge the performance of our component separation method, we utilize simulated datasets under the conditions of a LiteBIRD-like experiment~\cite{LiteBIRD_2022}, observing the sky over a large range of frequencies from 40 to 402 GHz. The simulations were made by coadding CMB and noise realizations with simulated foreground maps (specified in Section \ref{sec:fgs_model}) at each frequency. The CMB realizations in use are generated by the \texttt{synfast} function in \texttt{healpy} directly from the Planck 2018 best-fit power spectra. Here, the tensor-to-scalar ratio $r$ is taken to be zero; hence, the B-modes in our simulation contain only the lensing signal. The noise is simulated to be Gaussian, isotropic, and white based on the sensitivity parameters from Table 13 of~\cite{LiteBIRD_2022}. These assumptions are quite unrealistic; however, we do not expect changes in our conclusions nor additional bias on $r$ should this be untrue. To derive statistical constraints, we use 200 realizations of these two components. All of the maps in use follow the HEALPix\cite{Gorski_2005} pixelization scheme and are simulated at nside 256. The CMB and foreground maps are smoothed to the angular resolution of each frequency channel.
\subsection{Performance with increasing number of components}\label{subsec:ext_comps}
\begin{figure}[h]
    \centering
    \includegraphics[width=0.49\linewidth]{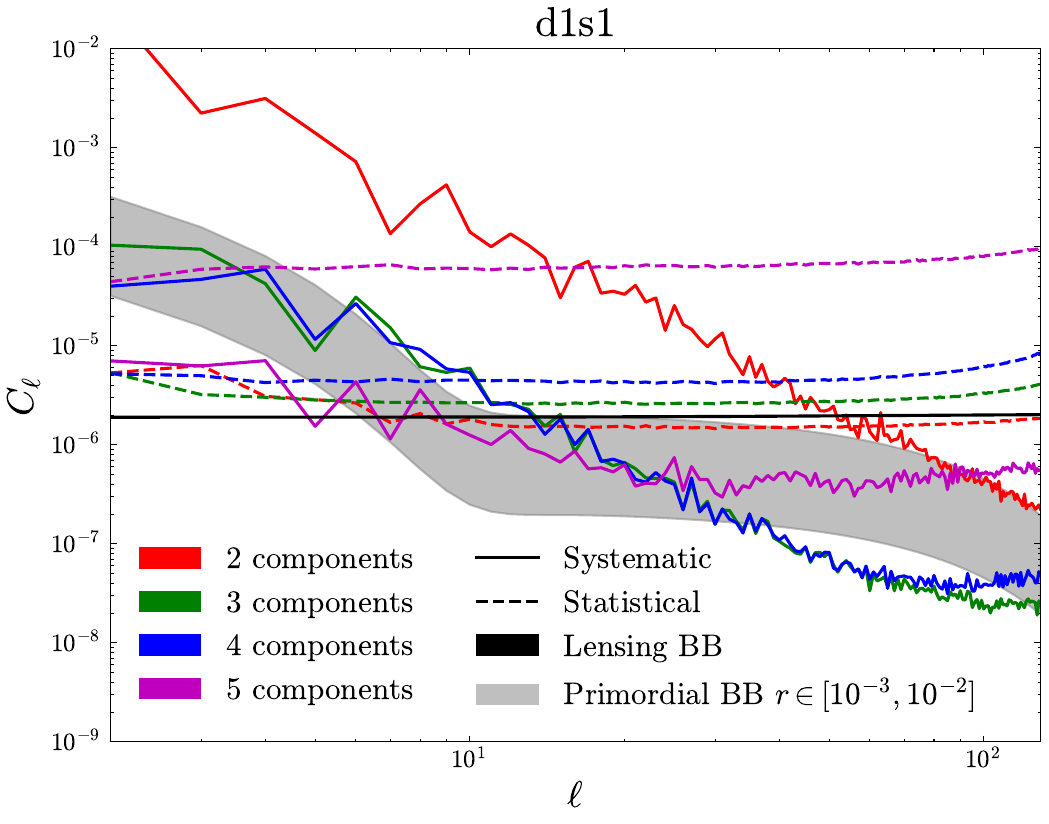}
    \includegraphics[width=0.49\linewidth]{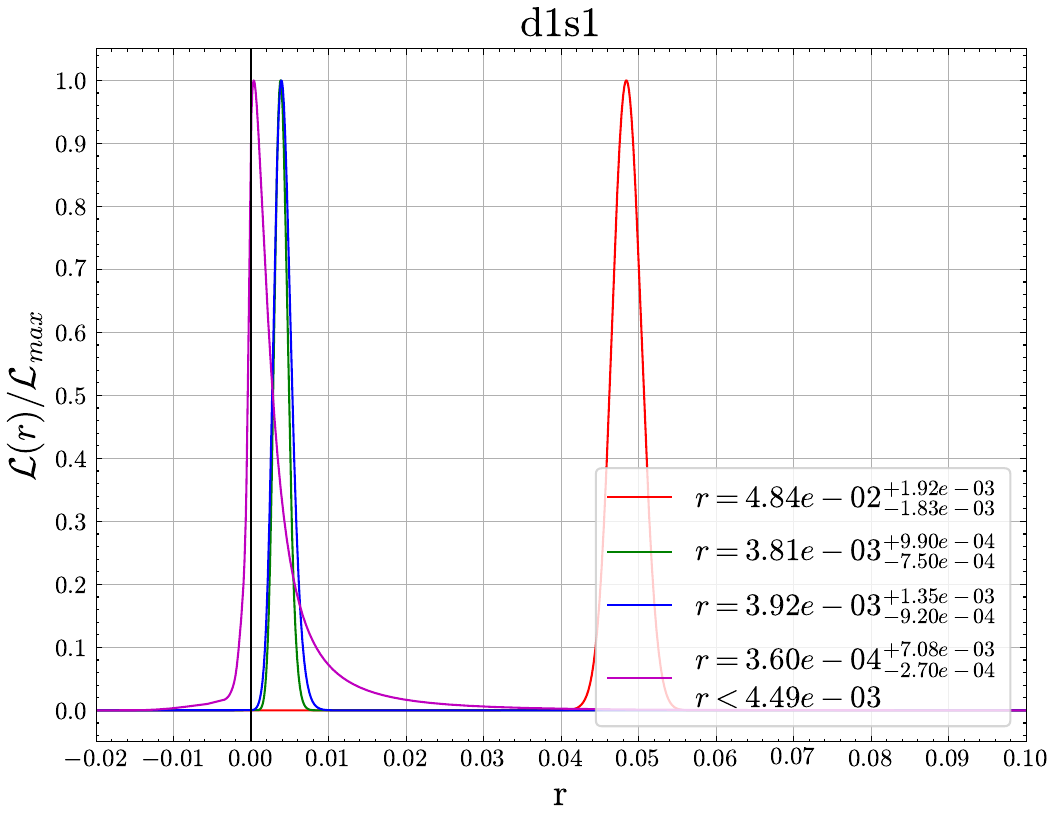}
    \includegraphics[width=0.49\linewidth]{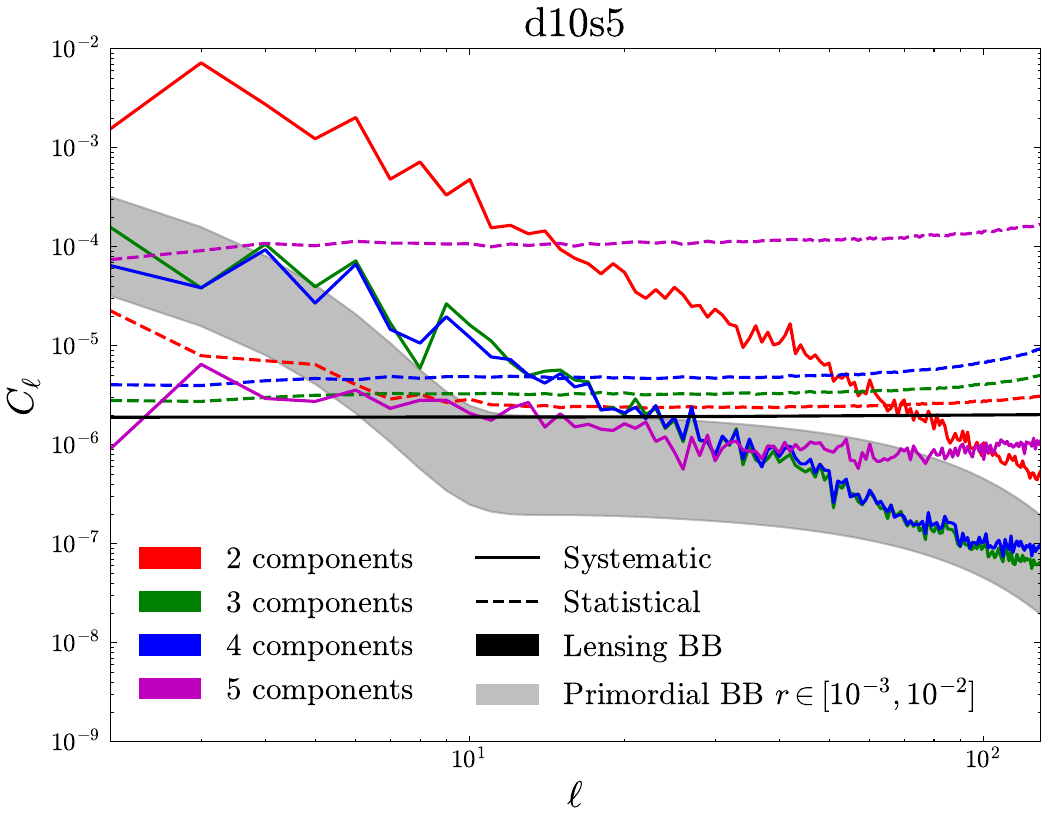}
    \includegraphics[width=0.49\linewidth]{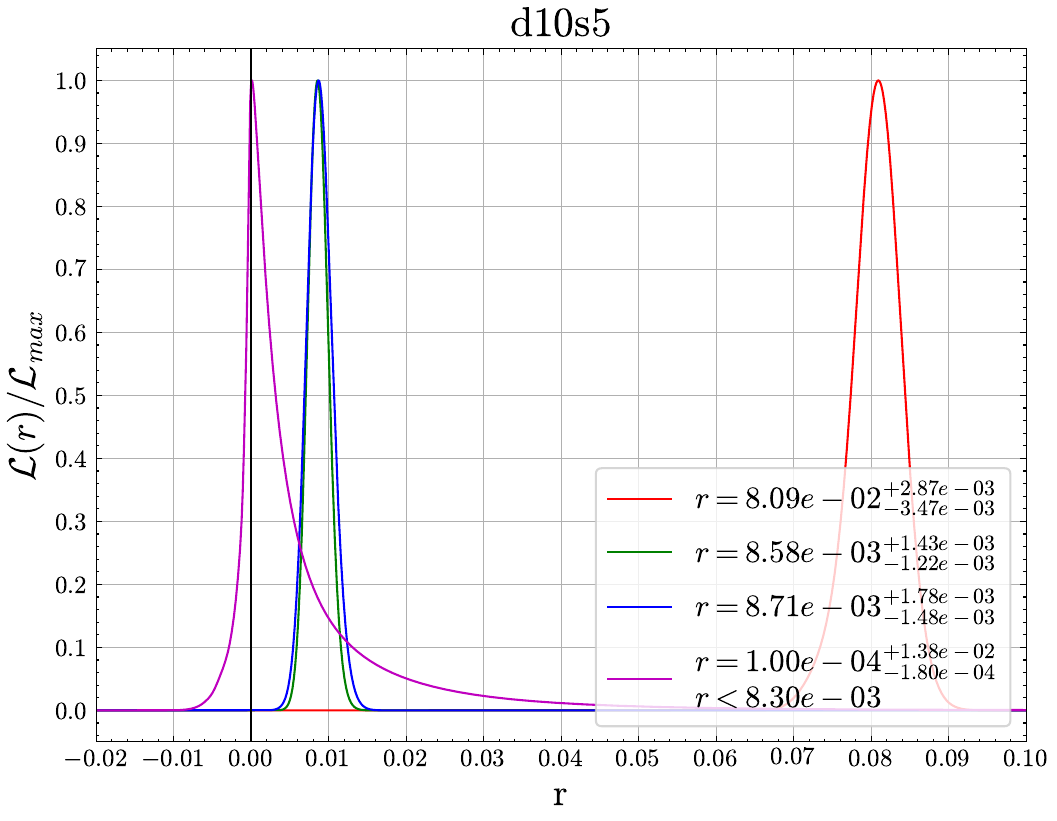}
    \caption{Results of the SMICA with extra components approach. Left: Residual after component separation, separated into systematic and statistical, overplotted for reference with the lensing B-mode and the primordial B-mode with $r \in [10^{-3}, 10^{-2}]$. Right: The average posterior of the tensor-to-scalar ratio when the residuals are propagated into the likelihood. The constraints $r_{\rm bias}$, $|r_{\rm bias} - r_{\rm lower}|$ and $|r_{\rm bias} - r_{\rm upper}|$ are reported. $r_{\rm 68\,(from\,0)}$ is given only when $r_{\rm true} = 0$ lies within $1\sigma$ of $r_{\rm bias}$. The upper row represents the results for \texttt{d1s1} while the lower is for \texttt{d10s5}.}
    \label{fig:ext_comps}
\end{figure}

We first explore the effect of increasing the number of foreground components' degrees of freedom within SMICA. The covariance in these cases is computed over the full sky with a GAL60 mask, without using any subsets. Figure~\ref{fig:ext_comps} shows the resulting systematic and statistical residuals, computed exactly from their definition in Eq.~\eqref{eq:sys_def} and~\eqref{eq:stat_def}, of the component-separated maps, together with the corresponding constraints on $r$, when SMICA with $N_{\mathrm{d}}=2$, 3, 4, and 5 foreground components is applied to the set of simulated observations using the two complex foreground models. For these runs, we apply a Planck GAL40 mask for the component-separated maps before computing the final spectra.\\
\begin{figure}
    \centering
    \includegraphics[width=1\linewidth]{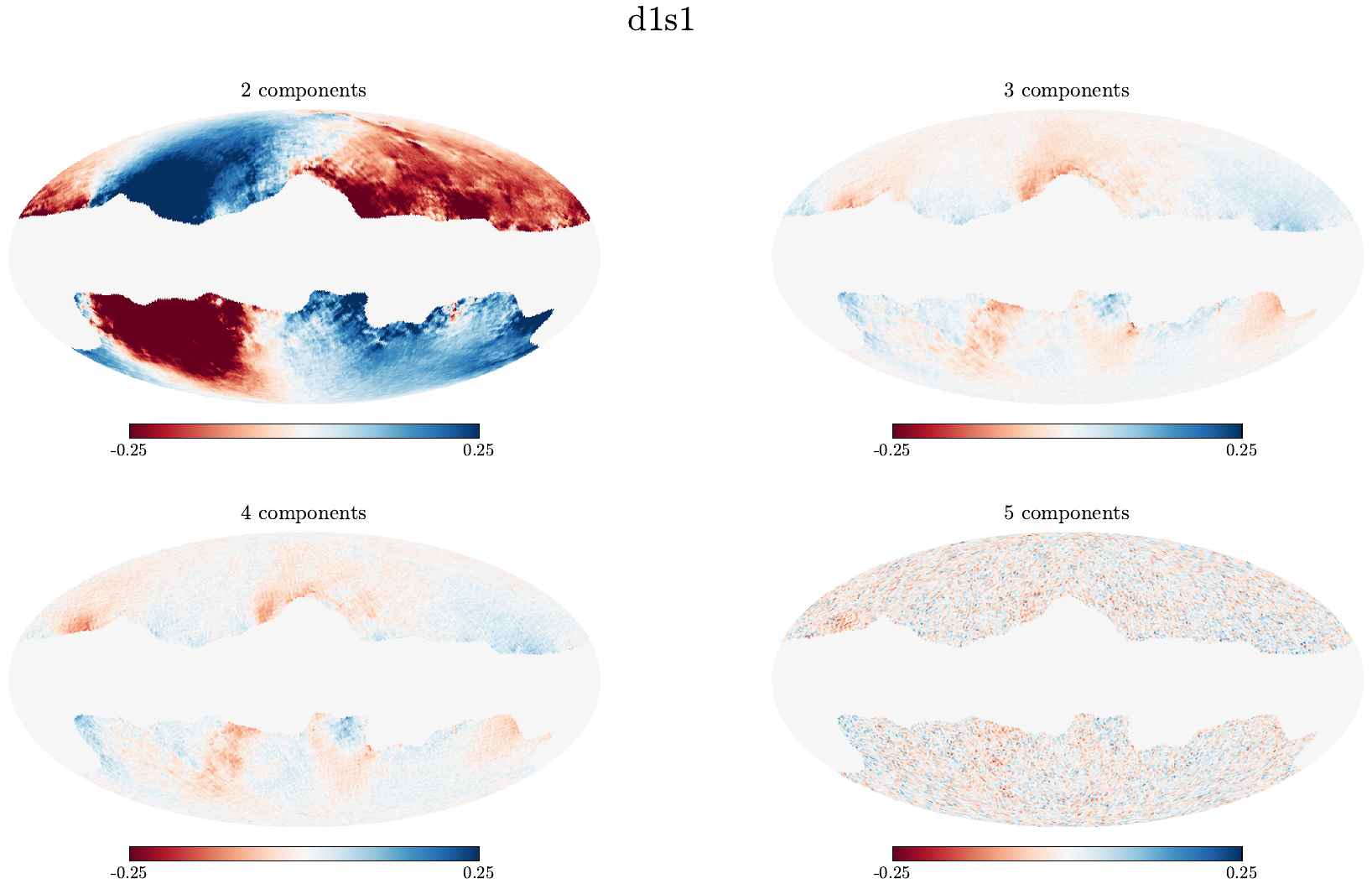}
    \includegraphics[width=1\linewidth]{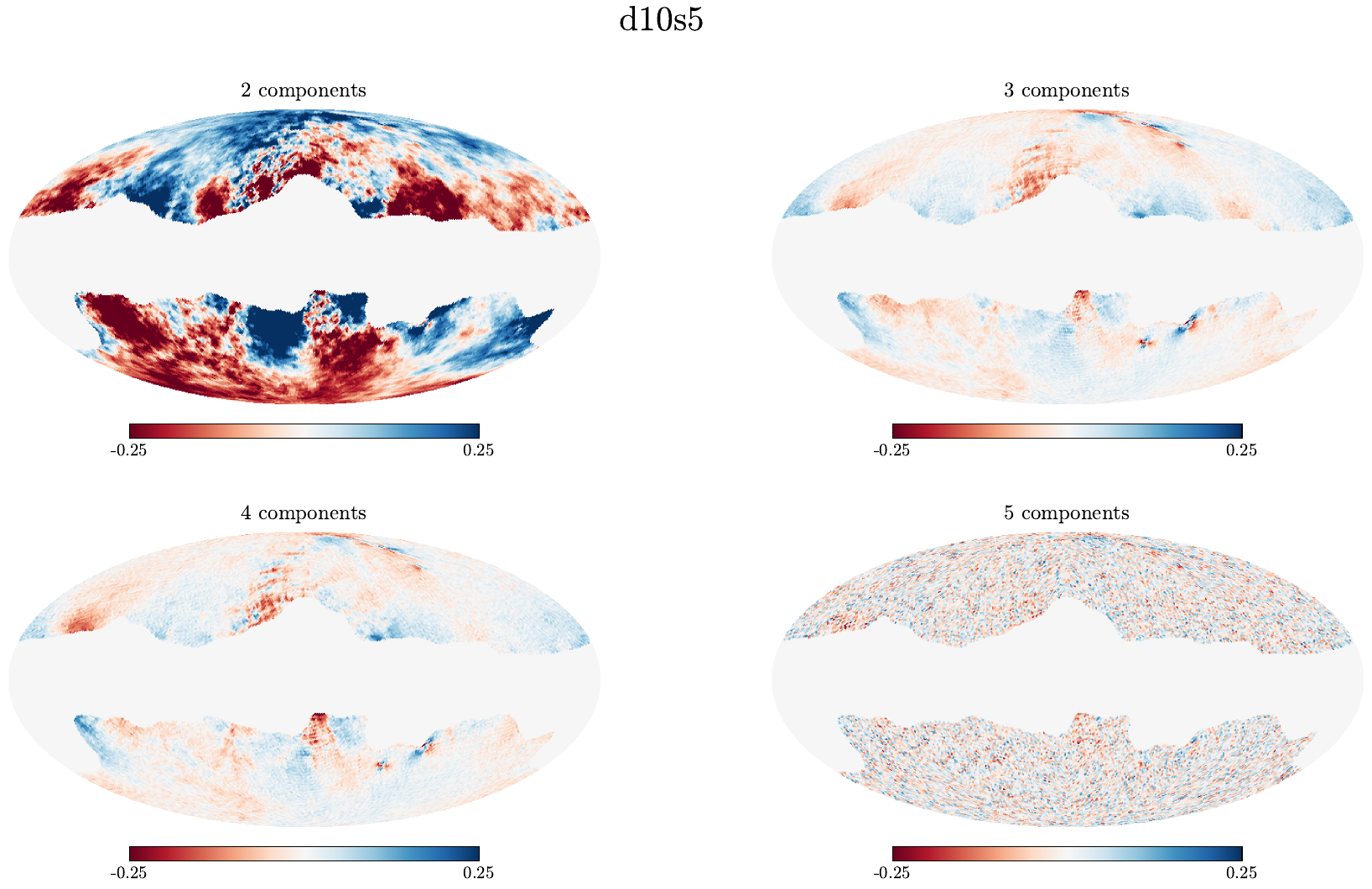}
    \caption{The systematic residual in pixel space for SMICA with 2, 3, 4 and 5 foreground components. The first set of 4 maps is for \texttt{d1s1} and the second set is for \texttt{d10s5}. A Planck GAL60 mask was applied to each of the maps to highlight the region where component separation is performed.}
    \label{fig:comps_resmap}
\end{figure}
\indent It is immediately apparent that using only the physical number of foreground components ($N_{\mathrm{d}}=2$) is insufficient for SMICA to accurately describe either foreground model. Although this configuration produces the smallest statistical residual in power spectrum space, its limited dimensionality leaves a substantial systematic residual. This translates into a large bias in the recovered tensor-to-scalar ratio, of approximately $4.8\cdot10^{-2}$ for \texttt{d1s1} and $8.0 \cdot 10^{-2}$ for \texttt{d10s5}. This behaviour reflects the increase in the effective dimensionality of the foreground covariance induced by the spatial variability of emission spectra. Note that despite the low statistical residual, the central $68\%$ region is rather large in this case, hovering around $3.7\cdot10^{-3}$ and $6.3\cdot10^{-3}$, driven by the large systematic bias.\\
\indent Introducing a third foreground component produces a substantial improvement. The systematic residual is strongly reduced, while the statistical residual increases moderately by roughly a factor of two in harmonic space. This reduces the bias on $r$ by roughly an order of magnitude, bringing it down to approximately $3.8\cdot10^{-3}$ for \texttt{d1s1} and roughly $8.6\cdot10^{-3}$ for \texttt{d10s5}. The corresponding central $68\%$ intervals have widths of approximately $1.7\cdot10^{-3}$ and $2.6\cdot10^{-3}$, respectively. This indicates that the additional component provides sufficient freedom to absorb a large fraction of the structure leaving a systematic bias in the 2-component model, consistent with the findings in Section~\ref{subsec:svd}.\\
\indent Interestingly, increasing the model further to four foreground components does not provide any noticeable improvement. The mean recovered $r$ remains essentially unchanged, at approximately $3.9\cdot10^{-3}$ for \texttt{d1s1} and $8.7\cdot10^{-3}$ for \texttt{d10s5}. At the same time, the central $68\%$ interval broadens moderately, with widths of approximately $2.3\cdot10^{-3}$ and $3.3\cdot10^{-3}$, due to the larger parameter space. This suggests that adding a fourth component has little impact on the CMB residuals, but instead primarily acts within the subspace spanned by the Galactic components, effectively providing a refinement of the foreground model in SMICA.\\
\indent Finally, increasing the model complexity to five foreground components produces a further reduction in the systematic residual, bringing the mean recovered $r$ down to approximately $3.6\cdot10^{-4}$ for \texttt{d1s1} and $1.0\cdot10^{-4}$ for \texttt{d10s5}. This places the recovered values in effectively unbiased territory for both foreground models. However, this improvement comes at a substantial cost in statistical uncertainty. The central $68\%$ interval widths increase to approximately $7.4\cdot10^{-3}$ for \texttt{d1s1} and $1.4\cdot10^{-2}$ for \texttt{d10s5}, corresponding to increases of roughly a factor of three to four relative to the three-component configuration. We also report the one-sided upper $68\%$ limit from $r_{\rm true} = 0$ of $r < 4.5\cdot10^{-3}$ and $r < 8.4\cdot10^{-3}$ for the two cases.\\
\indent The improvement in foreground cleaning can also be observed in pixel space, as demonstrated in Figure~\ref{fig:comps_resmap}. The maps are overlaid with the Planck GAL60 mask, which effectively defines the full region over which we perform component separation. Cleaning with a three-component foreground model is significantly more efficient than with two components, with many large-scale features removed. Similar to the harmonic-space results, the difference in systematic residual in CMB maps between the $N_{\mathrm{d}}=3$ and $N_{\mathrm{d}}=4$ cases is difficult to discern, while for the highest number of components tested, the remaining foreground features are heavily suppressed.\\
\begin{figure}
    \centering
    \includegraphics[width=0.49\linewidth]{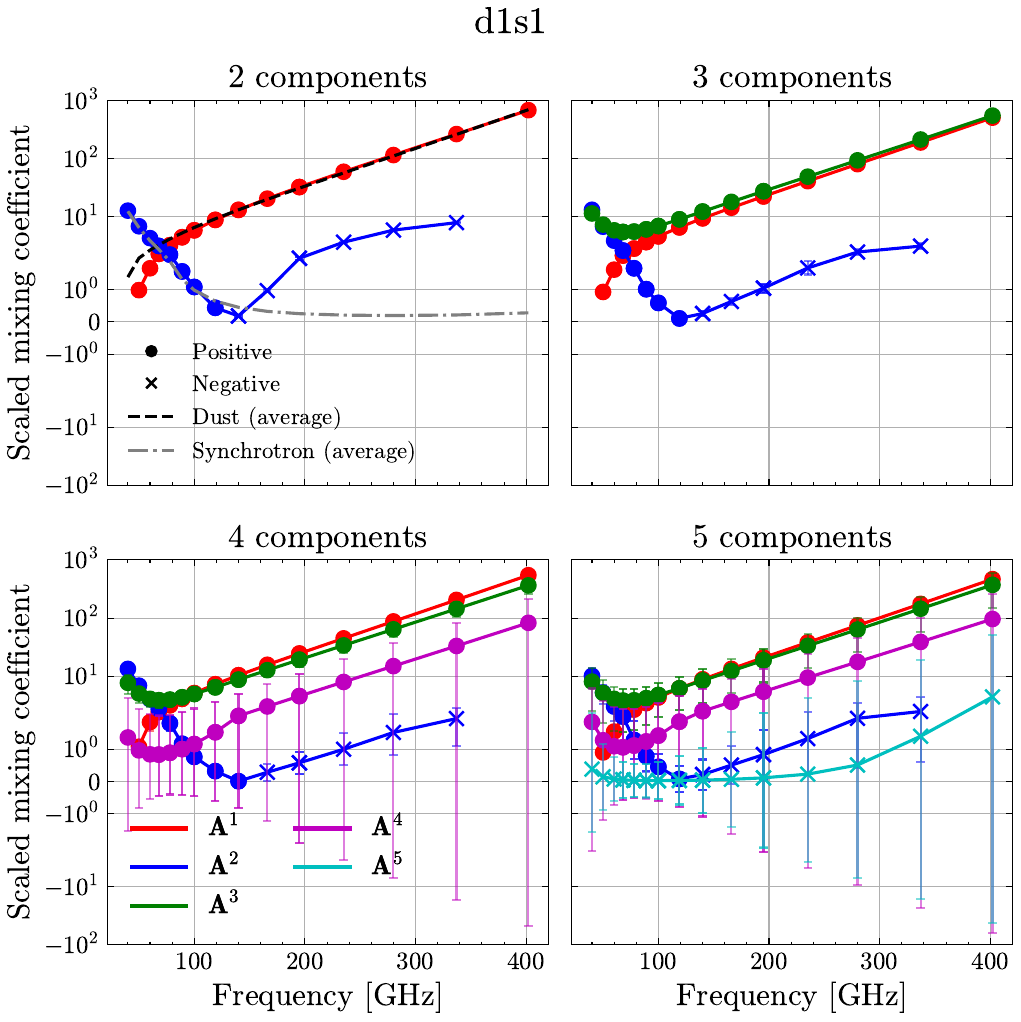}
    \includegraphics[width=0.49\linewidth]{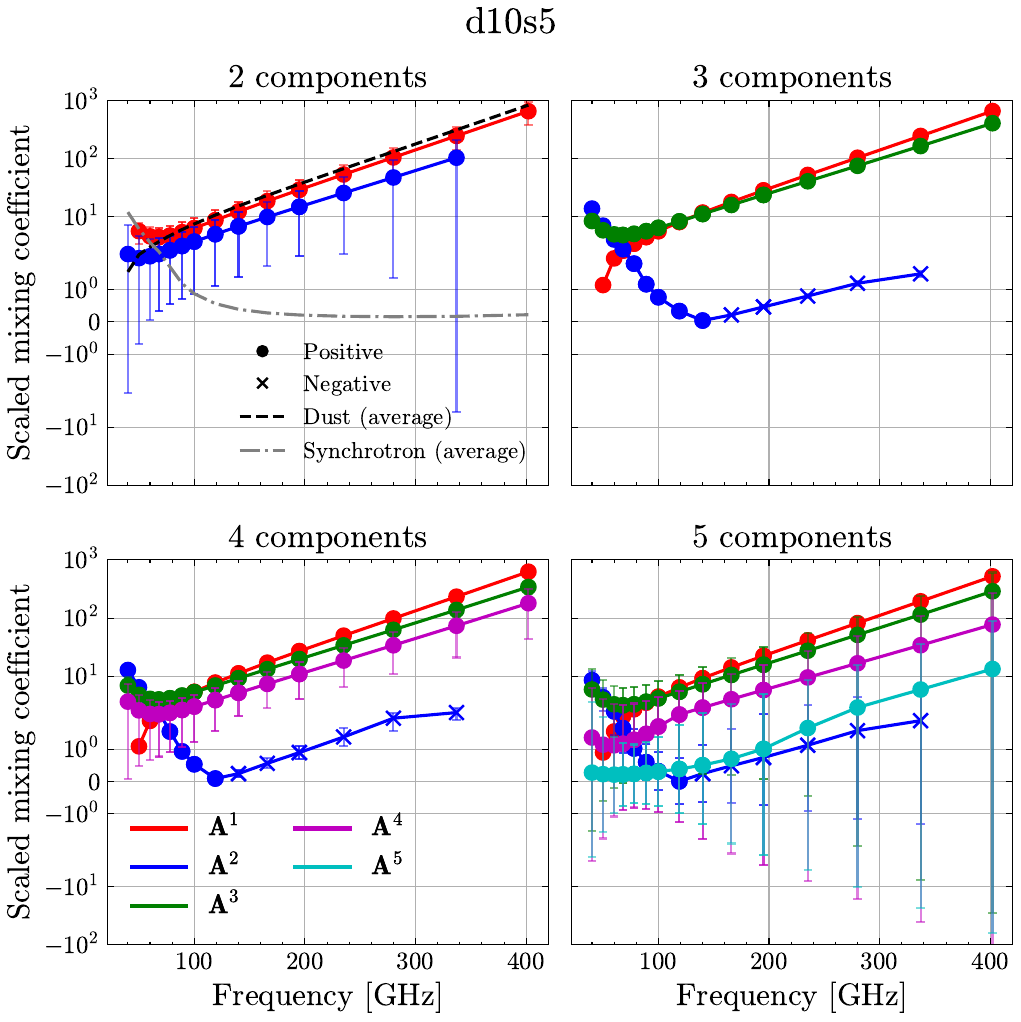}
    \caption{Realization-averaged columns of the best-fit SMICA mixing matrices for the \texttt{d1s1} (left) and \texttt{d10s5} (right) foreground models, for $N_{\rm d}=2,3,4,5$ components. The first two components satisfy the constraints of Eq.~\eqref{eq:mixmat_constraints}; the remaining, unconstrained components are not uniquely identified by the fit and are sorted by decreasing amplitude before averaging. Each column is scaled by $\sqrt{\sum_b n_b C_b^{ii}}$ to reflect its relative contribution to the total foreground power. For $N_{\rm d}=2$, the SED from the PySM3 dust and synchrotron templates (dashed, shaded by the spatial std.\ of $\beta$ and $T_{\rm d}$) is shown for comparison. Circles/crosses mark positive/negative mean coefficients; error bars show the standard deviations across realizations.}
    \label{fig:mixmat_ext_comps}
\end{figure}
\indent Another useful quantity to examine is the best-fit SMICA mixing matrix. Figure~\ref{fig:mixmat_ext_comps} shows the realization-averaged mixing laws for the four configurations with $N_{\rm d}=2$, $3$, $4$, and $5$, for both the \texttt{d1s1} and \texttt{d10s5} foreground models. The first two constrained components generally retain their expected dust- and synchrotron-like behavior due to the imposed constraints in Eq.~\eqref{eq:mixmat_constraints}: the first closely tracks the dust template, while the second follows the synchrotron template at low frequencies but turns over and crosses to negative values at high frequencies in most configurations, reflecting a partial cancellation against the dust contribution, capturing partially the imperfections of our modeling. The main exception is the $N_{\rm d}=2$ \texttt{d10s5} case, where the second component instead stays positive and bends upward with a dust-like slope at high frequency, suggesting that with only two degrees of freedom, SMICA uses both components to describe the more complex dust emission rather than cleanly separating dust and synchrotron. The additional, unconstrained components instead exhibit effective spectral shapes combining dust- and synchrotron-like behavior, as expected from the separability discussion in Sec.~\ref{subsec:smica}: when the underlying spectra are close to proportional, SMICA cannot uniquely resolve their individual directions, only the subspace they jointly span, up to an internal rotation. Since dust emission dominates, this rotation imprints a dust-like shape and leaks power onto the extra components even if they represent comparatively weak, sub-dominant modes. Their variance also differs: the third component remains stable across seeds, consistent with it absorbing the dominant share of the residual foreground power, whereas the fourth and fifth show considerably larger variance. This matches their interpretation as progressively sub-dominant directions in the foreground subspace, whose low power leaves their amplitude and orientation weakly constrained and thus more susceptible to noise.\\
\indent These results highlight the inherent trade-off of increasing the dimensionality of the SMICA model. While additional components provide greater freedom to describe the foreground covariance and reduce the systematic residual, this improvement eventually comes at the cost of substantially increased statistical uncertainty. The transition from $N_{\mathrm{d}}=2$ to $N_{\mathrm{d}}=3$ provides the most favourable improvement, reducing the bias by roughly an order of magnitude while retaining relatively tight constraints on $r$. Increasing to four components provides no improvement on the bias, whereas five components can reach an effectively unbiased regime but at the expense of a substantially broader posterior. \\
\indent We therefore consider $N_{\mathrm{d}}=3$ to provide the best compromise between systematic and statistical residuals and adopt it as our baseline configuration for the analyses presented in the following sections. From this point onward, we refer to this exact configuration as ``full-sky'' (despite not covering the full sky), in contrast to the subset/patches approach. However, note that this baseline configuration still leaves a residual bias on $r$ at the $10^{-3} - 10^{-2}$ level. \\
\indent An additional discussion on the nature of these extra components is provided in Section~\ref{subsec:moment_expansion}. There, we attempt to provide a physical rationale for the increase in the effective dimensionality of the foreground covariance in harmonic space and discuss how SMICA operates in this regime.
\subsection{The arising ILC-bias in the SMICA on subset approach}\label{subsec:ilcbias}
As simply increasing the number of components in the model could not fully account for the complexity of the foregrounds, leaving a bias on the order of $10^{-3} - 10^{-2}$, we resort to exploring the SMICA on subset approach. Before going to the approach itself, we examine the methodological ILC bias that arises and must be handled carefully. \\
\indent The ILC bias~\cite{Delabrouille_2008, Coulton_2024} is a well-known effect in blind component separation, often coming with the ILC family of methods, hence the name. As demonstrated in Section \ref{subsec:smica}, the effect is related to the spurious correlations between the sky signal and residuals after component separation that is propagated by the reconstruction operator $\mathbf{\hat{W}}$. Therefore, from Eq.~\ref{eq:cl_gls}, we can define the ILC bias as:
\begin{align}
    C_{\ell}^{\rm ILC-bias,\,(i)} =     2\mathcal{C}_\ell
    \left[
        \mathbf{s}^{(i)}\mathbf{m}_{\rm res}^{(i),\dagger}
    \right] = \tilde{C}_\ell^{(i),\rm GLS} - C_{\ell}^{\rm res,\,(i)} - C_\ell^{(i)}.
\end{align}
Ideally, this term should be small. \\
\indent Leveraging the fact that we are working with simulations, we can compute this term explicitly and evaluate its impact on the tensor-to-scalar ratio estimation by including it in the likelihood of Eq.~\eqref{eq:cosmo_like},
with the following modification:
\begin{align}
    C_\ell^{\rm total,\,(i)}
    =
    C_\ell^{\mathrm{lens}}
    +
    C_\ell^{\rm ILC-bias,\,(i)}
    +
    C_\ell^{\rm stat,\,(i)}.
\end{align}
Here, we omit the systematic residual $C_\ell^{\rm sys}$ and replace with $C_\ell^{\rm ILC-bias,\,(i)}$ in order to isolate the impact of the ILC bias on the $r$ estimation, as the systematic bias could partially compensate and obscure its effect. We evaluate the modified likelihood over 200 realizations and use the resulting ensemble-averaged posterior to quantify its impact on $r$. We further characterize the significance of this effect by expressing the
bias on $r$ in units of the standard error of the ensemble mean:
\begin{equation}
    S_{\rm ILC}
    =
    \frac{r_{\rm bias}^{\rm ILC}}
    {\left(r_{\rm upper}^{\rm ILC}-r_{\rm lower}^{\rm ILC}\right)/\sqrt{N_{\rm sim}}},
    \label{eq:ilc_bias_significance}
\end{equation}
where $N_{\rm sim}$ is the number of realizations. Ideally, $S_{\rm ILC}$ should be close to zero, and we deem the bias to be insignificant when $|S_{\rm ILC}|<1$, meaning it remains within one standard deviation of the true value. \\
\begin{figure}
    \centering
    \includegraphics[width=0.49\linewidth]{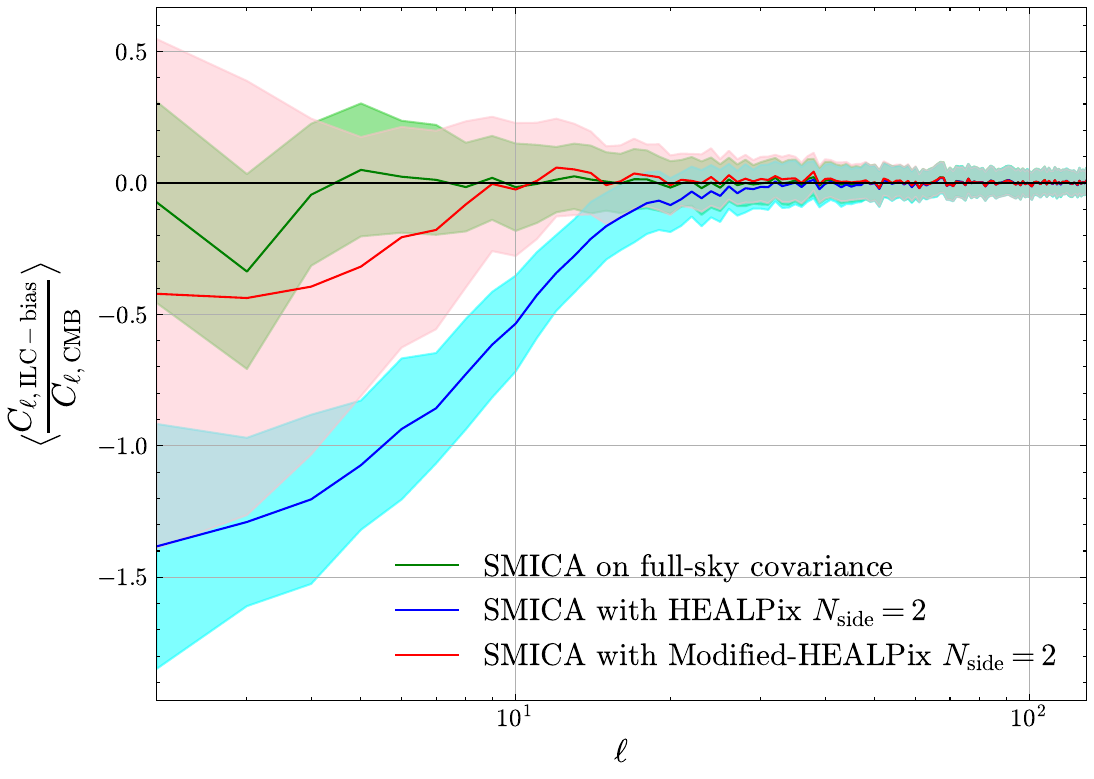}
    \includegraphics[width=0.49\linewidth]{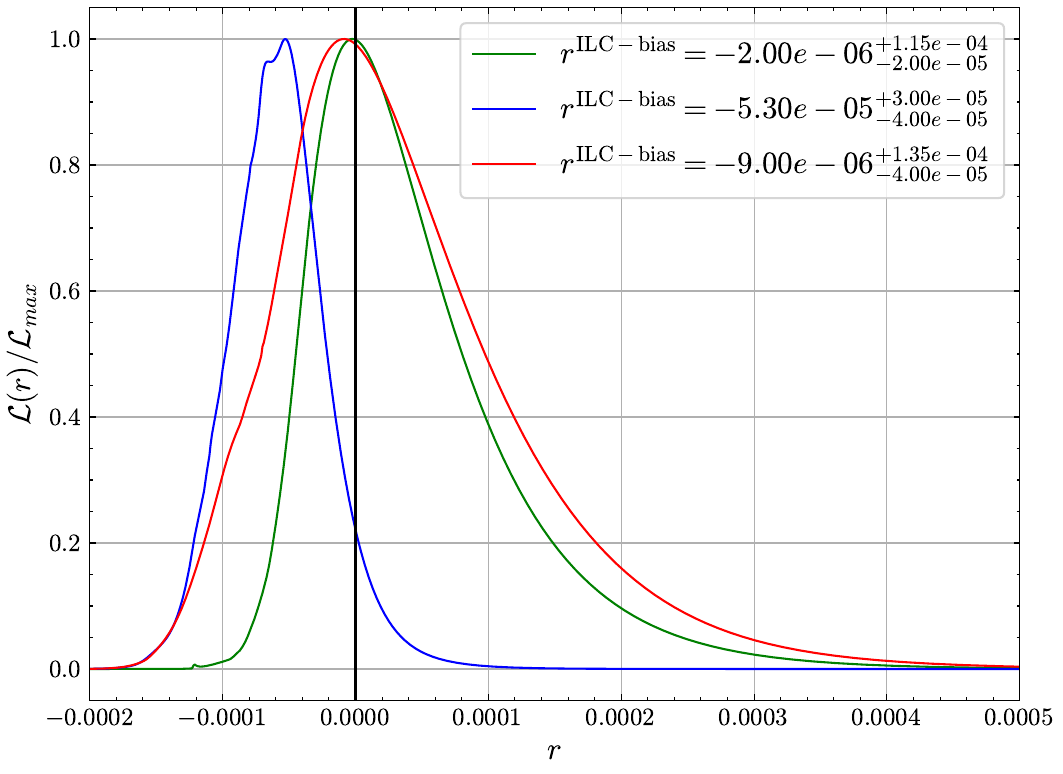}
    \caption{The arising ILC-bias in the SMICA on subset approach, demonstrated using results on the simple \texttt{d0s0} model. Left: The mean and variance over realizations of the ILC-bias signal $C_\ell^{\rm ILC-bias}$ are normalized to the lensing B-mode for 3 cases, full-sky, HEALPix and modified-HEALPix of $N_{\rm side} = 2$. Right: The average posterior on $r$ induced by the ILC-bias for each of the cases.}
    \label{fig:ilc_bias}
\end{figure}
\indent Figure~\ref{fig:ilc_bias} illustrates the ILC-like bias for the simple \texttt{d0s0} model with a 2-component SMICA run, comparing full-sky, HEALPix $N_{\rm side}=2$, and modified-HEALPix $N_{\rm side}=2$ covariance estimation. The bias spectra are computed on the full sky, highlighting that the effect is intrinsic to the component-separation procedure rather than introduced by a subsequent power-spectrum estimation step. With full-sky covariance, the large number of available modes suppresses spurious correlations, keeping the effect negligible, yielding $S_{\rm ILC}\simeq-0.21$. In contrast, the standard HEALPix patches produce a strong suppression of the largest-scale CMB power, corresponding to a clear shift of order $-5\cdot10^{-5}$ in $r$ that is highly significant given that $S_{\rm ILC}\simeq-10.7$. The modified-HEALPix approach substantially reduces this effect to a satisfactory level of $S_{\rm ILC}\simeq-0.72$, albeit with an increased statistical uncertainty. Note that the bias is only reduced, not fully cured, as the modes, especially large-scale ones, likely remain continuous despite the pixel domain cut. Therefore, a negative bias slightly larger than in the full-sky case remains in the final estimate. Nevertheless, we will still present results obtained with the standard HEALPix pipeline with the ILC bias removed alongside the modified-HEALPix approach to demonstrate the possible strength of the approach.\\
\subsection{SMICA on independent subsets of pixels}\label{subsec:patches}
\begin{figure}
    \centering
    \includegraphics[width=0.49\linewidth]{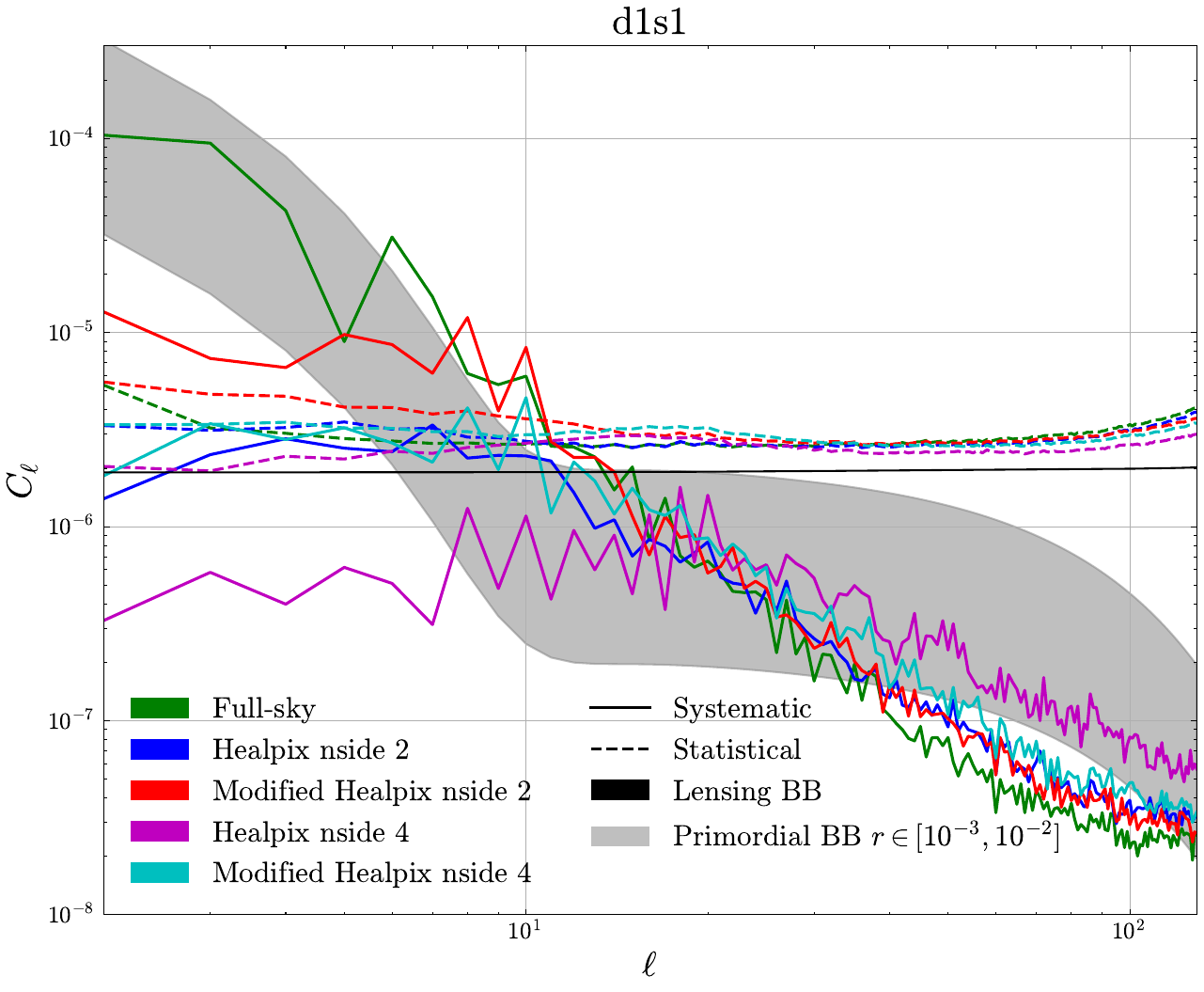}
    \includegraphics[width=0.49\linewidth]{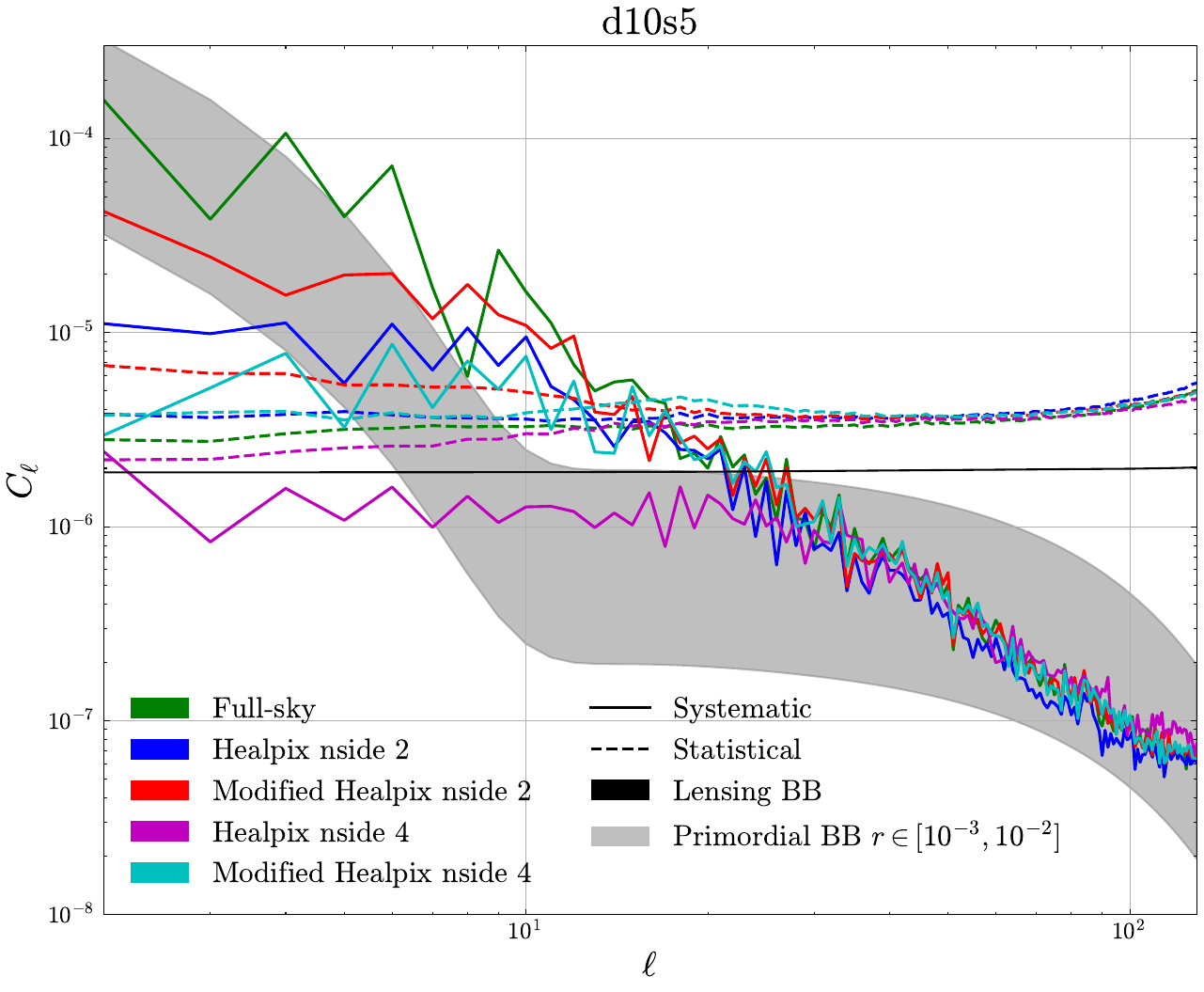}
    \caption{Systematic and mean statistical residuals obtained for the different pipeline configurations, shown for $\texttt{d1s1}$ (left) and $\texttt{d10s5}$ (right). Five different sky-partitioning strategies are considered: full-sky, HEALPix with $\texttt{Nside}=2$, HEALPix with $\texttt{Nside}=4$, modified HEALPix with $\texttt{Nside}=2$, and modified HEALPix with $\texttt{Nside}=4$.}
    \label{fig:patches_res}
\end{figure}

\begin{figure}
    \centering
    \includegraphics[width=1.0\linewidth]{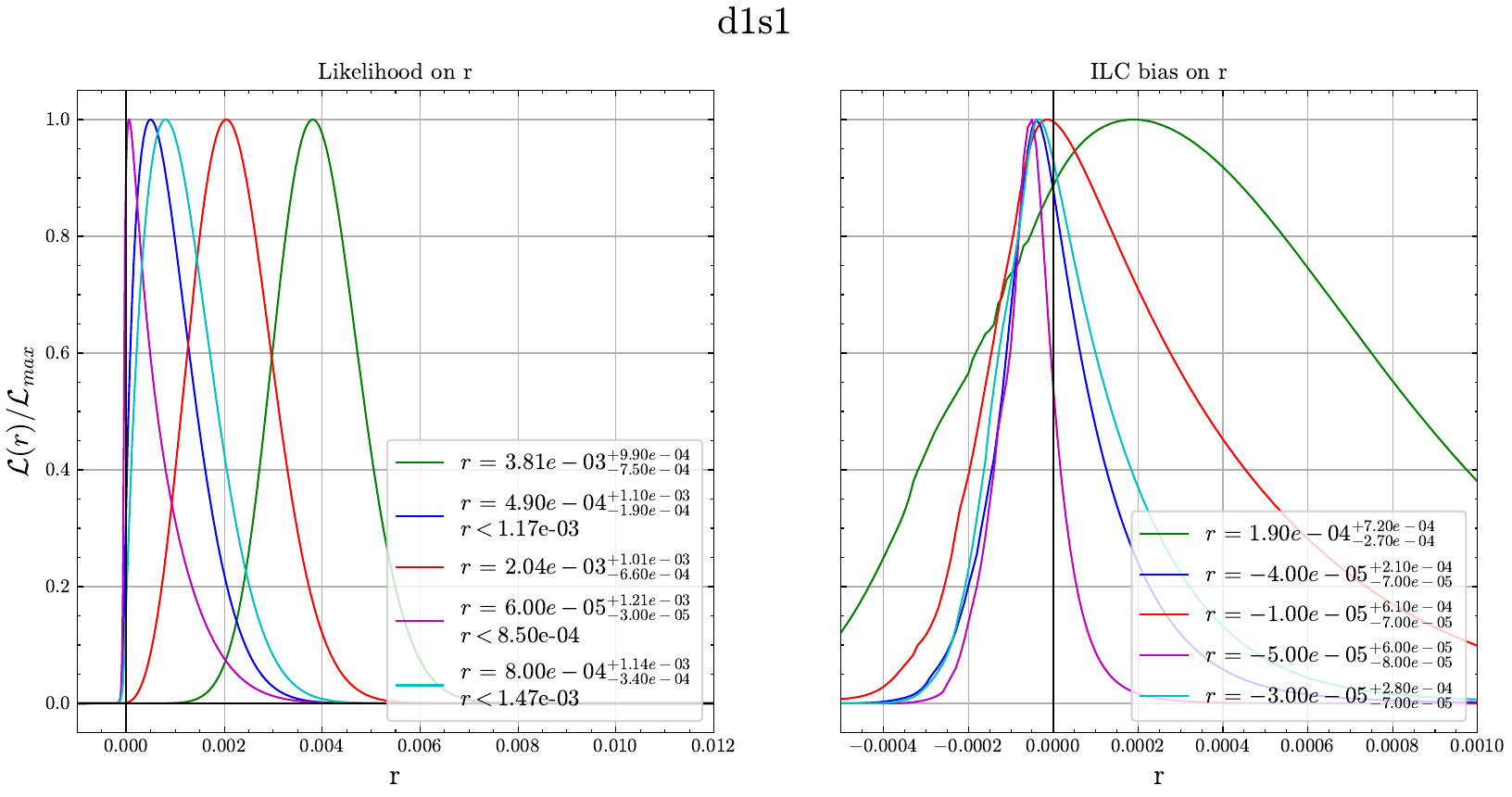}
    \includegraphics[width=1.0\linewidth]{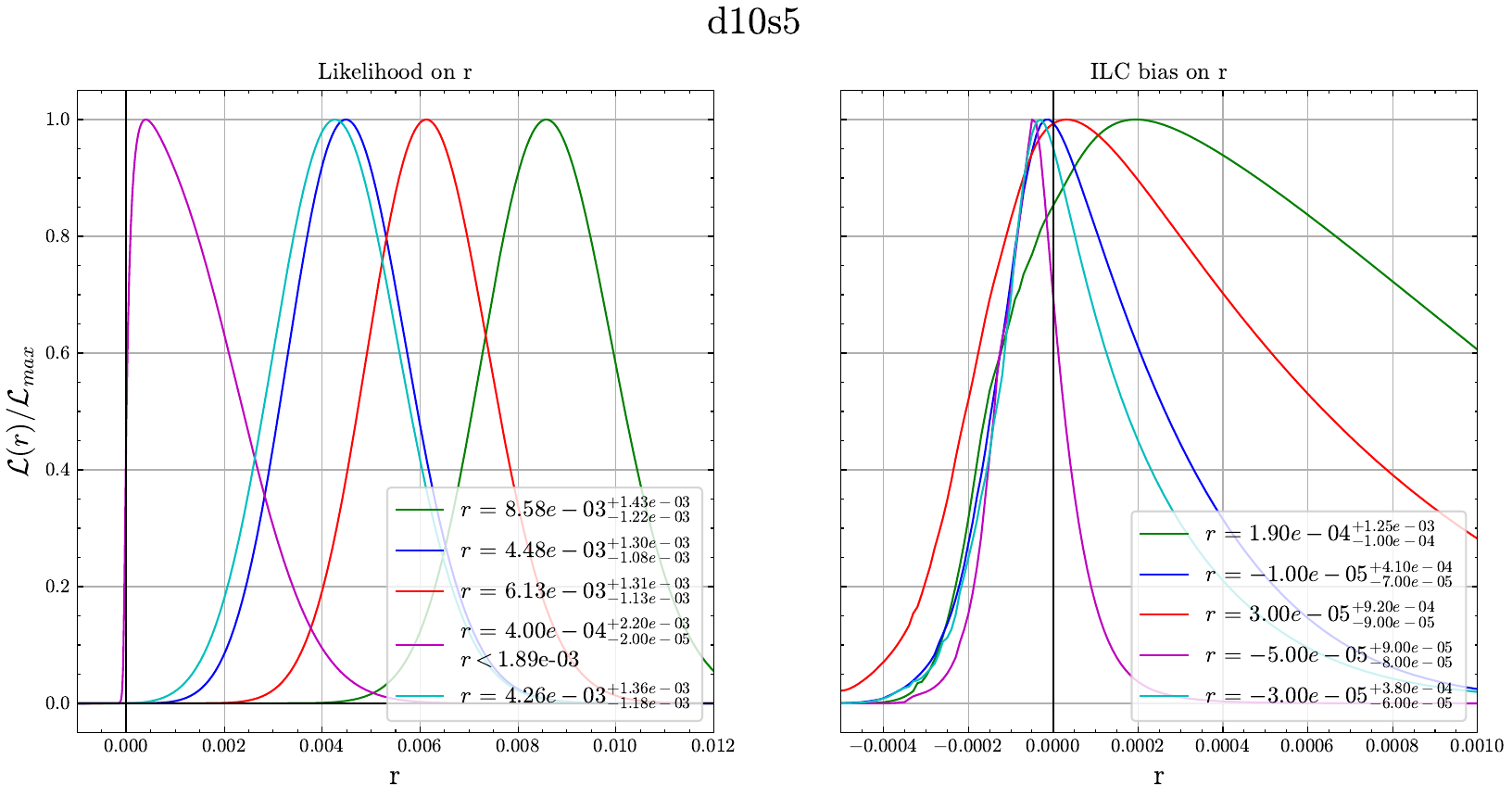}
    \caption{The average posterior (left) corresponding to the residuals in Figure \ref{fig:patches_res} (same color code for the different sky partitioning strategies) on $r$ and the associated ILC-bias (right) of the different modes of the pipeline on \texttt{d1s1} (top) and \texttt{d10s5} models (bottom).}
    \label{fig:patches_lklh}
\end{figure}

\begin{figure}
    \centering
    \includegraphics[width=1\linewidth]{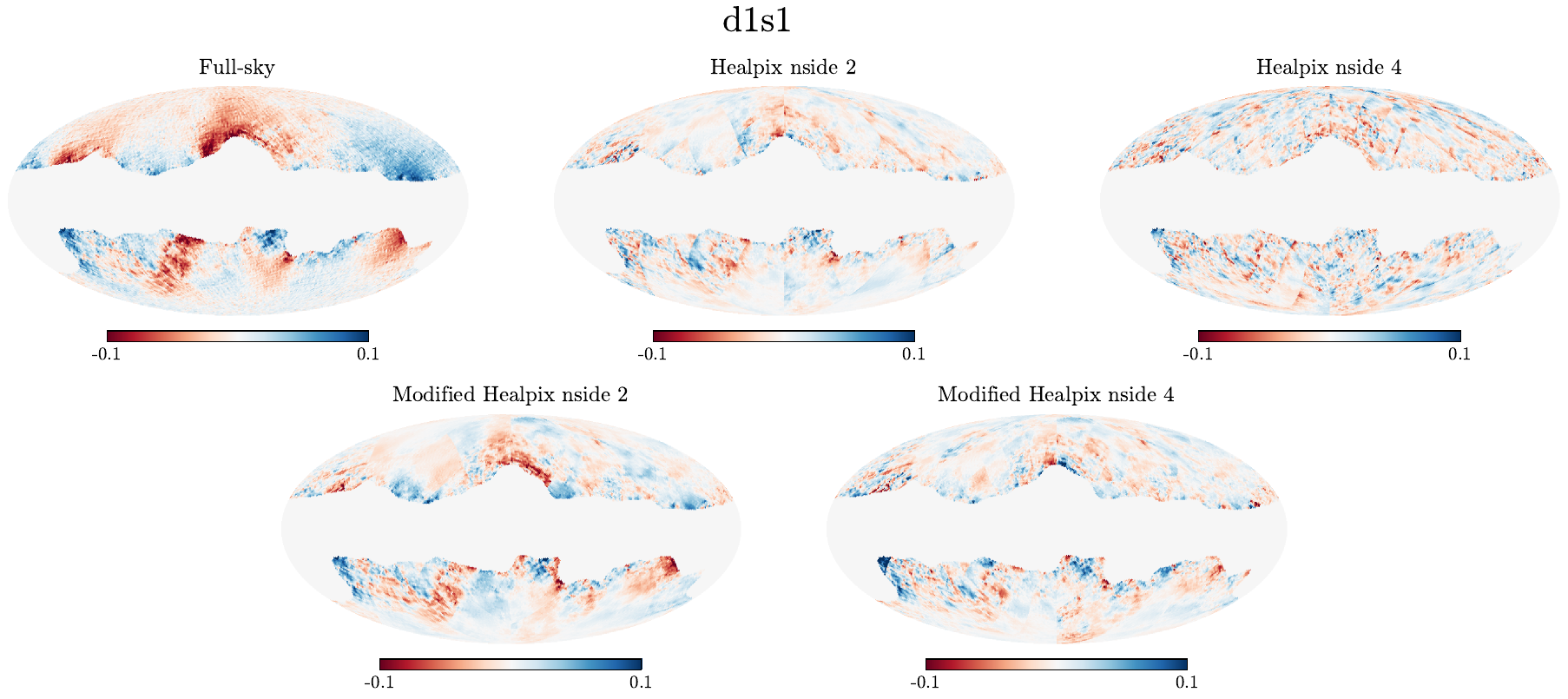}
    \includegraphics[width=1\linewidth]{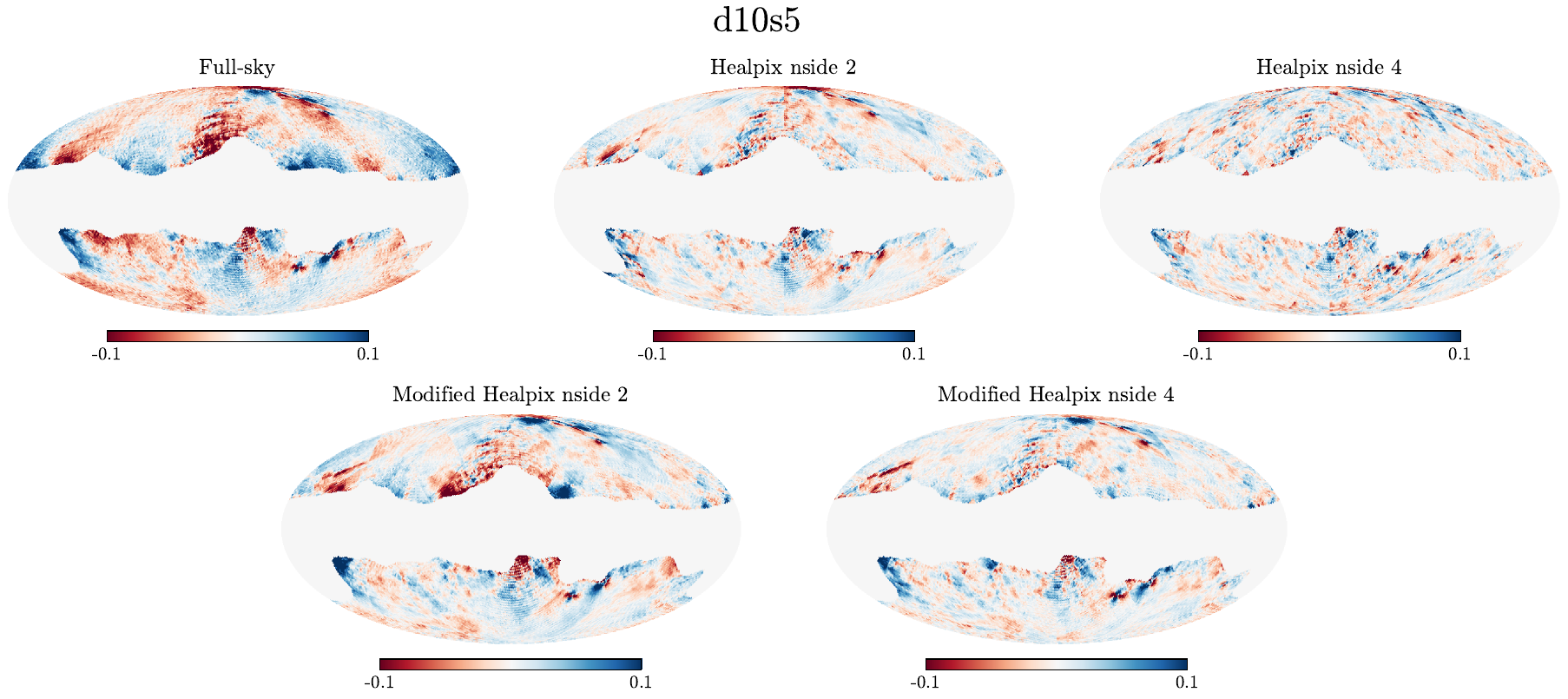}
    \caption{Systematic residual in pixel space for the different configuration of the pipeline. First set of 5 plots is for \texttt{d1s1} model while the second is for \texttt{d10s5}. Planck GAL60 is overlayed on each of the maps to highlight the region where component separation is effectively performed.}
    \label{fig:patches_resmap}
\end{figure}

We present the performance of the proposed SMICA-on-HEALPix-patches pipeline on simulated observations with \texttt{d1s1} and \texttt{d10s5} foreground models. The residuals, separated into systematic and statistical contributions, are shown in Figure~\ref{fig:patches_res}, together with the lensing $B$-mode spectrum and the primordial $B$-mode signal for $r\in[10^{-3},10^{-2}]$ for reference. Figure~\ref{fig:patches_lklh} shows two corresponding posterior distributions of the tensor-to-scalar ratio: one obtained after propagating the systematic and statistical residuals described in Section~\ref{subsec:likelihood_r} to the likelihood, and another isolating the contribution from the ILC bias, following the methodology of Section~\ref{subsec:ilcbias}. Table~\ref{tab:ilc_bias} further quantifies this effect by reporting the negative bias on $r$ associated with the isolated ILC bias, together with its significance, $S_{\rm ILC}$. Finally, Figure~\ref{fig:patches_resmap} presents the average foreground residual maps obtained with the different implementations.\\
\begin{table}
    \centering
    \begin{tabular}{llcc}
        \hline
        Foreground & Patches & $r_{\rm bias}^{\rm ILC\, bias}$ & $S_{\rm ILC}$ \\
        \hline
        \texttt{d1s1}  & $N_{\rm side}=2$             & $-4$ & $-2.02$ \\
        \texttt{d1s1}  & Modified $N_{\rm side}=2$ & $-1$ & $-0.21$ \\
        \texttt{d1s1}  & $N_{\rm side}=4$             & $-5$ & $-5.05$ \\
        \texttt{d1s1}  & Modified $N_{\rm side}=4$ & $-3$ & $-1.21$ \\
        \hline
        \texttt{d10s5} & $N_{\rm side}=2$             & $-1$ & $-0.29$ \\
        \texttt{d10s5} & Modified $N_{\rm side}=2$ & $3$ & $0.42$ \\
        \texttt{d10s5} & $N_{\rm side}=4$             & $-5$ & $-4.16$ \\
        \texttt{d10s5} & Modified $N_{\rm side}=4$ & $-3$ & $-0.97$ \\
        \hline
    \end{tabular}
    \caption{ILC-bias results for the patch-based SMICA configurations. All $r^{\rm ILC\,bias}$-quantities are given in units of $10^{-5}$. The $S_{\rm ILC}$ column is computed based on Eq.~\eqref{eq:ilc_bias_significance} and should be interpreted as number of $\sigma$ away from $r_{\rm true} = 0.$.}
    \label{tab:ilc_bias}
\end{table}
\indent Introducing spatial variability through standard HEALPix patches substantially improves the foreground cleaning, particularly on large angular scales in power spectrum space, where the model with a constant mixing matrix across the sky leaves significant foreground residuals. Imposing a finer patch resolution of $N_{\rm side}=4$ further reduces the systematic residual, translating into a corresponding reduction of the bias on $r$ to below $5\cdot10^{-4}$ for both foreground models. Therefore, this scheme provides a nearly unbiased recovery in all but the $N_{\rm side}=2$--\texttt{d10s5} case, with the improvement also visible through the suppression of large-scale foreground features in the reconstructed maps.\\
\indent This improved cleaning, however, comes at the cost of a significant ILC bias. Since the SMICA weights are estimated from the same sky region to which they are applied, they can adapt optimally to local foreground properties but become contaminated by realization-dependent spurious correlations. As shown in Table~\ref{tab:ilc_bias}, although the biases are only a few $10^{-5}$, they are statistically significant, and the effect becomes stronger for smaller patches as fewer modes are available for covariance estimation. In addition to this, the suppressed statistical residual observed in the HEALPix patch configurations should not be interpreted as a genuine gain in precision. This is because, as the weights contain correlations, the cross-spectrum between the two terms entering $C_\ell^{\rm stat}$ in Eq.~\eqref{eq:stat_def}, which was previously neglected, becomes non-negligible. This leads to artificially reduced uncertainty and, consequently, an underestimated uncertainty on $r$. Although this effect was not clearly visible in the simpler \texttt{d0s0} considered previously, it becomes apparent for the more complex models, where the larger residual contamination can induce stronger suppression, further illustrating the difficulty of accurately characterizing the ILC bias.\\
\indent These behaviours are avoided in the modified HEALPix-patch scheme, which was designed to mitigate the ILC-bias. The effectiveness of the correction is evident in the right panel of Fig.~\ref{fig:patches_lklh}, where the ILC-bias becomes consistent with zero for both foreground models in the $N_{\rm side}=2$ case, with the corresponding $S_{\rm ILC}$ values in Table~\ref{tab:ilc_bias} reduced to unity or below. We also observe that the statistical residual spectra are no longer artificially suppressed, as in the standard HEALPix-patch case. This substantial improvement in mitigating the ILC-bias, however, comes at the cost of some foreground-cleaning performance, as the available information is no longer used optimally. This trade-off first appears in harmonic space through the increased systematic residual spectra relative to the standard HEALPix-patch implementation. It is also reflected at the likelihood level, where the modified weighting scheme sacrifices much of the particularly low-bias solution obtained with the standard patches: the recovered bias is approximately $2.0\cdot10^{-3}$ and $8.0\cdot10^{-4}$ for \texttt{d1s1}, and $6.1\cdot10^{-3}$ and $4.2\cdot10^{-3}$ for \texttt{d10s5}, for $N_{\rm side}=2$ and $4$, respectively. Nevertheless, these values remain substantially below the corresponding full-sky biases, while the central $68\%$ interval widths are $1.7\cdot10^{-3}$ and $1.5\cdot10^{-3}$ for \texttt{d1s1}, and $2.4\cdot10^{-3}$ and $2.5\cdot10^{-3}$ for \texttt{d10s5}, for $N_{\rm side}=2$ and $4$, respectively. Thus, the modified scheme retains a substantial reduction in bias without significantly degrading the statistical constraining power, while sacrificing some of the cleaning efficiency of the standard HEALPix patches. This trade-off is also visible in the residual maps of Fig.~\ref{fig:patches_resmap}, where more foreground structure remains than in the standard patch configuration, but still substantially less than in the full-sky case.\\
\indent For the modified-HEALPix $N_{\rm side}=4$, a rather significant ILC-bias remains despite the correction, with $S_{\rm ILC}$ of order unity for both foreground models. Further suppression may be possible by extending the covariance-estimation region, increasing the exclusion cuts, or removing selected $\ell$ modes from the covariance estimation, but these modifications would come at the cost of further degrading the foreground-cleaning performance. This therefore represents a current limitation of the approach: ideally, increasingly fine spatial resolutions would allow the component-separation model to better capture more complex foreground variations, but the correction becomes progressively more difficult to maintain at small patch sizes.\\
\indent One noteworthy feature of the different component-separation configurations is that most of the improvement is concentrated at large angular scales. This behaviour is expected, as the SMICA-on-subsets approach allows the mixing matrix to vary between patches, thereby accounting for large-scale spatial variations of the foreground spectral properties. At smaller angular scales, however, the residual power spectra of the different configurations become much more similar. At these scales, specifically those that are smaller than the size of the patch, in addition to the foregrounds being weaker, all configurations rely on the same underlying SMICA model with three foreground components, and therefore possess a similar ability to describe the foreground emission.\\
\indent Overall, the patch-wise SMICA approach substantially improves the treatment of spatially varying foregrounds, but is affected by the ILC bias. The modified-HEALPix weighting scheme mitigates this effect while retaining much of the improvement in foreground cleaning for both foreground models. Despite these improvements, a significant systematic residual remains, producing a bias on the order of $2$ to $4
\cdot10^{-3}$ for \texttt{d1s1} and $6$ to $8\cdot10^{-3}$ for \texttt{d10s5} in the recovered tensor-to-scalar ratio. To further reduce this systematic bias, we consider two additional optimizations: improving the mask applied to the post-component-separation maps and marginalizing over foreground templates. To avoid cluttering the presentation, we focus only on two representative configurations: SMICA with a full-sky covariance and three foreground components, and SMICA on modified-HEALPix patches with $N_{\rm side}=2$.
\subsection{Effect of optimized post-component-separation mask}\label{subsec:ideal_mask}
\begin{figure}
    \centering
    \includegraphics[width=1\linewidth]{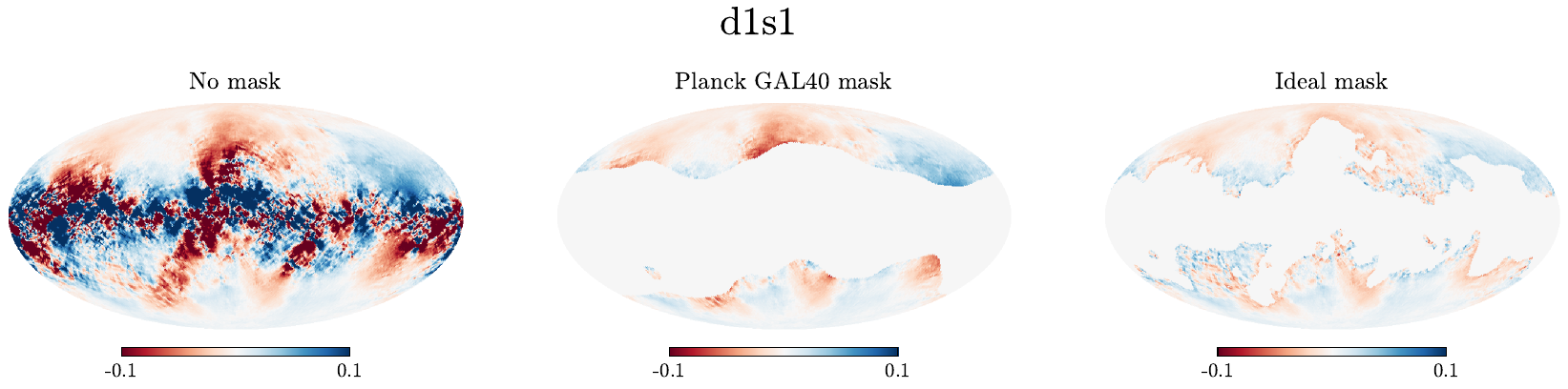}
    \includegraphics[width=1\linewidth]{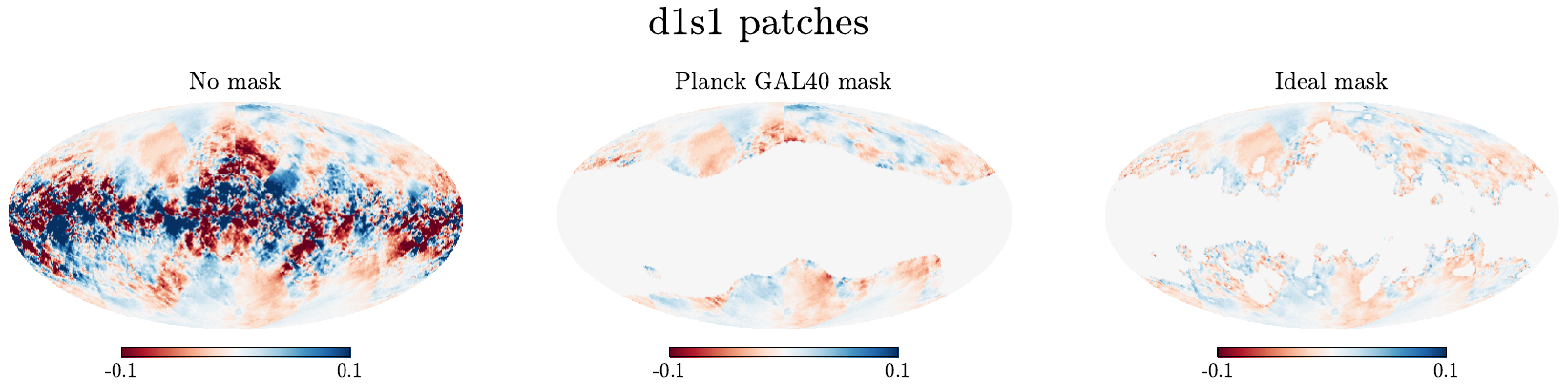}
    \includegraphics[width=1\linewidth]{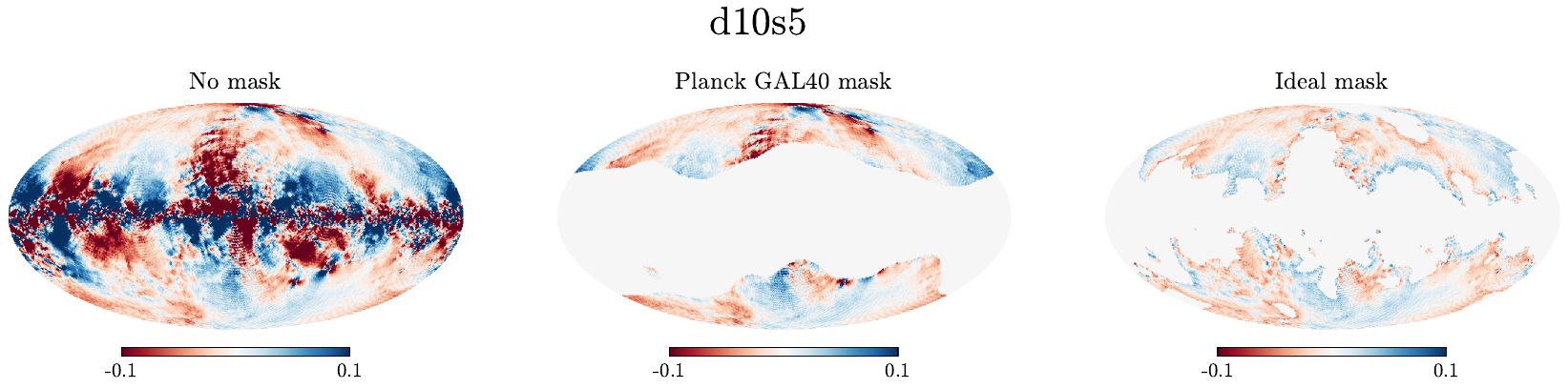}
    \includegraphics[width=1\linewidth]{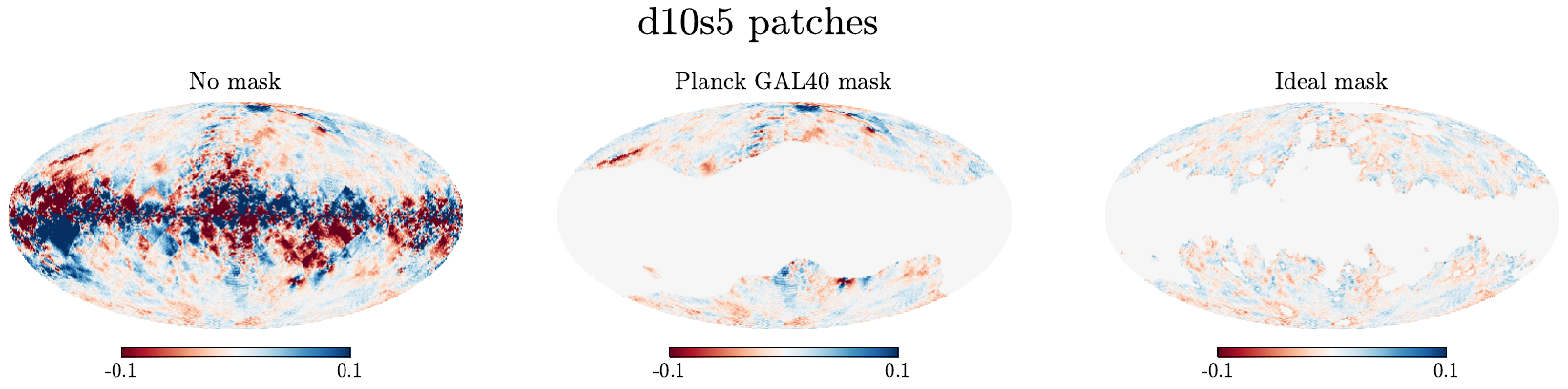}
    \caption{Comparison of the unmasked systematic residual with the Planck
    GAL40 and ideal masks for \texttt{d1s1} full-sky (first row),
    \texttt{d1s1} modified-HEALPix patches $N_{\rm side}=2$ (second row),
    \texttt{d10s5} full-sky (third row), and \texttt{d10s5} modified-HEALPix
    patches $N_{\rm side}=2$ (fourth row).}
    \label{fig:mask_comparison}
\end{figure}
\indent In this section, we'll shortly demonstrate the impact of optimizing the post-component-separation mask on the foreground residuals and the resulting constraints on $r$. As discussed in Section~\ref{subsec:masking}, the Planck GAL40 mask used throughout the previous sections is not optimal for polarized component separation. A more suitable strategy would be to construct a mask that traces the morphology of the residual foreground contamination. We therefore illustrate the maximum potential gain by thresholding the recovered systematic residual $\mathbf{m}_{\rm sys}$ to retain $f_{\rm sky}=0.5$, defining an "ideal" mask, shown in Figure~\ref{fig:mask_comparison}. Using these masks, we observe a reduction in the bias on $r$ of order $10^{-3}$ for \texttt{d1s1} and around $2$ to $3 \cdot10^{-3}$ for the more complex \texttt{d10s5} model, while the statistical uncertainty remains largely unchanged. Although such a construction is not possible with real data, it illustrates the potential of a residual-informed mask.\\
\indent Constructing such a mask from real observations is more challenging,
since the true residual foreground is unknown. In practice, one would
therefore need to rely on accurate models or external foreground products to
identify regions where component separation is expected to perform poorly.
Such approaches may provide a useful compromise between the idealized mask
considered here and a generic Galactic mask.
\subsection{Marginalizing over internal foreground templates}\label{subsec:marg}
\begin{figure}[tbp]
    \centering
    \includegraphics[width=0.48\linewidth]{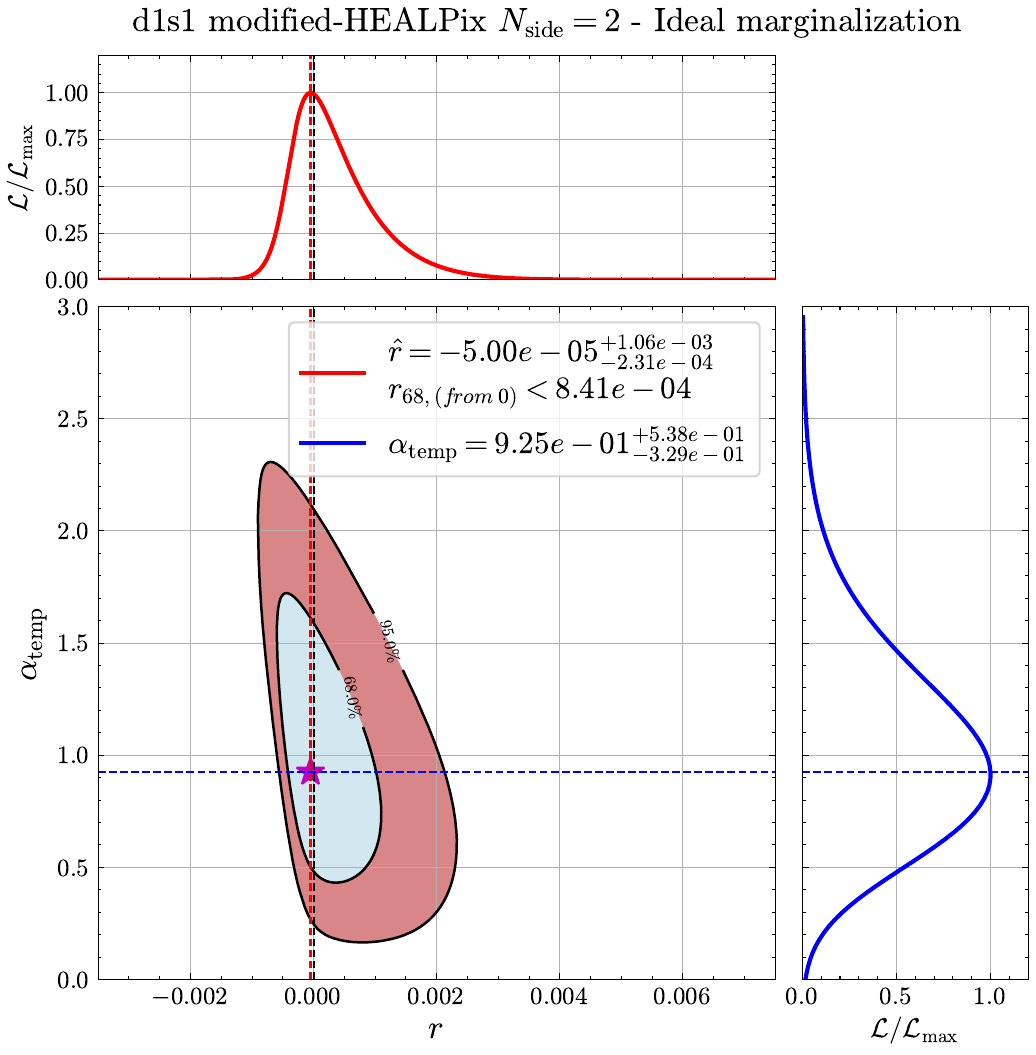}
    \includegraphics[width=0.48\linewidth]{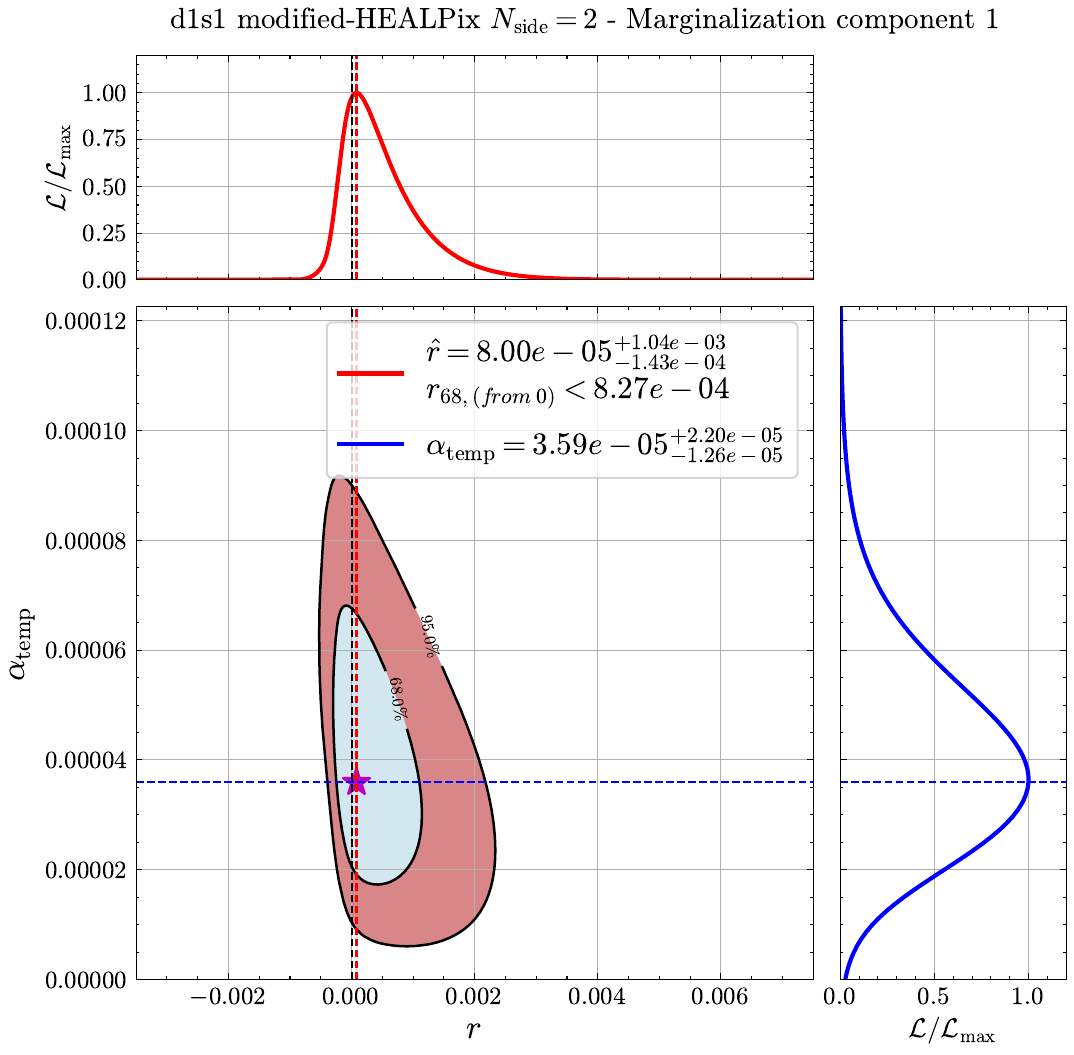}
    \includegraphics[width=0.48\linewidth]{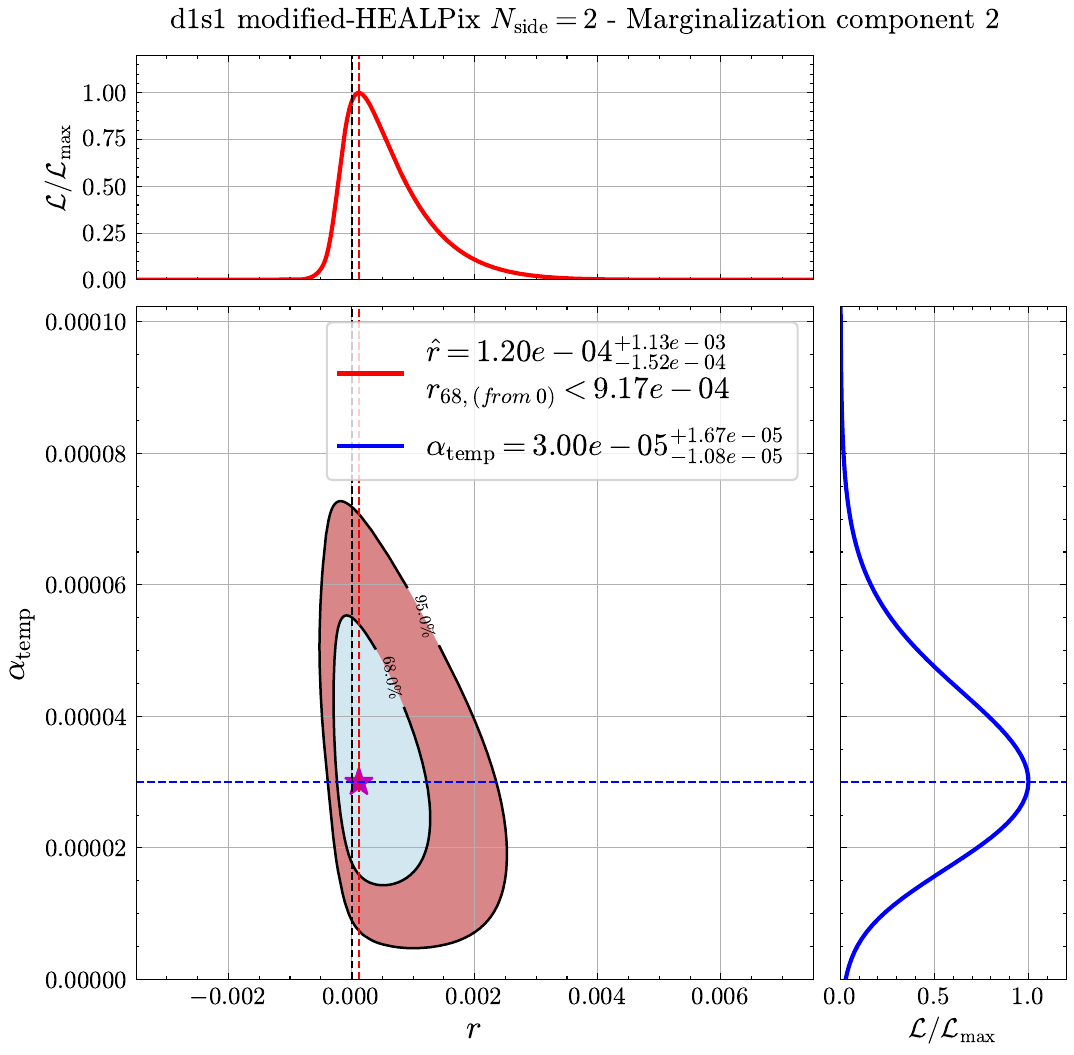}
    \includegraphics[width=0.48\linewidth]{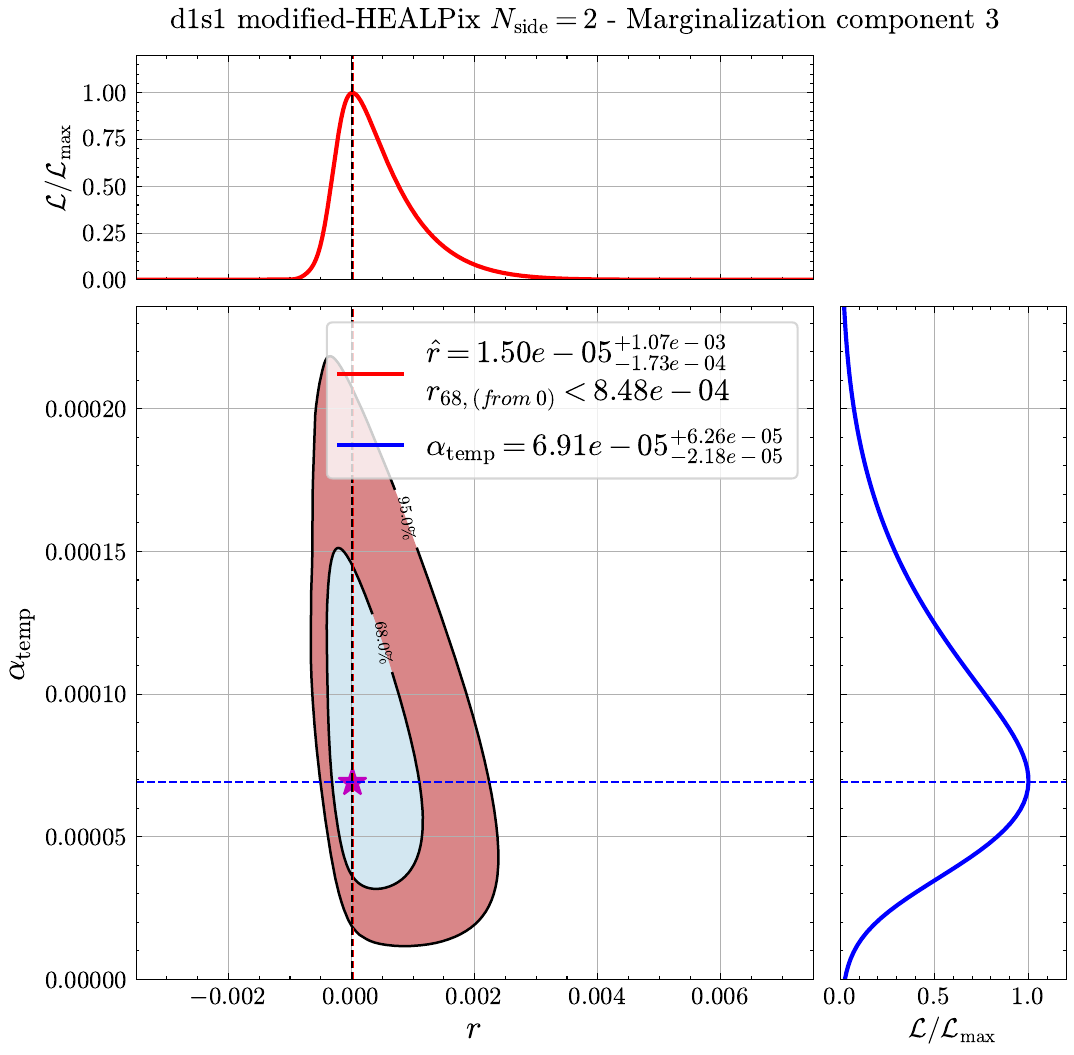}
    \caption{Result of marginalization on the modified-HEALPix $N_{\rm side} = 2$ SMICA running mode when applied to the \texttt{d1s1} model. The set of displayed plots is subsequently ideal (top left), realistic over component 1 (top right), component 2 (bottom left) and component 3 (bottom right).}
    \label{fig:patches_d1s1_marg}
\end{figure}

\begin{figure}[p]
    \centering
    \includegraphics[width=0.48\linewidth]{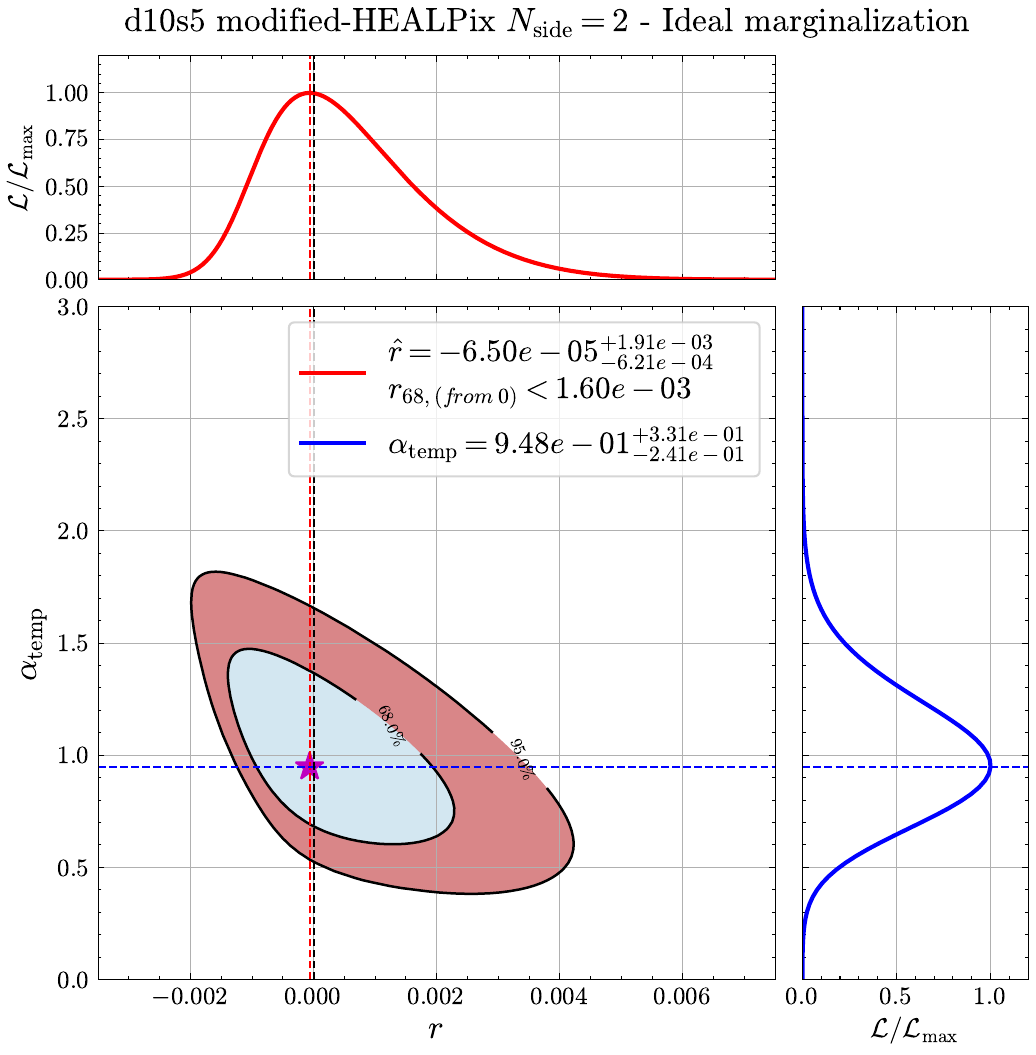}
    \includegraphics[width=0.48\linewidth]{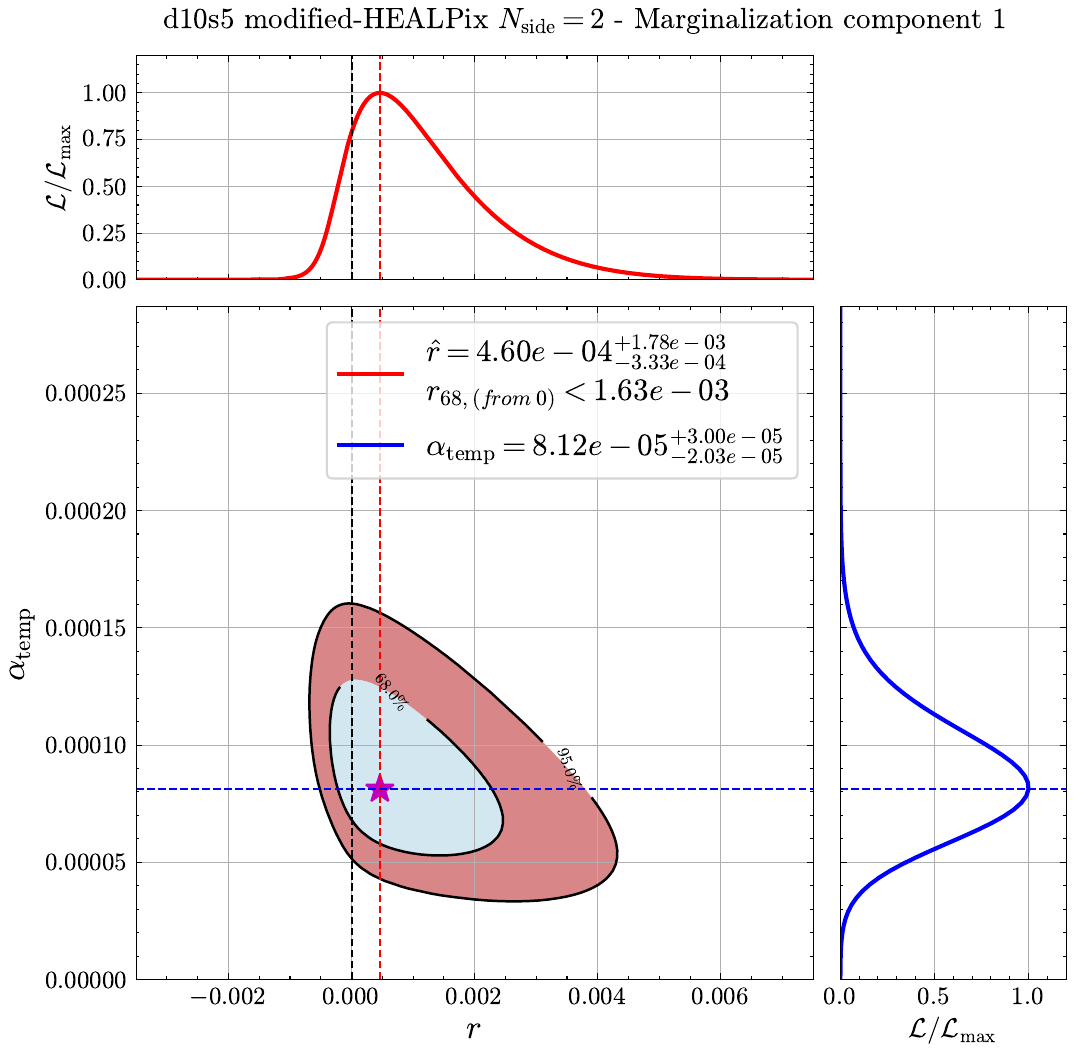}
    \includegraphics[width=0.48\linewidth]{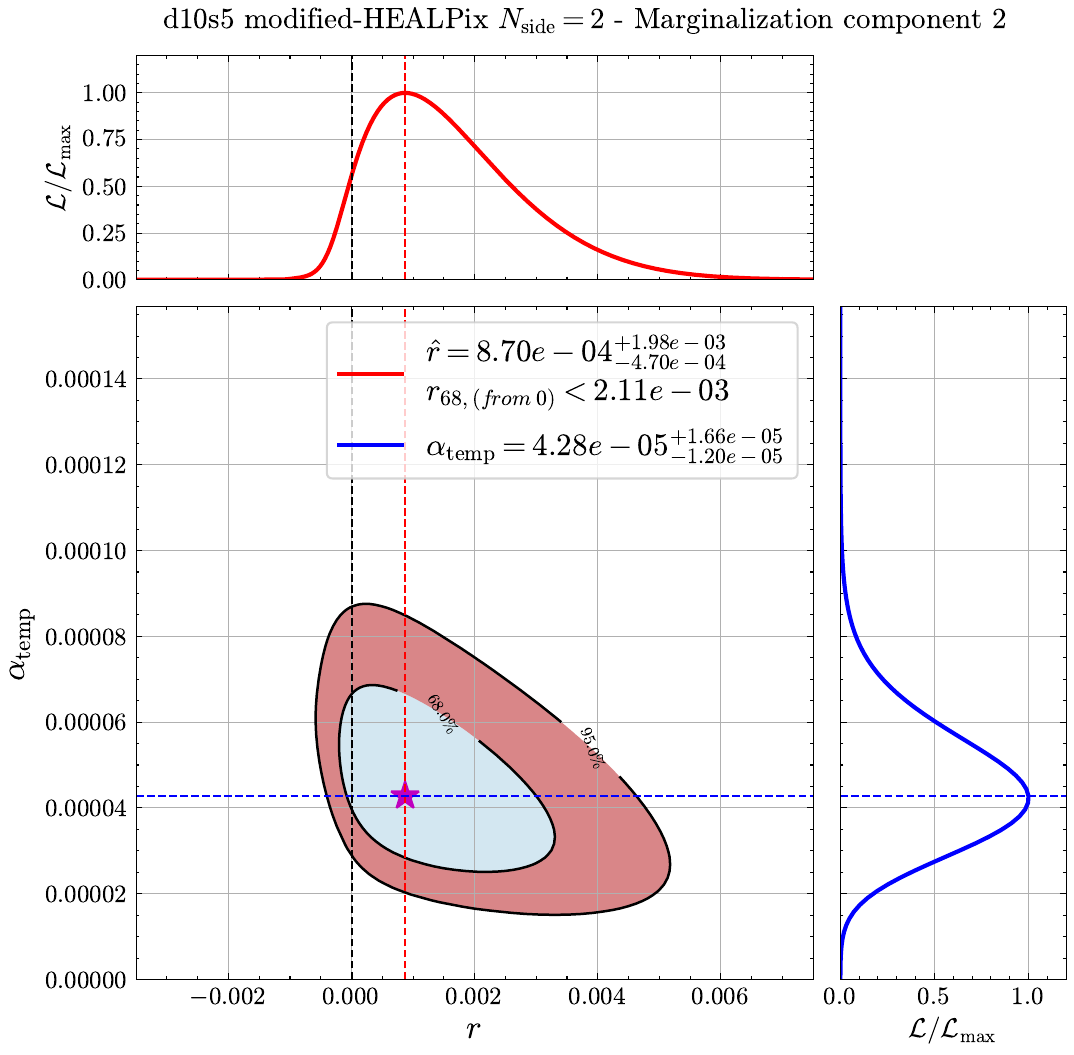}
    \includegraphics[width=0.48\linewidth]{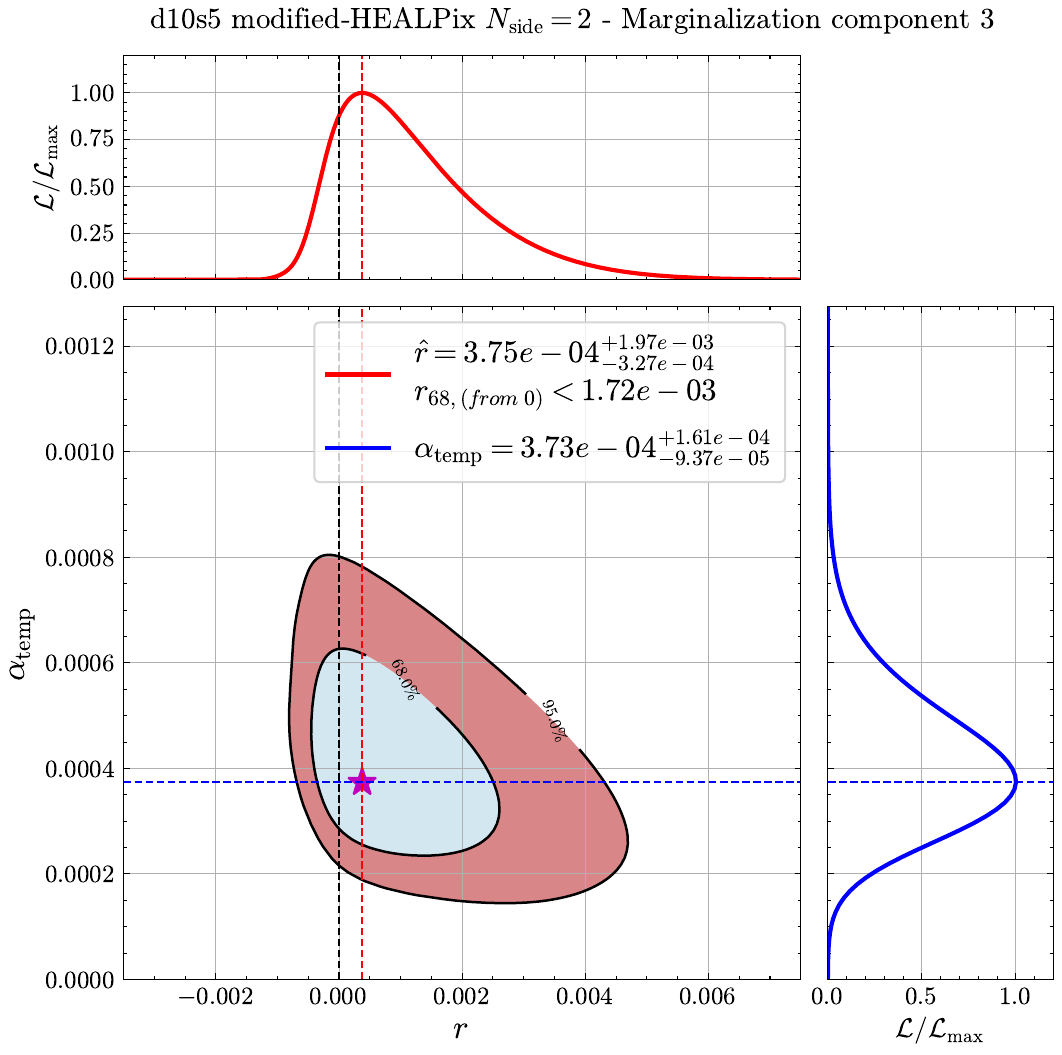}
    \caption{Result of marginalization on the modified-HEALPix $N_{\rm side} = 2$ SMICA running mode when applied to the \texttt{d10s5} model. The set of displayed plots is subsequently ideal (top left), realistic over component 1 (top right), component 2 (bottom left) and component 3 (bottom right).}
    \label{fig:patches_d10s5_marg}
\end{figure}

\begin{figure}
    \centering
    \includegraphics[width=0.49\linewidth]{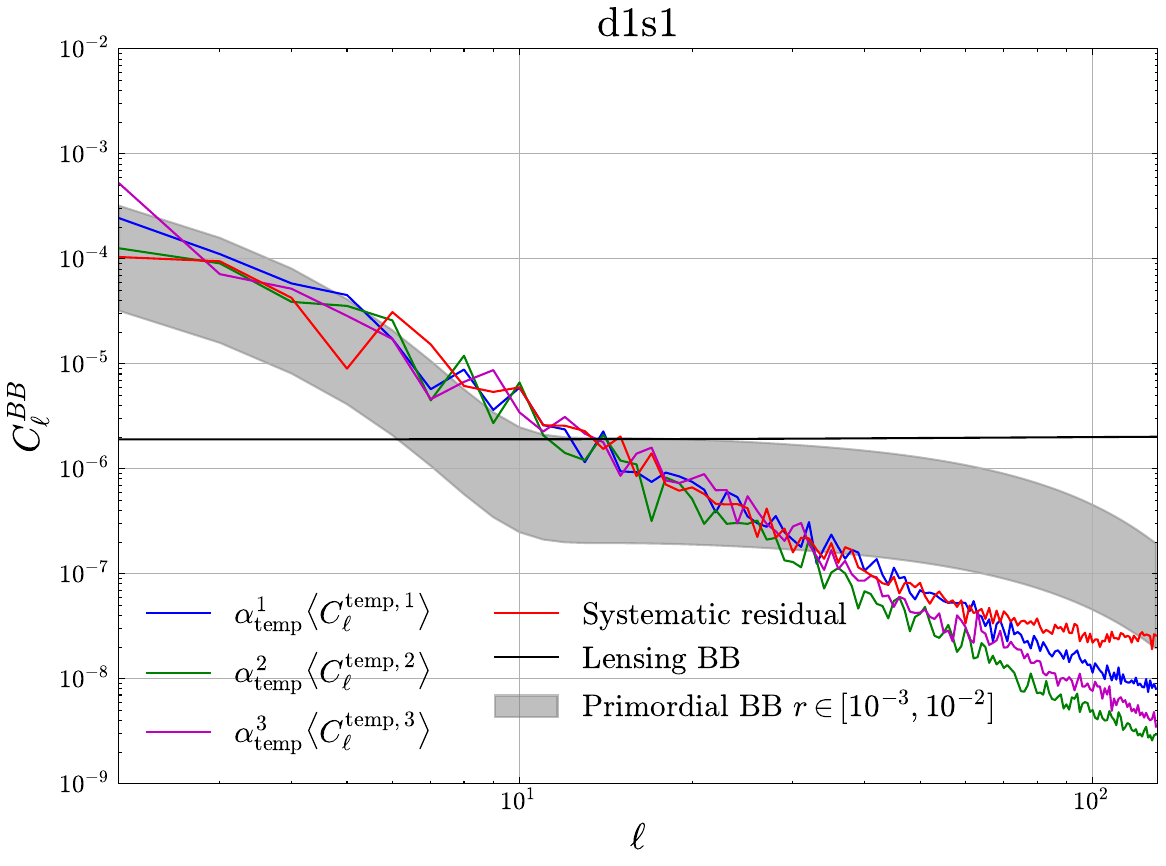}
    \includegraphics[width=0.49\linewidth]{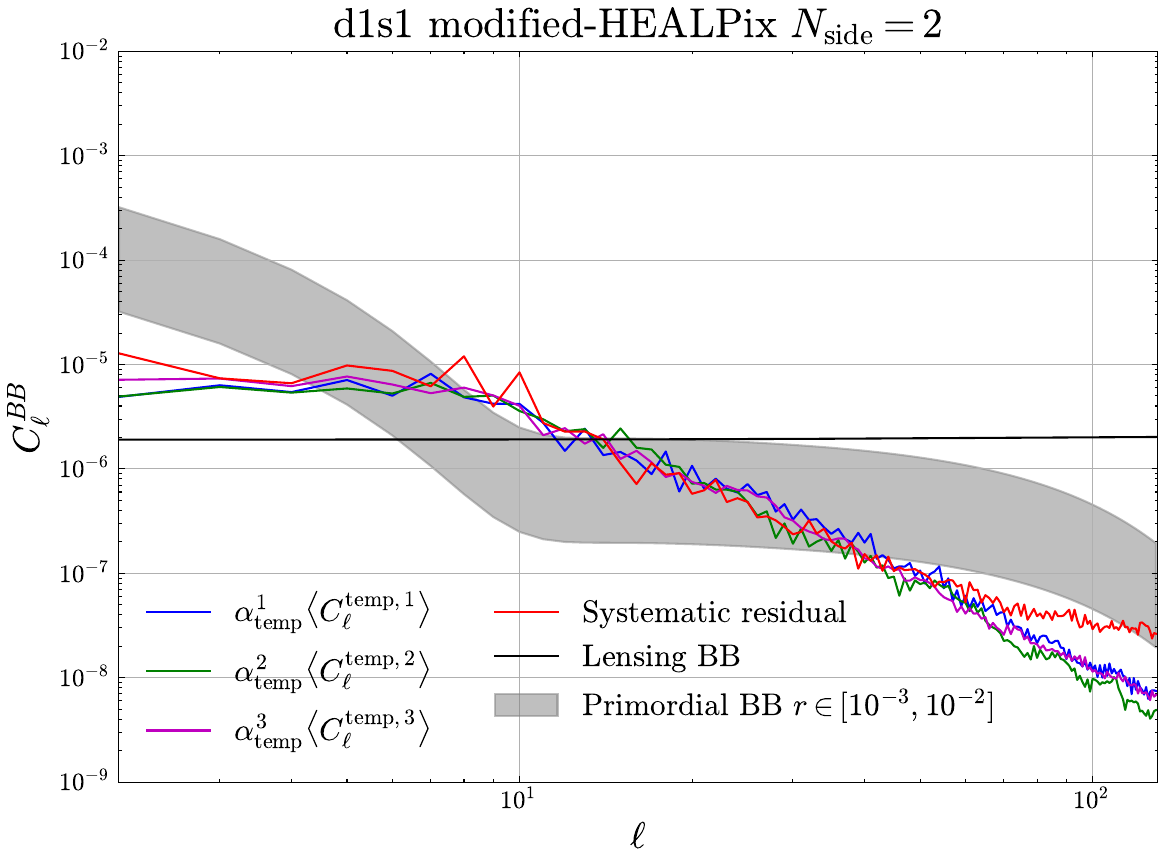}
    \includegraphics[width=0.49\linewidth]{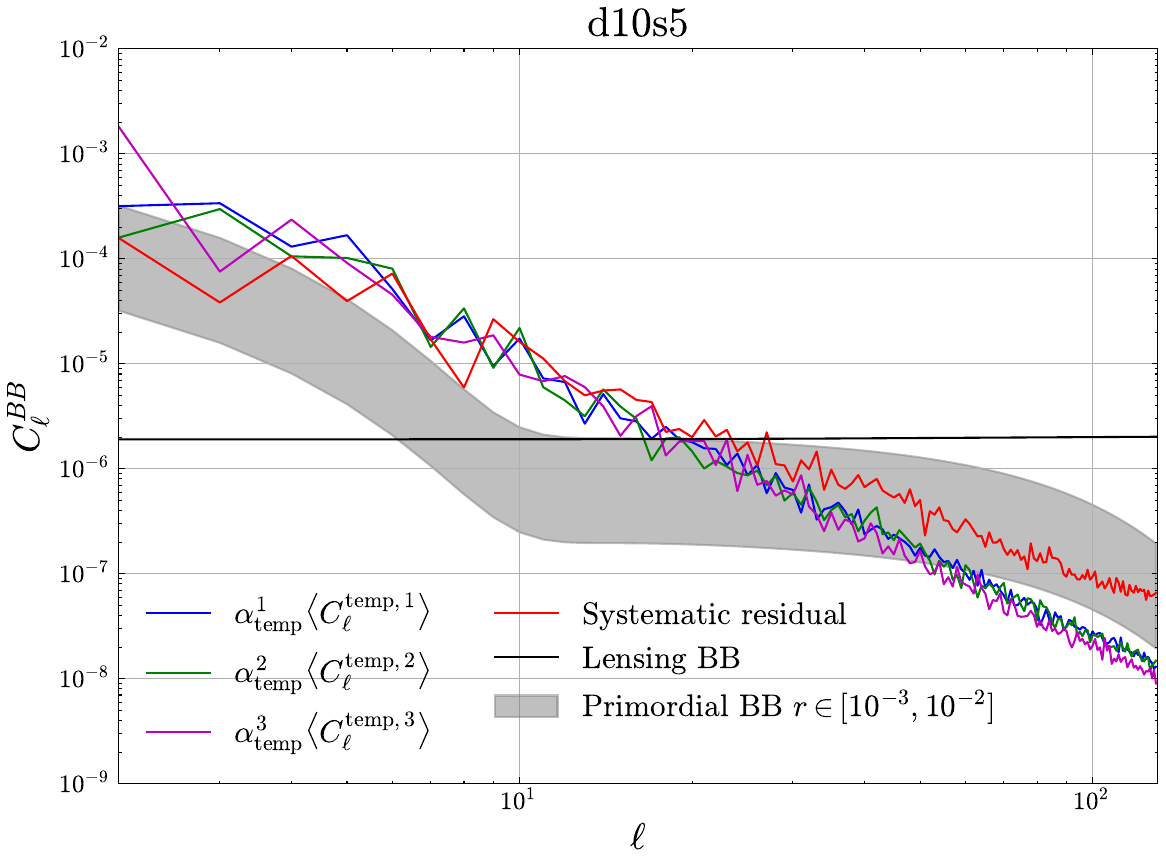}
    \includegraphics[width=0.49\linewidth]{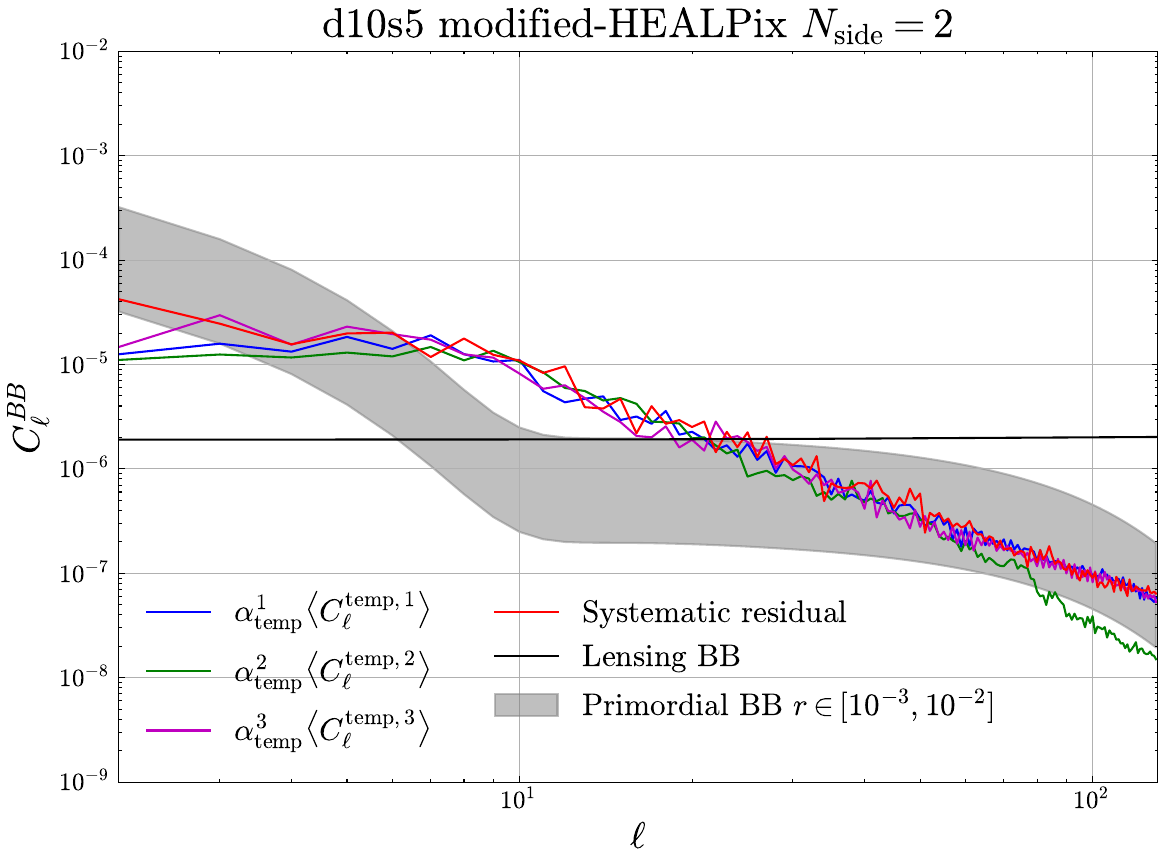}
    \caption{Comparision between the systematic residual spectra and foreground template spectra of each reconstructed component times the corresponding bestfit $\alpha_{\rm temp}$ in each configurations of marginalization. The set of displayed plots is subsequently \texttt{d1s1} full-sky (top left), \texttt{d1s1} modified-HEALPix $N_{\rm side} = 2$ (top right), \texttt{d10s5} full-sky (bottom left) and \texttt{d10s5} modified-HEALPix $N_{\rm side} = 2$ (bottom right).}
    \label{fig:marg_spectra}
\end{figure}

After component separation, a final debiasing procedure can be applied to further mitigate the impact of residual foreground contamination on the recovered tensor-to-scalar ratio. As outlined in Section~\ref{subsec:likelihood_r}, this is achieved at the likelihood level by introducing a template that captures the spectral shape of the residual contamination.\\
\indent Here, we investigate a self-contained alternative in which the template is derived directly from the foreground components reconstructed by SMICA, inspired by~\cite{Errard_2019}. While such templates are not expected to perform perfectly, they provide a practical option that avoids the need for an external pipeline. Their effectiveness relies on the residual foreground contamination shape remaining sufficiently close to the reconstructed foreground signal. This assumption may gradually break down as the residual becomes dominated by features not captured by the power spectra of the reconstructed components.\\
\indent For the baseline pipeline considered throughout this paper, SMICA reconstructs three foreground components: a dust-dominated component (column $\mathbf{A}^1$), a synchrotron-dominated component (column $\mathbf{A}^2$), and a third component (column $\mathbf{A}^3$) that absorbs the remaining foreground complexity. Their maps are reconstructed through the generalized least-squares solution of Eq.~\eqref{eq:gls} and are therefore affected by a reconstruction noise bias (similar to $C_\ell^{\rm stat, (i)}$ for the CMB). We define the template spectrum using the $j^{th}$ foreground component as:
\begin{equation}
    C_\ell^{\mathrm{temp},\;j,\,(i)}
    =
    \frac{
    C_\ell^{j,\,(i)}
    -
    \left\langle C_\ell^{\mathrm{noise},\;j,\,(i)} \right\rangle
    }{
    \max\!\left(
    C_\ell^{j,\,(i)}
    -
    \left\langle C_\ell^{\mathrm{noise},\;j,\,(i)} \right\rangle
    \right)
    },
\end{equation}
where the mean reconstruction noise bias $ \left\langle C_\ell^{\mathrm{noise},\;j,\,(i)} \right\rangle$ is subtracted before normalization. Similar to before, the (i) notation indicates that the template is constructed independently for each realization, introducing additional statistical variance. Finally, the normalization removes the arbitrary amplitude since only the spectral shape is fitted through $\alpha_{\rm temp}$.\\
\indent We conduct the same analysis for the modified-HEALPix $N_{\rm side}=2$ pipeline; however, an additional modification is required when applying the template marginalization. Since the approach models foreground spectral variability on a patch-by-patch basis, it efficiently absorbs variations occurring on scales larger than the patch size. So we expect a suppression of residual power at low multipoles. We account for this by flattening the template spectrum below the characteristic multipole of the patches while preserving its small-scale structure. The modified template is defined as:
\begin{equation}
    C_\ell^{\rm temp} =
    C_{\ell_t}^{\rm temp}+ \frac{1}{2}\left(C_\ell^{\rm temp}-C_{\ell_t}^{\rm temp}\right)\left[1+\tanh\left(\frac{\ell-\ell_t}{\Delta\ell}\right)\right].
\end{equation}
Here, $\ell_t = 6$ corresponds approximately to the angular scale of a modified-HEALPix patch of $N_{\rm side} = 2$, while $\Delta\ell = 2$ controls the smoothness of the transition. The $\tanh$ function was chosen arbitrarily.\\
\begin{figure}
    \centering
    \includegraphics[width=0.49\linewidth]{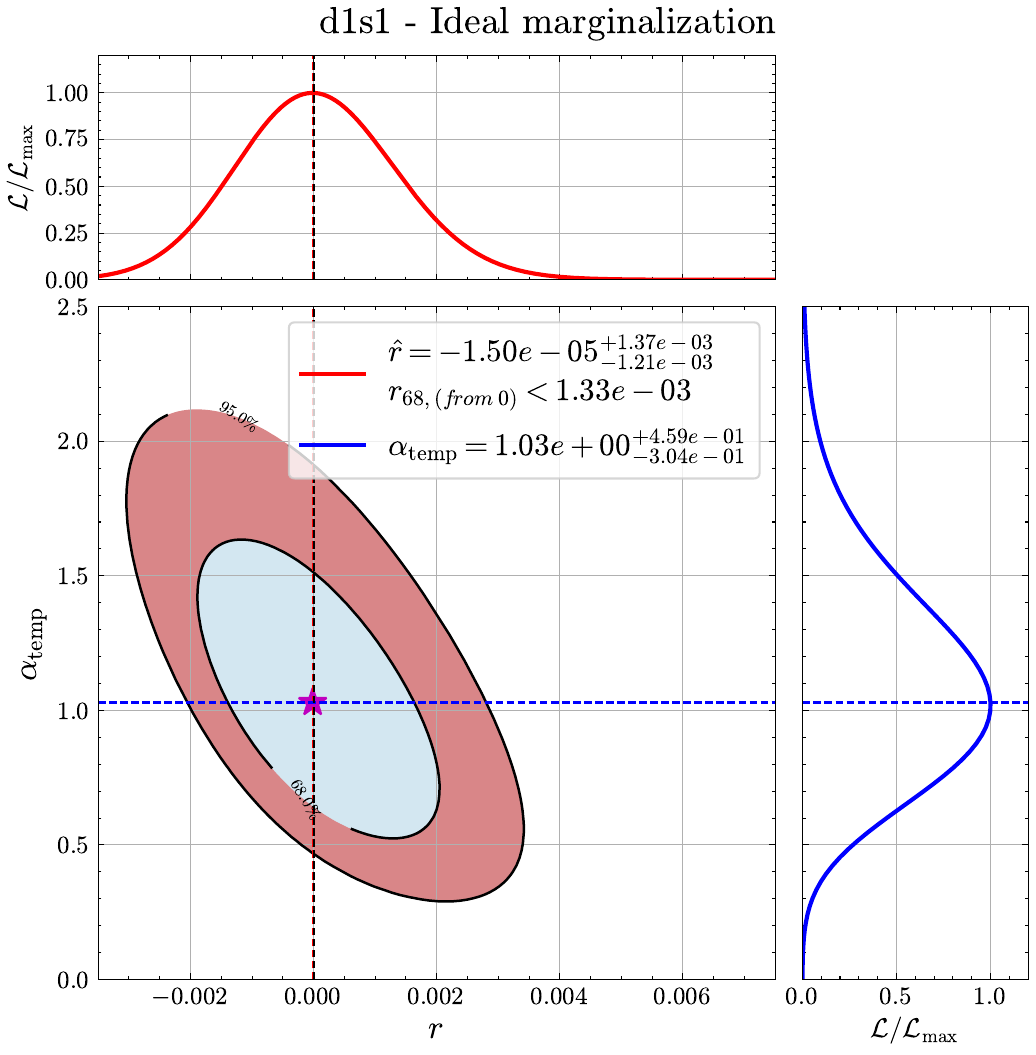}
    \includegraphics[width=0.49\linewidth]{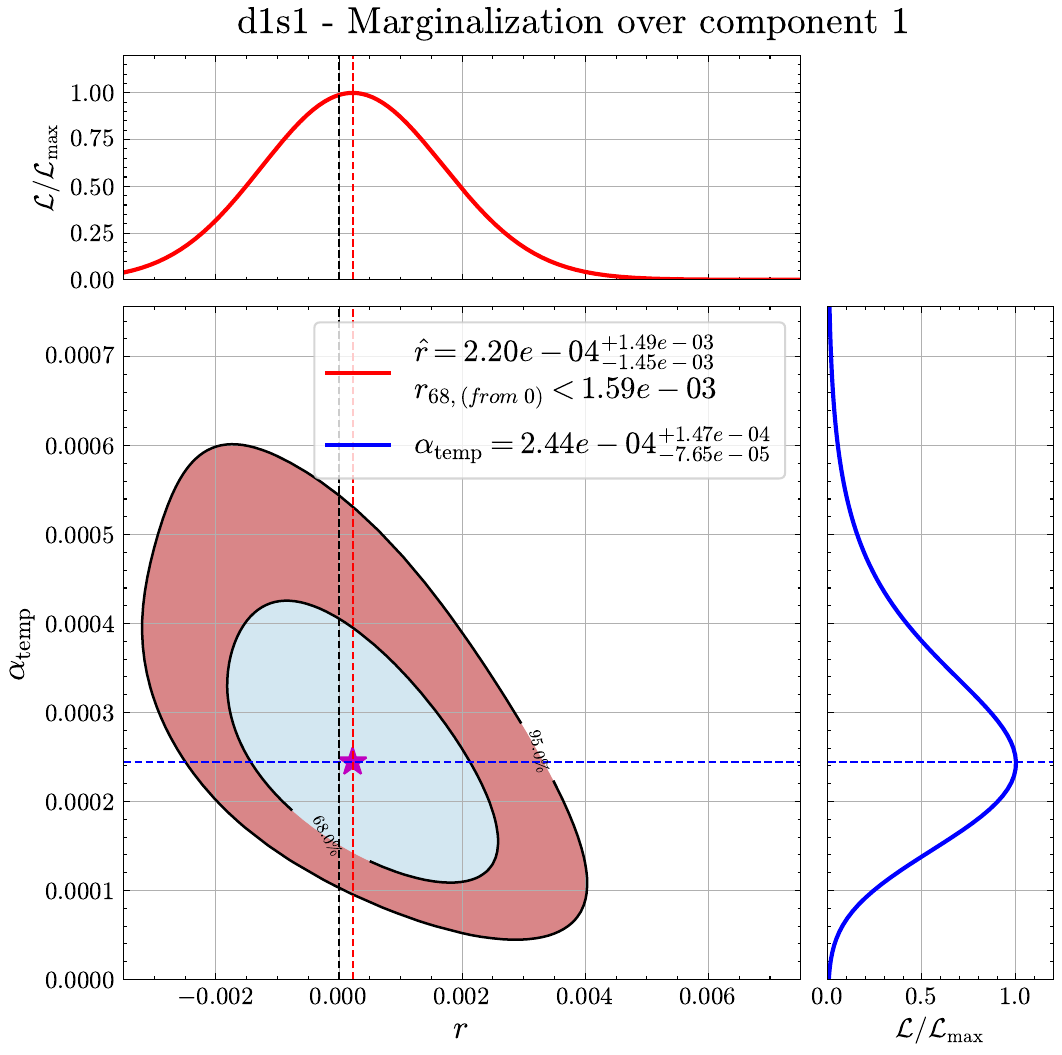}
    \includegraphics[width=0.49\linewidth]{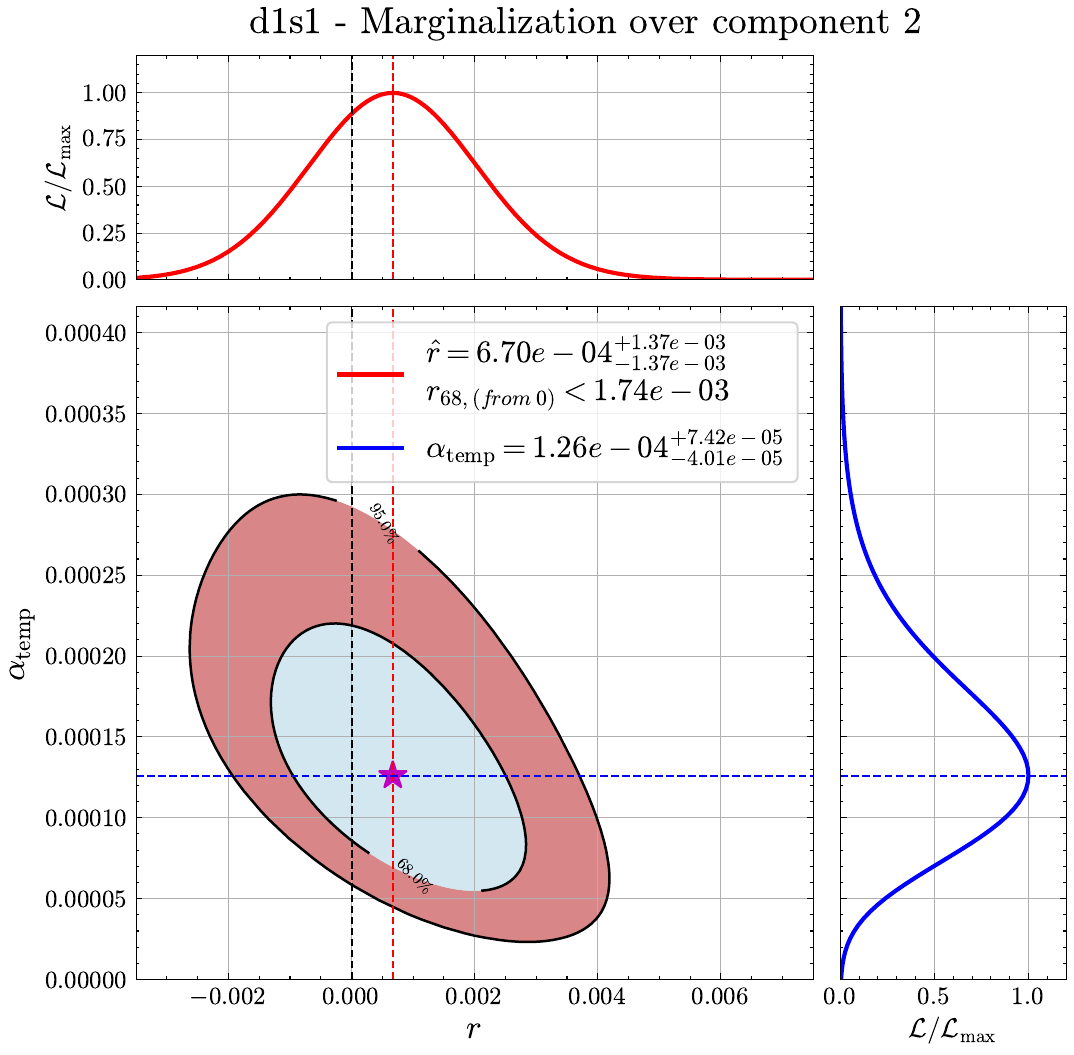}
    \includegraphics[width=0.49\linewidth]{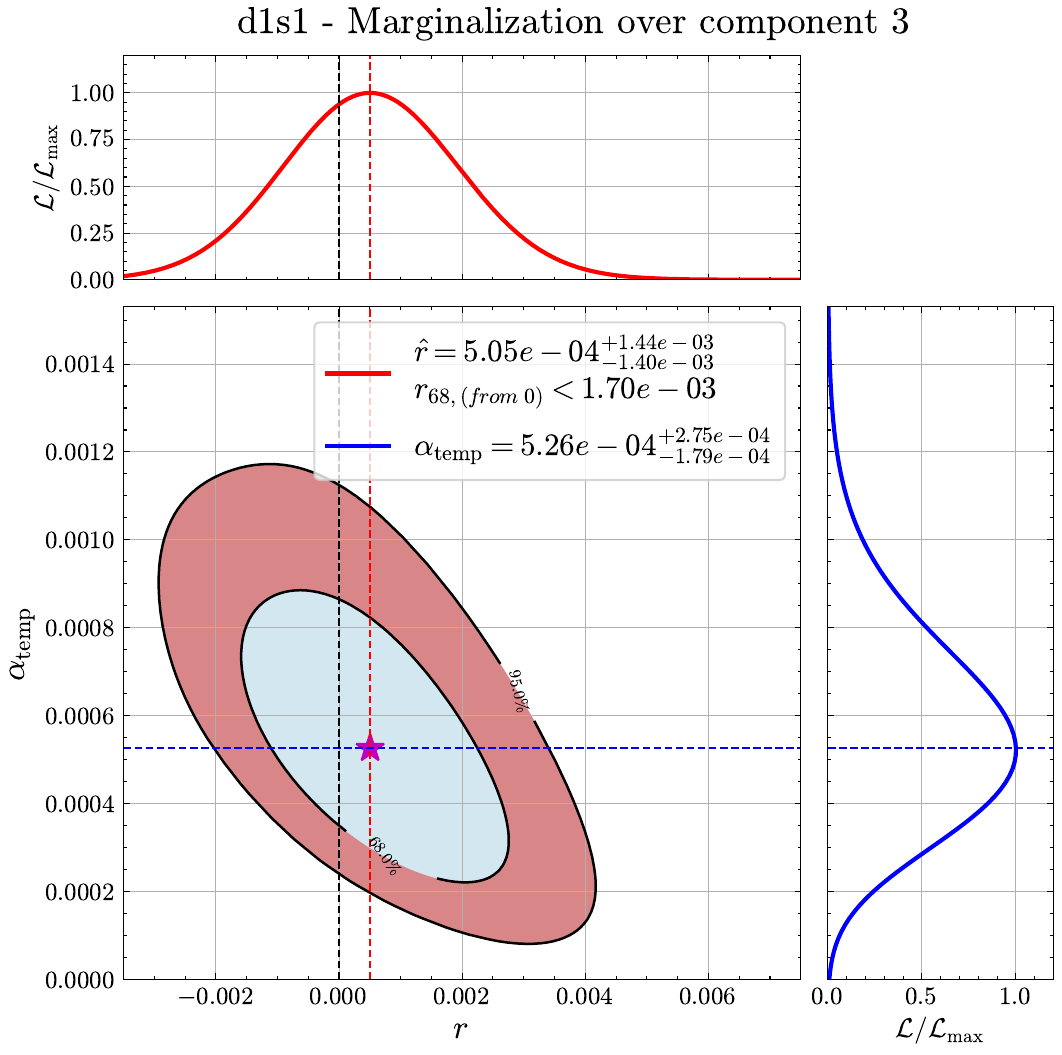}
    \caption{Result of marginalization on the full-sky pipeline running mode when applied to \texttt{d1s1} foreground. The set of displayed plots is subsequently ideal (top left), realistic over component 1 (top right), component 2 (bottom left) and component 3 (bottom right).}
    \label{fig:fs_d1s1_marg}
\end{figure}
\begin{figure}
    \centering
    \includegraphics[width=0.49\linewidth]{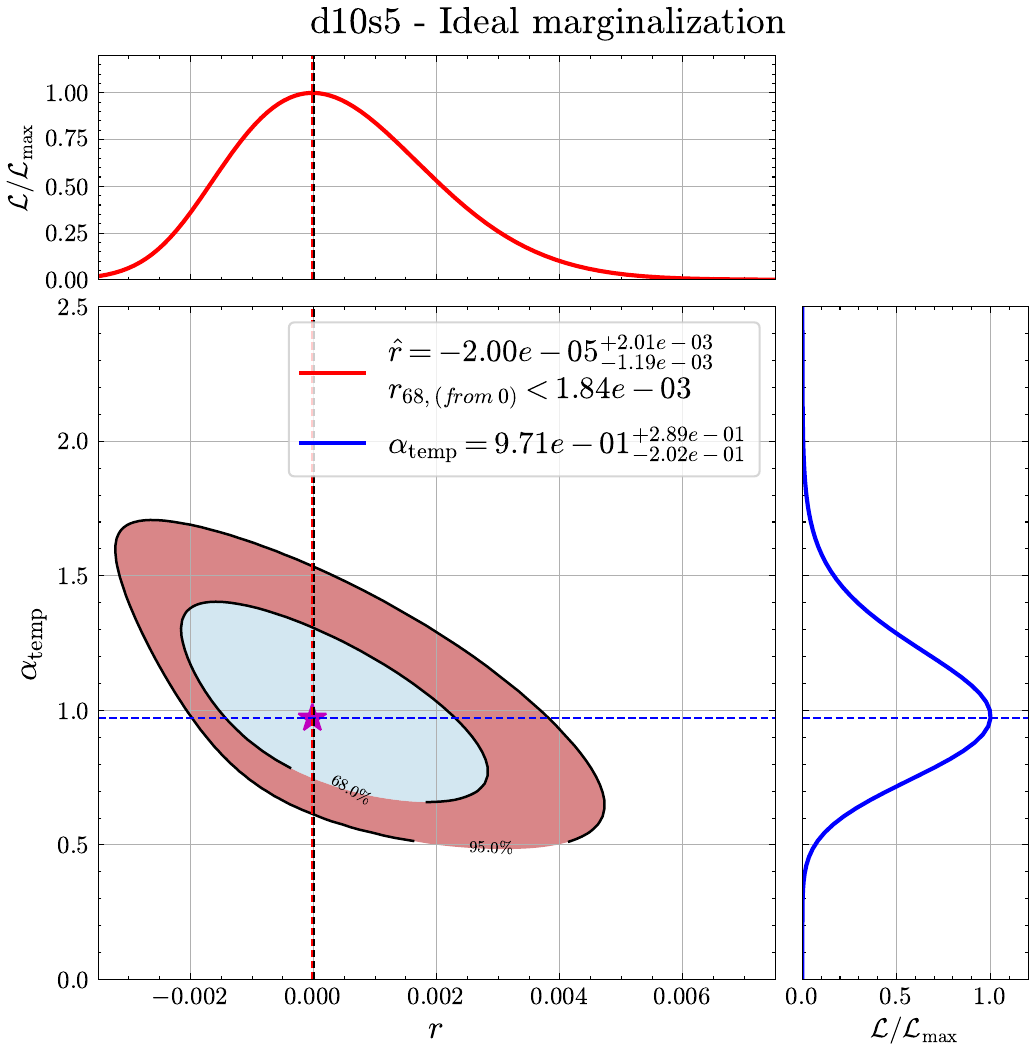}
    \includegraphics[width=0.49\linewidth]{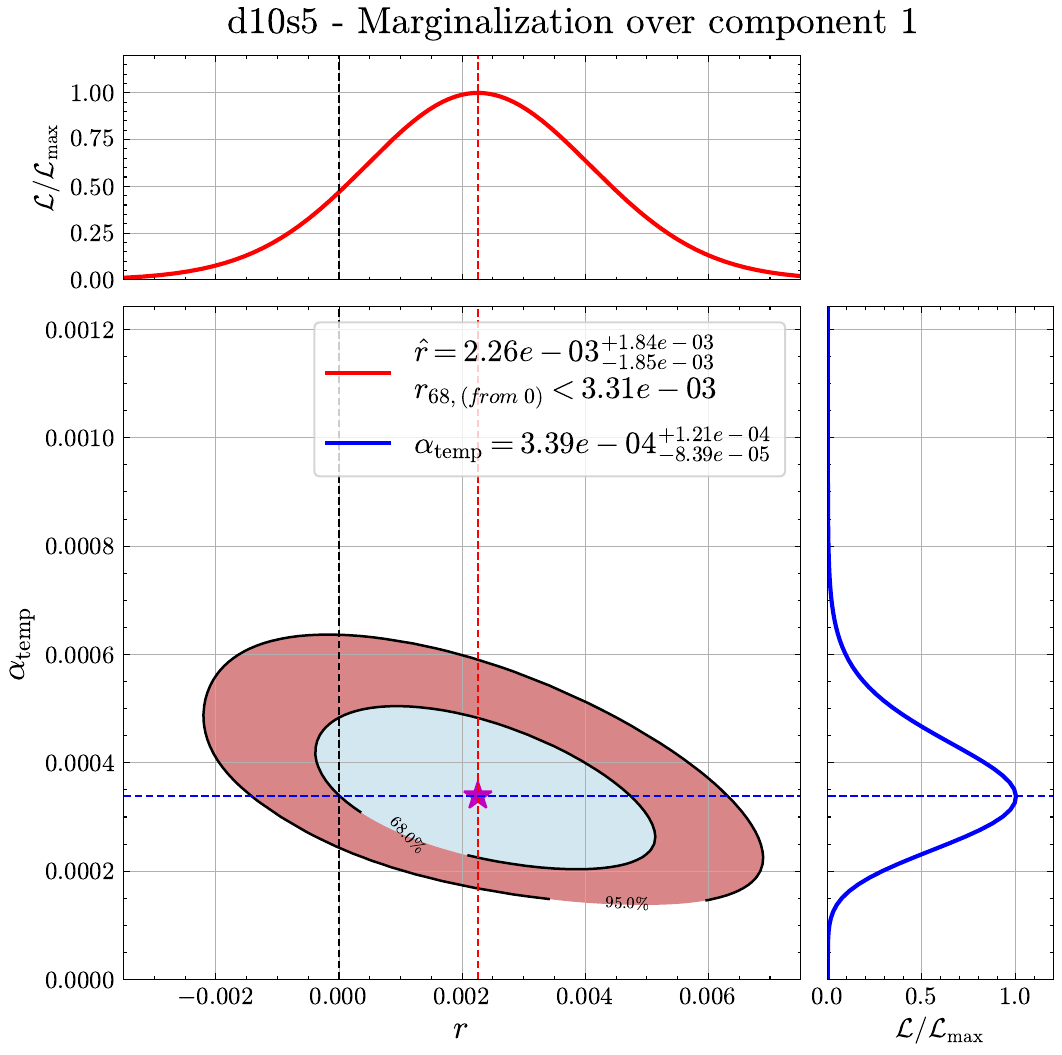}
    \includegraphics[width=0.49\linewidth]{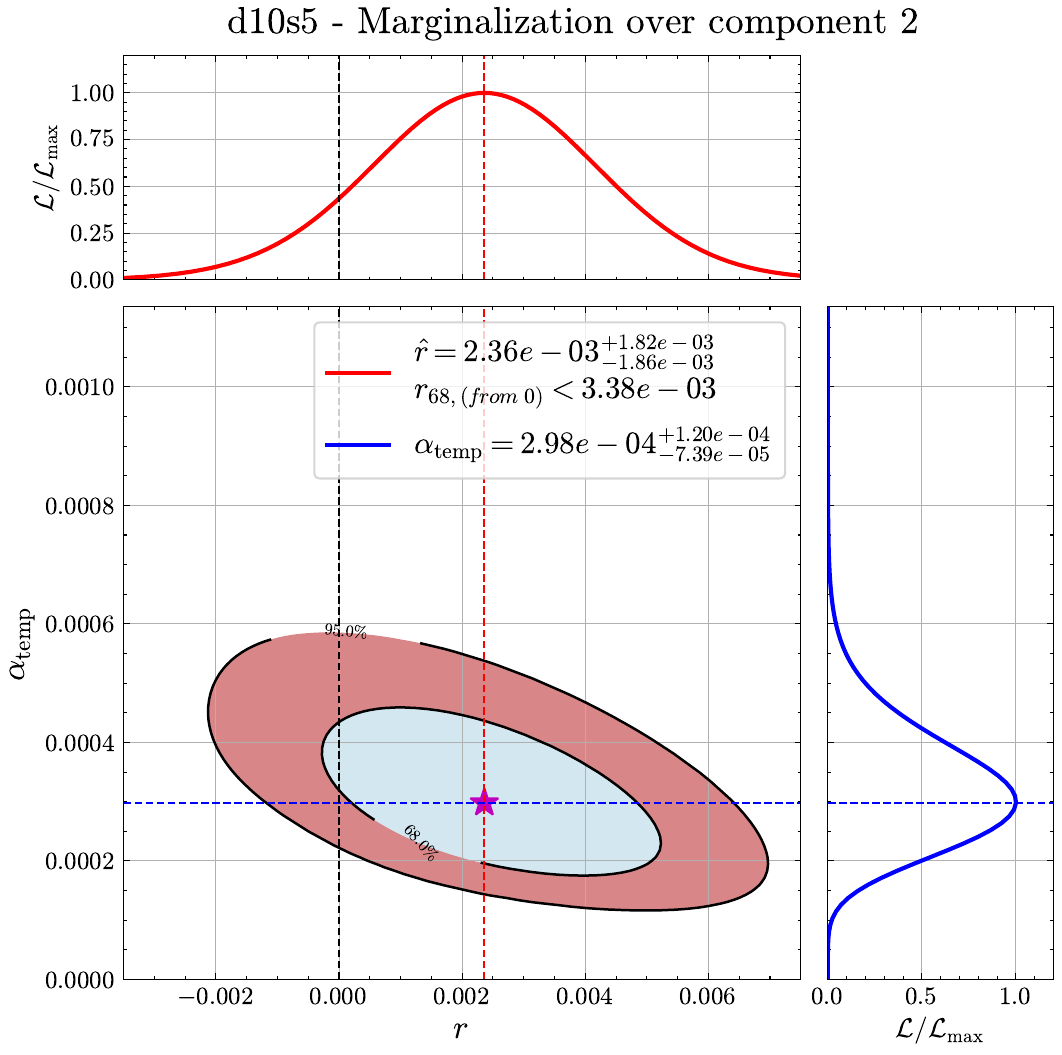}
    \includegraphics[width=0.49\linewidth]{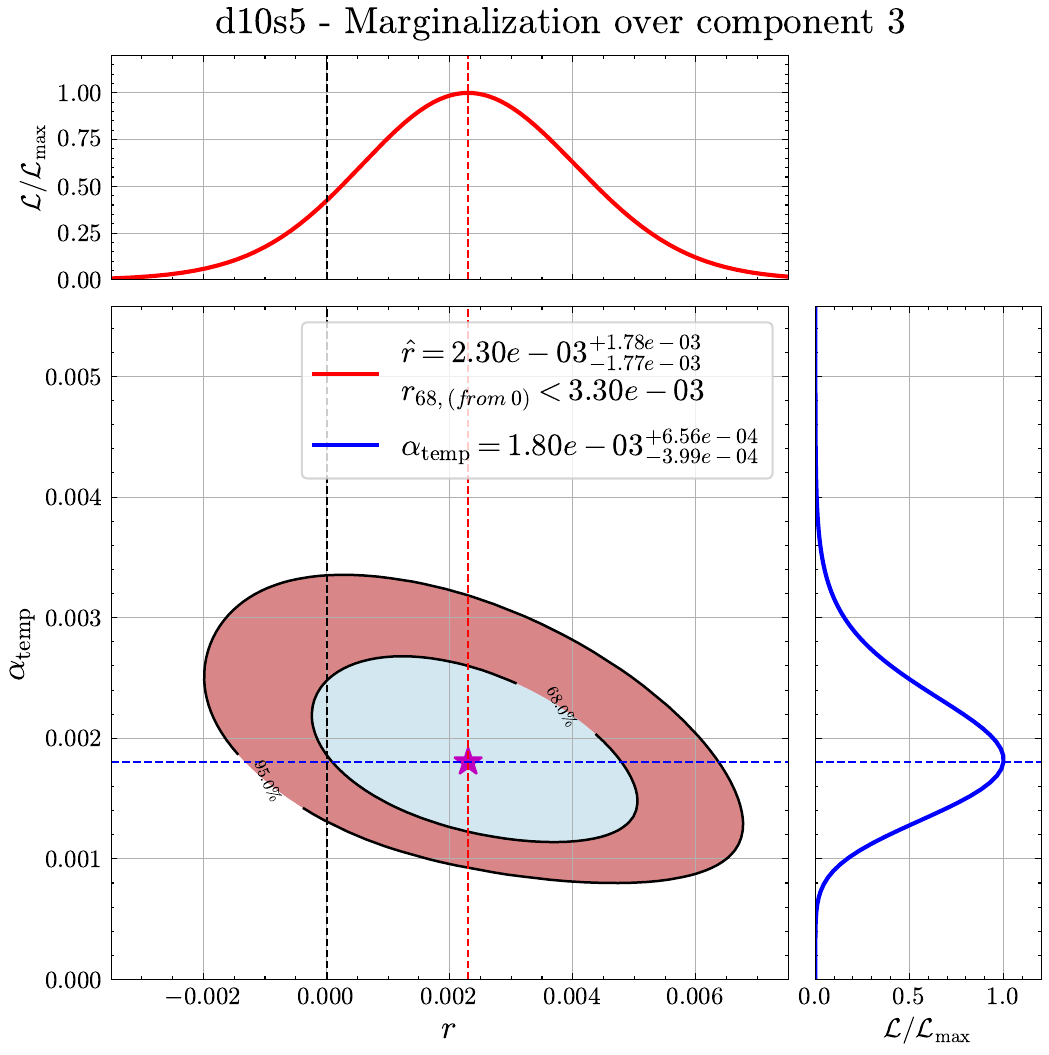}
    \caption{Result of marginalization on the full-sky SMICA running mode when applied to the \texttt{d10s5} model. The set of displayed plots is subsequently ideal (top left), realistic over component 1 (top right), component 2 (bottom left) and component 3 (bottom right).}
    \label{fig:fs_d10s5_marg}
\end{figure}
\indent Figures~\ref{fig:fs_d1s1_marg} and \ref{fig:fs_d10s5_marg} present the joint posterior distributions of the parameters obtained from the template-marginalization procedure applied to the full-sky covariance SMICA pipeline for the \texttt{d1s1} and \texttt{d10s5} foreground models, respectively. The corresponding results for the modified-HEALPix $N_{\rm side}=2$ pipeline are shown in Figures~\ref{fig:patches_d1s1_marg} and \ref{fig:patches_d10s5_marg}. Table~\ref{tab:r_constraints} summarizes the resulting statistical constraints on the tensor-to-scalar ratio before and after marginalization, while Figure~\ref{fig:marg_spectra} compares the systematic residual spectra with the best-fit template spectra.\\
\indent We consider an idealized benchmark in which $C_\ell^{\rm temp}=C_\ell^{\rm sys}$, corresponding to a template that exactly matches the residual contamination, shown in the top left plot of Figures~\ref{fig:fs_d1s1_marg} and \ref{fig:fs_d10s5_marg}. In this case, we verify that $\alpha_{\rm temp}$ is consistent with unity, and the recovered tensor-to-scalar ratio is unbiased. Interestingly, the procedure induces only a modest increase in the statistical uncertainty. This therefore represents the best performance achievable by the marginalization procedure.\\
\indent With the internally reconstructed templates, we observe that, across all configurations, components 1 and 3 consistently provide the best performance and give broadly similar results. This is consistent with their predominantly dust-like character, as discussed in Section~\ref{subsec:ext_comps}, suggesting that the residual contamination is largely traced by the dust-dominated modes.\\
\indent For \texttt{d1s1}, the full-sky configuration already benefits substantially from template marginalization. Using component 1, the bias is reduced from $3.81\cdot10^{-3}$ to $2.20\cdot10^{-4}$, with a central $68\%$ interval width of $2.94\cdot10^{-3}$ and an upper limit of $1.59\cdot10^{-3}$. Component 3 gives a slightly larger bias of $5.05\cdot10^{-4}$ but a similar interval width of $2.84\cdot10^{-3}$ and upper limit of $1.70\cdot10^{-3}$. The modified-HEALPix $N_{\rm side} = 2$ configuration provides better foreground cleaning, so that the templates need only account for a smaller remaining systematic contribution. The bias after marginalization in this case is then reduced to $8.00\cdot10^{-5}$ for component 1 and $1.50\cdot10^{-5}$ for component 3, with interval widths of $1.18\cdot10^{-3}$ and $1.25\cdot10^{-3}$, respectively. In both cases, the result is effectively unbiased, with tight upper limits of $r<8.27\cdot10^{-4}$ and $r<8.48\cdot10^{-4}$ at $68\%$ confidence for components 1 and 3, respectively. The effectiveness of the procedure can be understood from the top row of Figure~\ref{fig:marg_spectra}: for \texttt{d1s1}, the templates in both the full-sky and modified-HEALPix cases closely match the shape of the systematic residual. \\
\indent For \texttt{d10s5}, where the foreground model is more complex, the benefit of reducing the residual contamination before marginalization becomes even more apparent. In the full-sky case, marginalization with components 1 and 3 can only reduce the bias to $2.25\cdot10^{-3}$ and $2.30\cdot10^{-3}$, respectively, which remains significant, while the constraints are relatively broad, with central $68\%$ interval widths of $3.70\cdot10^{-3}$ and $3.55\cdot10^{-3}$. In contrast, applying the same procedure to the modified-HEALPix $N_{\rm side}=2$ results in biases of $4.60\cdot10^{-4}$ and $3.75\cdot10^{-4}$ for components 1 and 3, respectively, with narrower interval widths of $2.11\cdot10^{-3}$ and $2.29\cdot10^{-3}$. These correspond to effectively unbiased estimates, with upper limits of $r<1.63\cdot10^{-3}$ and $r<1.72\cdot10^{-3}$ at $68\%$ confidence level, respectively. The difference between the two configurations in \texttt{d10s5} is largely driven by how well the reconstructed templates capture the shape of the remaining systematic residual. As shown in the bottom row of Figure~\ref{fig:marg_spectra}, in the full-sky case, the templates fail to trace the residual accurately, overestimating it on large scales and underestimating it on small scales, limiting the effectiveness of the marginalization. In contrast, the modified-HEALPix templates follow the residual very closely, particularly for components 1 and 3.\\
\indent In addition to the internal templates, there are alternative approaches, for example, as demonstrated in~\cite{Carones_2025}, where the method GNILC~\cite{Remazeilles_2011} was used. This case highlights that an independent dedicated component separation pipeline for foreground-reconstruction can be used to build the templates for likelihood level marginalization.

\begin{table}[t]
\centering
\begin{tabular}{llcccc}
\hline
Model & Patches & Marginalization &
$r_{\rm bias}$ & $\Delta r_{68}$ & $r_{68}$ \\
\hline
\multirow{6}{*}{\texttt{d1s1}}
 & \multirow{3}{*}{Full-sky}
 & No marg. & $3.81\cdot10^{-3}$ & $1.74\cdot10^{-3}$ & $4.31\cdot10^{-3}$ \\
 % & & Ideal    & $-1.50\cdot10^{-5}$ & $2.58\cdot10^{-3}$ & $1.33\cdot10^{-3}$ \\
 & & Comp. 1  & $2.20\cdot10^{-4}$ & $2.94\cdot10^{-3}$ & $1.59\cdot10^{-3}$ \\
 % & & Comp. 2  & $6.70\cdot10^{-4}$ & $2.74\cdot10^{-3}$ & $1.74\cdot10^{-3}$ \\
 & & Comp. 3  & $5.05\cdot10^{-4}$ & $2.84\cdot10^{-3}$ & $1.70\cdot10^{-3}$ \\
\cline{2-6}
 & \multirow{3}{*}{\makecell[l]{Modified-HEALPix\\$N_{\rm side}=2$}}
 & No marg. & $2.04\cdot10^{-3}$ & $1.67\cdot10^{-3}$ & $2.57\cdot10^{-3}$ \\
 % & & Ideal    & $-5.00\cdot10^{-5}$ & $1.29\cdot10^{-3}$ & $8.41\cdot10^{-4}$ \\
 & & Comp. 1  & $8.00\cdot10^{-5}$ & $1.18\cdot10^{-3}$ & $8.27\cdot10^{-4}$ \\
 % & & Comp. 2  & $1.20\cdot10^{-4}$ & $1.28\cdot10^{-3}$ & $9.17\cdot10^{-4}$ \\
 & & Comp. 3  & $1.50\cdot10^{-5}$ & $1.25\cdot10^{-3}$ & $8.48\cdot10^{-4}$ \\
\hline
\multirow{6}{*}{\texttt{d10s5}}
 & \multirow{3}{*}{Full-sky}
 & No marg. & $8.58\cdot10^{-3}$ & $2.64\cdot10^{-3}$ & $9.28\cdot10^{-3}$ \\
 % & & Ideal    & $-2.00\cdot10^{-5}$ & $3.20\cdot10^{-3}$ & $1.84\cdot10^{-3}$ \\
 & & Comp. 1  & $2.25\cdot10^{-3}$ & $3.70\cdot10^{-3}$ & $3.31\cdot10^{-3}$ \\
 % & & Comp. 2  & $2.36\cdot10^{-3}$ & $3.68\cdot10^{-3}$ & $3.38\cdot10^{-3}$ \\
 & & Comp. 3  & $2.30\cdot10^{-3}$ & $3.55\cdot10^{-3}$ & $3.30\cdot10^{-3}$ \\
\cline{2-6}
 & \multirow{3}{*}{\makecell[l]{Modified-HEALPix\\$N_{\rm side}=2$}}
 & No marg. & $6.13\cdot10^{-3}$ & $2.44\cdot10^{-3}$ & $6.77\cdot10^{-3}$ \\
 % & & Ideal    & $-6.50\cdot10^{-5}$ & $2.53\cdot10^{-3}$ & $1.60\cdot10^{-3}$ \\
 & & Comp. 1  & $4.60\cdot10^{-4}$ & $2.11\cdot10^{-3}$ & $1.63\cdot10^{-3}$ \\
 % & & Comp. 2  & $8.70\cdot10^{-4}$ & $2.45\cdot10^{-3}$ & $2.11\cdot10^{-3}$ \\
 & & Comp. 3  & $3.75\cdot10^{-4}$ & $2.29\cdot10^{-3}$ & $1.72\cdot10^{-3}$ \\
\hline
\end{tabular}
\caption{Final constraints on the tensor-to-scalar ratio $r$ for the
\texttt{d1s1} and \texttt{d10s5} foreground models. We consider five marginalization configurations: no marginalization, the ideal template, and templates constructed from each of the three reconstructed foreground components. These are evaluated for two component-separation pipelines: the full-sky configuration with $N_{\rm d}=3$ and the modified-HEALPix configuration with $N_{\rm side}=2$. We report the recovered bias $r_{\rm bias}$, the width of the central 68\% interval $\Delta r_{68}=r_{\rm upper}-r_{\rm lower}$, and the 68\% upper limit on
$r$ from $r_{\rm true} = 0$.}
\label{tab:r_constraints}
\end{table}
% \clearpage
\clearpage
\section{Discussion}\label{sec:discussions}
In this section, we discuss two aspects that are particularly relevant to this study. First, we examine the relationship between the SMICA likelihood and the sequential SMICA+GLS pipeline, highlighting the key differences between the two. We then turn to the physical interpretation of the additional foreground components required by SMICA, and discuss how the effective dimensionality of the foreground covariance can arise from spatial variations in the foreground spectral parameters.
\subsection{The SMICA likelihood}\label{subsec:smica_gls_lklh}
\begin{figure}
    \centering
    \includegraphics[width=0.75\linewidth]{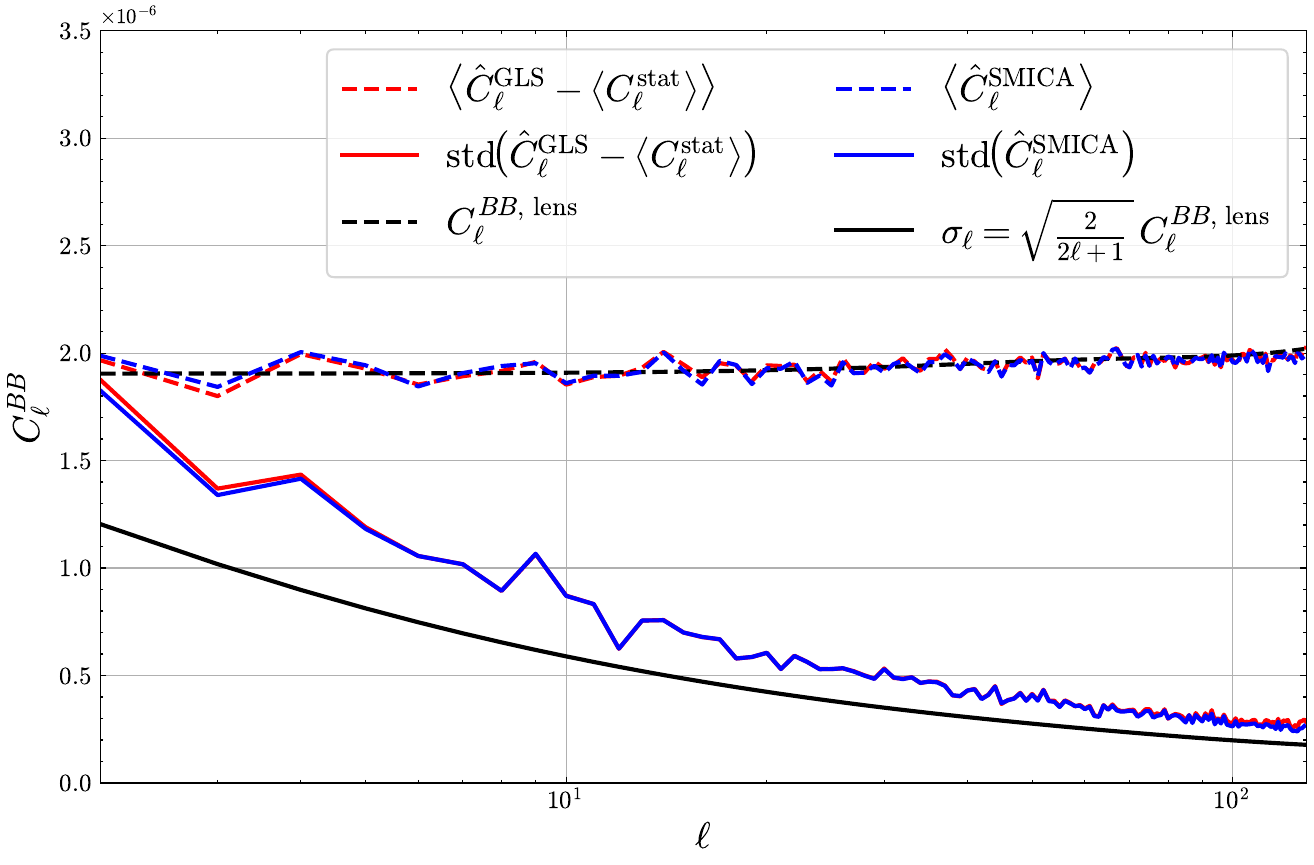}
    \caption{Comparison between the mean and standard deviation of the output CMB power from the full SMICA likelihood and the debiased GLS (with SMICA parameters) filtered CMB map in an ideal context: LiteBIRD-like configuration, full-sky and \texttt{d0s0} foregrounds.}
    \label{fig:d0s0_comparison}
\end{figure}

The SMICA framework can be interpreted as a multi-component maximum-likelihood estimator of the data. We can derive it from the linear mixture model in Eq.~\eqref{eq:linmod_real} by treating the component amplitudes $\mathbf{s}$ as latent Gaussian random fields with physics encoded in the covariance $\mathbf{R_S}(C_\ell)$. The SMICA likelihood can also be obtained by analytically marginalizing over all possible realizations of the sky,
\begin{equation}
\mathcal{L}_{\rm SMICA}(\mathbf{y}|\mathbf{A},\mathbf{R}_N,\mathbf{R_S}(C_\ell))
=
\int
\mathcal{L}(\mathbf{y},\mathbf{s}|\mathbf{A},\mathbf{R}_N,\mathbf{R_S}(C_\ell))
\,{\rm d}\mathbf{s}.
\end{equation}
The unknown parameters are then jointly estimated through
\begin{equation}
(\hat{\mathbf{A}},\hat{\mathbf{R}}_N,\mathbf{\widehat{R}_S}(\hat{C}_\ell^{\rm SMICA}))
=
\arg\max_{\mathbf{A},\mathbf{R}_N,\mathbf{R_S}(C_\ell)}
\mathcal{L}_{\rm SMICA}
(\mathbf{y}|\mathbf{A},\mathbf{R}_N,\mathbf{R_S}(C_\ell)).
\end{equation}
In this formulation, the latent sky realization is never explicitly reconstructed, but instead remains a variable that was analytically marginalized over in the formalism. The underlying physic is therefore inferred directly from the likelihood of the observed multifrequency data, represented by $\hat{C}_\ell^{\rm SMICA}$.\\
\indent The SMICA+GLS pipeline considered in this paper begins with exactly the same maximum-likelihood fit of the component model. The estimated quantities $\hat{\mathbf{A}},\,\hat{\mathbf{R}}_N$ are subsequently extracted to construct the generalized least-squares estimator, $\hat{\mathbf{s}}=\mathbf{W}(\hat{\mathbf{A}},\hat{\mathbf{R}}_N)\mathbf{y}
$, before inference is performed on the power spectrum of the reconstructed maps:
$\tilde{C}_\ell^{\rm GLS} = \mathcal{C}_\ell\left[\hat{\mathbf{s}}\hat{\mathbf{s}}^\dagger\right]$
through an external likelihood, for example, $\mathcal{L}_{\mathbf{IW}}$ (Eq.~\eqref{eq:cosmo_like}).\\
\indent The distinction between the two approaches therefore does not lie in the estimation of the component model, which is identical in both cases, but in how the fitted model is subsequently used. In the SMICA approach, the physics are inferred directly within the marginalized likelihood of the observations. In this case, as long as the underlying model is correct, the recovered $\hat{C}_\ell^{\rm SMICA}$ is reflective of the true $C_\ell$ up to some statistical fluctuations and contains no further biases. In contrast, the sequential pipeline first replaces the latent sky with a point estimate $\hat{\mathbf{s}}$ before evaluating a separate likelihood. Therefore, the overall procedure is not obtained by maximizing a single global likelihood over all unknown quantities simultaneously, but by a sequence of conditional optimizations connected through the reconstructed map. And as derived in Section \ref{subsec:smica}, the deviation from maximum likelihood causes several methodological biases to arise when assessing the final $\tilde{C}_\ell^{\rm GLS}$, most notably the added statistical residuals and the ILC bias.\\
\indent Nevertheless, there are several practical reasons for adopting the latter in this work. As discussed previously, the map-based reconstruction naturally allows us to introduce spatial variations in the component-separation model that works in the harmonic domain. Furthermore, the harmonic covariance-based treatment in the maximum-likelihood approach becomes less straightforward in the presence of a cut sky, especially at low multipoles. More importantly, the non-optimality of the sequential approach can remain limited if the reconstruction biases are carefully controlled. In particular, the cross-term ILC bias can be strongly suppressed by ensuring that the reconstruction weights are sufficiently independent of the sky realization, as motivated by the modified-HEALPix approach, while the positive contribution from the statistical residual can be accurately propagated into the final likelihood. Under these conditions, the reconstructed-map approach should remain close to the optimal end-to-end result. \\
\indent We demonstrate this in the idealized full-sky \texttt{d0s0} case, where the assumptions underlying the maximum-likelihood approach are exact. A comparison between the output $\hat{C}_\ell^{\rm SMICA}$ and $\hat{C}_\ell^{\rm GLS} - \langle C_\ell^{\rm stat}\rangle$ over 200 realizations in this idealized setting is presented in Figure~\ref{fig:d0s0_comparison}. For the GLS pipeline, the full-sky coverage ensures that a sufficiently large number of modes contribute to the covariance estimation, thus suppressing the ILC bias. In this regime, we find that the outputs of the two pipelines are nearly identical, with GLS showing only a slight increase in variance at some multipoles. This demonstrates that, when the methodological biases are properly controlled, the GLS pipeline can achieve performance comparable to that of the full maximum-likelihood approach.\\
\indent The discussion above naturally motivates a full maximum-likelihood approach, in which the component model and cosmological parameters are inferred simultaneously in an optimal way. As the component-reconstruction step is avoided, the associated methodological biases are no longer present. This removes an important limitation of the current approach, in which mitigating the ILC-bias requires corrections that compromise the foreground-cleaning performance. For example, the spatial resolution of the subsets cannot be increased arbitrarily as the ILC-bias becomes increasingly significant. However, extending this approach to spatially varying foregrounds is quite challenging. If the mixing matrix varies across the sky, $\mathbf{A}=\mathbf{A}(\hat n)$, statistical isotropy is broken, and the harmonic-space covariance $\mathbf{R}_Y$ is no longer diagonal in $(\ell,m)$. Instead of taking the simple form $\mathbf{R}_{\mathbf{Y},\,\ell}$, it becomes a dense operator coupling the full set of harmonic coefficients. Evaluating the marginalized likelihood then requires operations such as the inverse and log-determinant of this full $(\ell,m)$-space covariance, rather than independently working with frequency--frequency covariance matrices at each multipole or multipole bin, as usually done in SMICA. \\
\indent We are therefore investigating an alternative Gibbs-sampling-based methodology, following the MICS~\cite{Leloup_2023} framework and its implementation, MICMAC~\cite{Morshed_2024}, in which the marginalization over the component maps $\mathbf{s}$ is performed numerically. This provides a potential route to retain the advantages of the full maximum-likelihood approach while addressing the limitations associated with the harmonic-space treatment and the GLS pipeline. This ongoing work will be presented in a forthcoming publication.
\subsection{Making physical sense of extra components in SMICA}\label{subsec:moment_expansion}
We investigate a physically motivated interpretation for the required increase in foreground components beyond the physical ones in SMICA. We consider the polarized thermal dust and synchrotron model from Eq.~\eqref{eq:dust_model} and~\eqref{eq:sync_model}:
\begin{align}
    y_{d, \nu} &= s_{d}(\mathbf{\hat{p}})  f_\nu(\beta_d(\mathbf{\hat{p}}), T_d(\mathbf{\hat{p}})),\\
    y_{s, \nu} &= s_{s}(\mathbf{\hat{p}})  g_\nu(\beta_s(\mathbf{\hat{p}})).
\end{align}
Similarly to the moment expansion framework~\cite{Chluba_2017, Vacher_2023}, we can write the spatially varying spectral parameters as:
\begin{equation}
\beta_d(\mathbf{\hat{p}})
=
\bar{\beta_d}
+
\delta\beta_d(\mathbf{\hat{p}}),
\qquad
T_d(\mathbf{\hat{p}})
=
\bar{T_d}
+
\delta T_d(\mathbf{\hat{p}}),
\qquad
\beta_s(\mathbf{\hat{p}})
=
\bar{\beta_s}
+
\delta\beta_s(\mathbf{\hat{p}}),
\end{equation}
and perform a first-order Taylor expansion around the mean values:
\begin{align}
y_{\mathrm{fgs},\nu}
&\simeq
f_\nu(\bar{\beta}_d, \bar{T}_d) s_d(\mathbf{\hat{p}})
+
\left.\frac{\partial f_\nu}{\partial\beta_d}\right|_{\bar{\beta}_d}
(s_d\delta\beta_d)(\mathbf{\hat{p}})
+
\left.\frac{\partial f_\nu}{\partial T_d}\right|_{\bar{T}_d}
(s_d\delta T_d)(\mathbf{\hat{p}})\nonumber \\
&+
g_\nu(\bar{\beta}_s) s_s(\mathbf{\hat{p}}) 
+ 
\left.\frac{\partial g_\nu}{\partial\beta_s}\right|_{\bar{\beta}_s}
(s_s\delta\beta_s)(\mathbf{\hat{p}}).
\label{eq:moment_expansion}
\end{align}
If each of these terms is treated as a separate ``component'', the harmonic foreground covariance in SMICA is simply  $\mathbf{R}_{\mathbf{f},\ell}=\mathbf{A_f}\mathbf{P}_{\mathbf{f},\ell}\mathbf{A_f}^T$,
where the stationary foreground mixing matrix is given by
\begin{equation}
\mathbf{A}_f^\nu =
\begin{pmatrix}
f_\nu(\bar{\beta}_d, \bar{T}_d)  &
\left.\dfrac{\partial f_\nu}{\partial\beta_d}\right|_{\bar{\beta}_d} &
\left.\dfrac{\partial f_\nu}{\partial T_d}\right|_{\bar{T}_d} &
g_\nu(\bar{\beta}_s)  &
\left.\dfrac{\partial g_\nu}{\partial\beta_s}\right|_{\bar{\beta}_s}
\end{pmatrix},
\end{equation}
and corresponding source covariance is given by
\begin{equation}
\mathbf{P}_{\mathbf{f},\ell} = 
\mathcal{C}_\ell
\left[
\begin{pmatrix}
s_d \\
s_d\delta\beta_d \\
s_d\delta T_d \\
s_s \\
s_s\delta\beta_s
\end{pmatrix}
\begin{pmatrix}
s_d^\dagger &
(s_d\delta\beta_d)^\dagger &
(s_d\delta T_d)^\dagger &
s_s^\dagger &
(s_s\delta\beta_s)^\dagger
\end{pmatrix}
\right].
\end{equation}
From this perspective, each additional moment introduces an additional degree of freedom in both the mixing matrix and the foreground covariance, providing a possible explanation for why increasing the number of foreground components in SMICA can improve its performance. Interestingly, when using five components, corresponding to the total number of moments up to first order of the canonical foreground SEDs, we recover an unbiased estimate of the tensor-to-scalar ratio. \\
\begin{figure}
    \centering
    \includegraphics[width=0.49\linewidth]{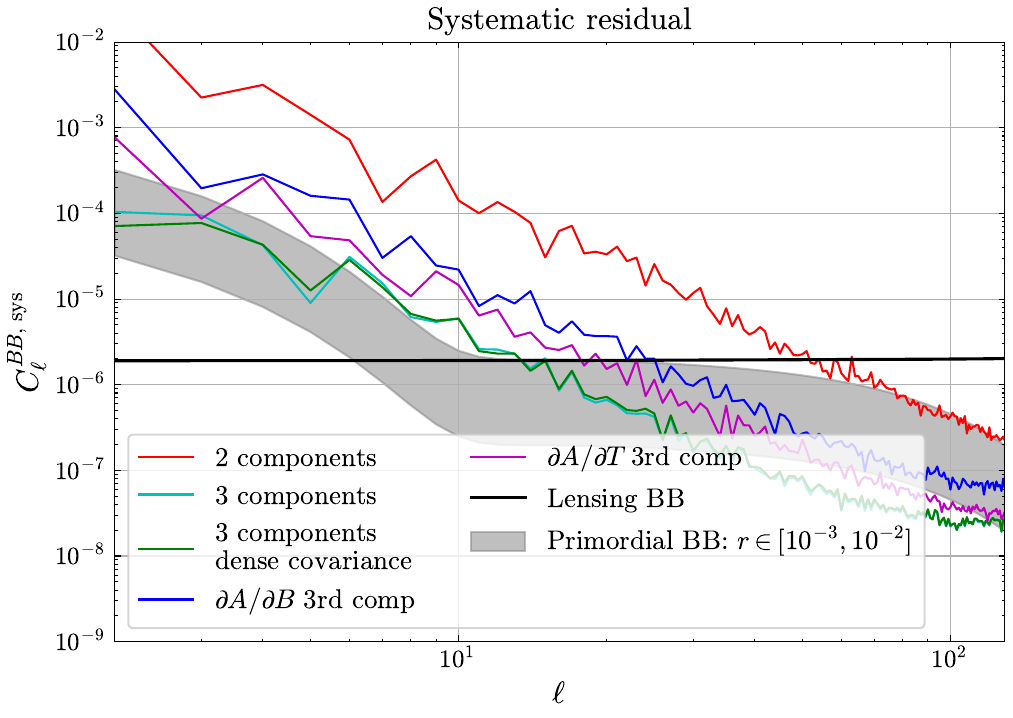}
    \includegraphics[width=0.49\linewidth]{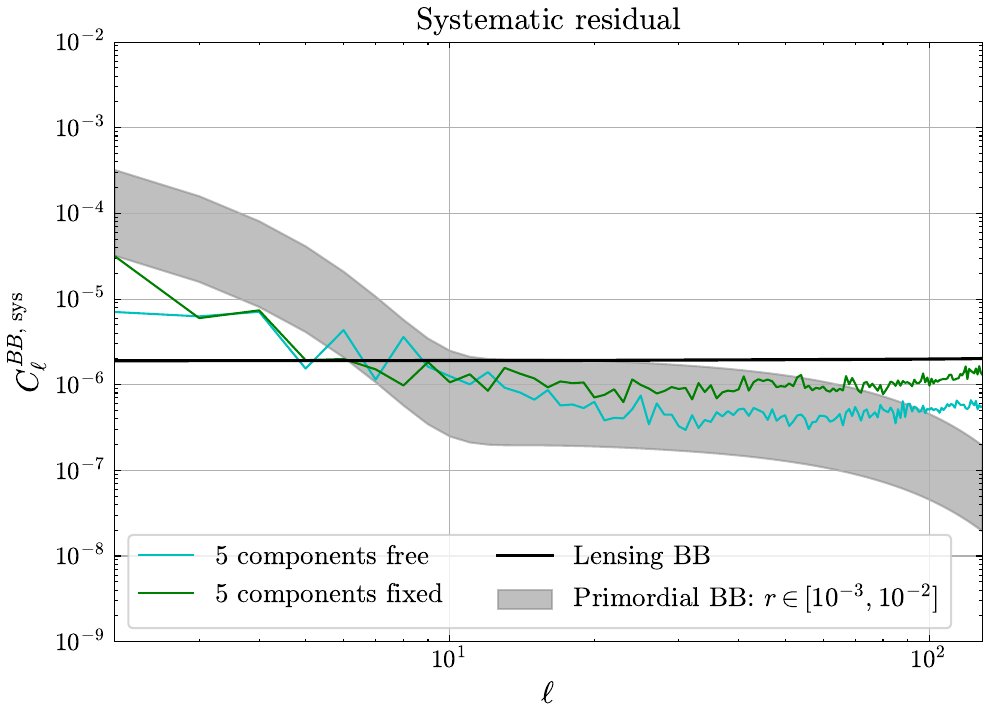}
    \caption{Comparison of the systematic residual spectra for the different configurations considered in the three- and five-component tests. The left panel shows the three-component comparison, while the right panel shows the five-component comparison. The lensing $B$-mode spectrum and primordial $B$-mode spectra for $r\in[10^{-3},10^{-2}]$ are shown as references.}\label{fig:d1s1_moment_exp_compare}
\end{figure}
\indent We implement a simple test to investigate the origin of this additional effective dimensionality. First, we compare four three-component configurations: a free third component with diagonal foreground covariance (used as a baseline in this paper), the same configuration with a dense covariance (allowing for cross correlations), and two cases in which the third component is fixed to either $\left.\partial f_\nu/\partial\beta_d\right|_{\bar{\beta}_d}$ or $\left.\partial f_\nu/\partial T_d\right|_{\bar{T}_d}$. These configurations test whether the additional degree of freedom identified by SMICA can be associated with the first-order moment expansion induced by spatial variations of the dust spectral parameters, and whether modelling correlations between the components play an important role. We then perform a second comparison using five components, contrasting the fully free configuration with one in which the three additional components are fixed to $\left.\partial f_\nu/\partial\beta_d\right|_{\bar{\beta}_d}$, $\left.\partial f_\nu/\partial T_d\right|_{\bar{T}_d}$, and $\left.\partial f_\nu/\partial\beta_s\right|_{\bar{\beta}_s}$, respectively. In this configuration, the three additional degrees of freedom exhaust the first-order contributions associated with spatial variations of the dust and synchrotron spectral parameters. The resulting systematic residual spectra are shown in Figure~\ref{fig:d1s1_moment_exp_compare}. Although the normalization and detailed form of the fixed laws depend on the choice of reference frequency $\nu_0$, we have verified that this choice has little impact on the resulting performance.\\
\indent For both fixed mixing laws, introducing the additional component noticeably reduces the systematic residual relative to the two-component model, confirming that part of the increased effective dimensionality can be associated with the spectral degrees of freedom generated by spatial variations in the dust parameters. However, neither fixed component performs as well as the freely fitted third component. This is expected, since SMICA does not impose a physical parameterization of the foreground variability, but instead identifies the spectral subspace that best describes the observed covariance. The optimal direction therefore need not coincide with either $\partial f_\nu/\partial\beta_d$ or $\partial f_\nu/\partial T_d$ individually, but can combine these and other correlated spectral variations. This is also consistent with the small difference between the diagonal and dense covariance cases: once the mixing matrix is free to adapt to the data, SMICA can already absorb correlated moment contributions into an appropriate spectral basis, leaving limited additional benefit from explicitly modelling their correlations.\\
\indent For the five-component comparison, fixing the three additional components to the $f_\nu$ derivatives produces systematic residuals similar to those obtained when all five components are left free. The agreement is particularly strong at large angular scales. Together with the results of Section~\ref{subsec:ext_comps}, where this five-component configuration provides an effectively unbiased estimate of $r$ despite its larger statistical uncertainty, this close agreement supports the interpretation of the additional SMICA components as effective first-order spectral moments associated with spatial variations of the physical foreground laws.
\section{Conclusion}\label{sec:conclusions}
Measuring the primordial CMB $B$-mode polarization requires component-separation methods capable of suppressing polarized Galactic foregrounds to an unprecedented level. In this work, we presented a complete component-separation framework based on Spectral Matching Independent Component Analysis (SMICA) for estimating the tensor-to-scalar ratio from multi-frequency polarization observations on a large sky fraction, and introduced two complementary strategies to address the limitations of its standard implementation.\\
\indent First, we investigated the effective number of foreground components required by SMICA to describe realistic polarized dust and synchrotron emission using the Planck sky model. A data-driven procedure probing the covariance structure of the foreground consistently identifies three dominant components beyond the physical dust and synchrotron emission. Extending the SMICA foreground model accordingly, we found that increasing the number of components reduces the systematic bias but rapidly degrades statistical sensitivity, demonstrating a fundamental trade-off between modelling flexibility and the uncertainty on the recovered cosmological parameters. The three-component model offers the best overall compromise, consistent with the covariance analysis. We further provided a physical motivation for this increased effective dimensionality, showing that the extra components can be associated with a first-order moment expansion of the spatial variations in the dust and synchrotron spectral parameters.\\
\indent Second, to better capture the spatial variability of the foreground spectral energy distributions, we implemented a version of SMICA in which the model parameters are estimated independently on subsets of the sky, defined by HEALPix pixels of $N_{\mathrm{side}} = 2$ and $4$, before being re-projected into pixel space through the Generalized Least Squares weights. This localized approach substantially reduces foreground residuals compared to the full-sky implementation, but introduces an ILC bias. We devise and implement a mitigation strategy, denoted modified-HEALPix, by estimating the covariance of a given subset from a surrounding region rather than the subset itself. Although this correction leads to some loss in component-separation performance, the pipeline still achieves a substantial reduction in residual foreground contamination relative to previous cases, particularly on the largest angular scales.\\
\indent Beyond the component-separation stage, we showed that further gains can be achieved through other optimizations. A post-separation mask tailored to the component-separation residual can help retain more of the sky while reducing more contamination. At the likelihood level, marginalizing over internally reconstructed foreground templates, which avoids the need for an external foreground reconstruction pipeline, consistently reduces the residual bias. Combined with the modified-HEALPix pipeline, this brings the recovered tensor-to-scalar ratio to unbiased levels, while maintaining a statistical uncertainty of a few $10^{-3}$ or below for simulated LiteBIRD-like observations implementing complex foreground models.\\
\indent Overall, these results demonstrate that the proposed pipeline offers a promising framework for measuring the tensor-to-scalar ratio in the presence of realistic polarized Galactic foregrounds, with successive reduction of residual contamination while maintaining competitive statistical sensitivity. Several aspects of the pipeline could nonetheless be further improved: the choice of HEALPix pixel subsets is arbitrary and could be optimized to better match Galactic morphology, while the treatment of power-spectrum estimation and the cosmological likelihood remains approximate.\\
\indent Lastly, we discussed how many of the complications can, in fact, be understood as consequences of departing from a fully maximum-likelihood treatment. These are methodological biases that make the overall procedure statistically sub-optimal and limit our ability to push component separation further. Motivated by this, we are developing a new framework that retains the full philosophy of SMICA while subjecting it to a global maximum-likelihood analysis, built on the MICS~\cite{Leloup_2023} formalism and its implementation MICMAC~\cite{Morshed_2024}. This forthcoming pipeline aims to perform component separation and cosmological inference within a single statistically consistent framework, eliminating many of the inaccuracies and limitations present in the current approach.

\acknowledgments
The authors would like to thank Cl\'{e}ment Leloup, Alessandro Carones, Magdy Morshed, L\'{e}o Vacher, Wassim Kabalan, Josquin Errard and Gilles Weymann-Despres for many interesting discussions and helpful inputs on the work.\\
\indent This work has received support from France 2030 through the project named Académie Spatiale d’Île-de-France (https://academiespatiale.fr/) managed by the National Research Agency under bearing the reference ANR-23-CMAS-0041. \\
\indent The project was done under funding by the CMB-Inflate program: the European Union’s Horizon 2020 research and innovation program under the Marie Skłodowska-Curie grant agreement No 101007633.\\
\indent Hoang Viet Tran and Benjamin Beringue acknowledge support from the SciPol project (scipol.in2p3.fr), funded by the European Research Council (ERC) under the European Union’s Horizon 2020 research and innovation program (Principal Investigator: Josquin Errard, Grant agreement No. 101044073).\\
\indent We acknowledge the IN2P3 Computer Centre (https://cc.in2p3.fr)
for providing the computing resources and services needed for the analysis. \\
\indent The authors acknowledge the use of OpenAI’s ChatGPT to assist with improving the clarity, organization, and language of selected portions of the text. The authors critically reviewed and revised all AI-assisted content and take full responsibility for the accuracy, integrity, and originality of the final manuscript.

\appendix
\section{Foreground complexity diagnostic in the case of unknown CMB covariance}\label{appendix:a}
\begin{figure}
    \centering
    \includegraphics[width=1\linewidth]{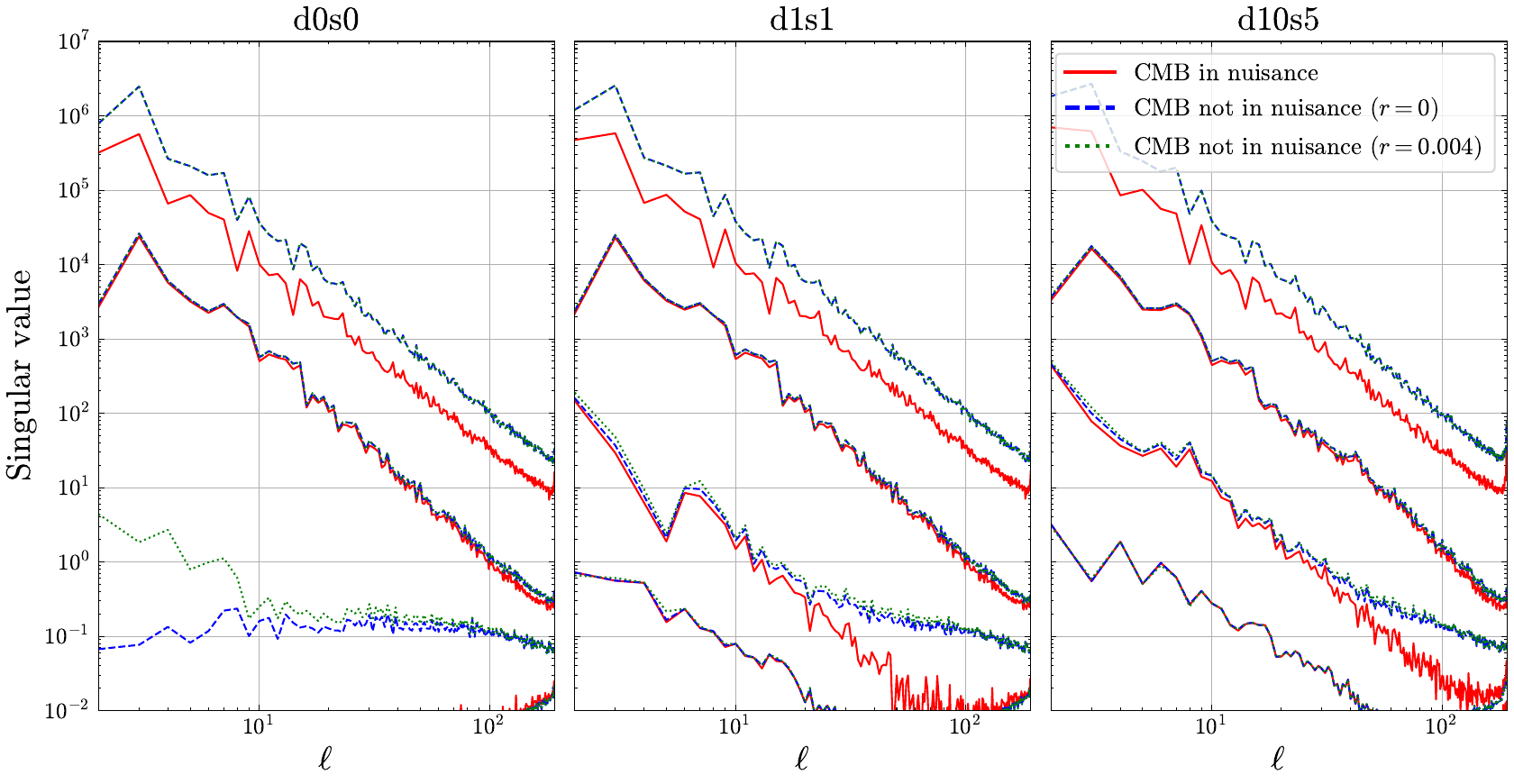}
    \caption{Singular values from performing decomposition on the realistic whitened covariance $\mathbf{\tilde{R}_Y^{real}}$. The curves are plotted with an offset of 1 to highlight the structures close to unity.}
    \label{fig:svd_cmb}
\end{figure}

\indent As noted in Section~\ref{subsec:svd}, the previous procedure assumes knowledge of the CMB covariance, which is unavailable for real data. In practice, we therefore whiten only by the noise covariance,
\begin{align}
    \mathbf{\widetilde{R}_Y^{real}}
    &=
    \mathbf{R_N}^{-1/2}\mathbf{R_Y}\mathbf{R_N}^{-1/2} \nonumber\\
    &=
    \mathbf{R_N}^{-1/2}(\mathbf{R_f}+\mathbf{R_C})
    \mathbf{R_N}^{-1/2}+\mathbb{I},
    \label{eq:real_whit_cov}
\end{align}
and perform the same SVD. Figure~\ref{fig:svd_cmb} shows the resulting singular-value spectra for $r=0$ and $r=0.004$.\\
\indent For \texttt{d0s0}, the inclusion of the CMB produces one additional significant mode, while the two dominant foreground modes remain essentially unchanged. For \texttt{d1s1} and \texttt{d10s5}, the dominant and weakest modes remain largely unchanged, while the intermediate modes are substantially affected at high multipoles, where the CMB contribution is significant. Rather than producing a distinct additional mode, the CMB power appears to be absorbed into the existing intermediate modes. Therefore, in these more complex cases, the number of significant SVD modes should not be interpreted simply as the number of foreground components plus one CMB component. Nevertheless, the SVD provides a useful first indication of the dimensionality to consider for the subsequent SMICA model, which must then be refined through iterative tests and combined with other strategies.

\end{document}